\documentclass[nofootinbib]{revtex4}
\usepackage{amsmath}
\usepackage{amsfonts}
\usepackage{amssymb}
\usepackage{graphicx}
\newcommand{\Dslash}[1] { \setbox0=\hbox{$#1$}     
\dimen0=\wd0   \setbox1=\hbox{/} \dimen1=\wd1  \ifdim\dimen0>\dimen1        
 \rlap{\hbox to \dimen0{\hfil/\hfil}}  #1 \else \rlap{\hbox to \dimen1{\hfil$#1$\hfil}}  /  \fi  }
\newcommand{\Dn}{\Dslash{n} }
\newcommand{\Dbn}{\Dslash{\bar n}}
\newcommand{\ns}{\Dslash{n}}
\newcommand {\nbs}{\Dslash{\bar n}}
\newcommand{\nbn}{\frac{\nbs\ns}{4}}

\begin{document}

\noindent
MKPH-T-10-28\\
HIM-2010-05
\bigskip

\title{
Soft spectator scattering in the nucleon form factors at large $Q^2$ within the SCET approach
}

\author{Nikolai Kivel}
\affiliation{Institut f\"ur Kernphysik, Johannes Gutenberg-Universit\"at, D-55099 Mainz, Germany} 
\affiliation{Helmholtz Institut Mainz, Johannes Gutenberg-Universit\"at, D-55099 Mainz, Germany} 
\affiliation{Petersburg Nuclear Physics Institute, Gatchina, 188350, Russia}
\author{Marc Vanderhaeghen}
\affiliation{Institut f\"ur Kernphysik, Johannes Gutenberg-Universit\"at, D-55099 Mainz, Germany}

\begin{abstract}
The proton form factors at large momentum transfer are dominated  by  two contributions which are associated 
with the hard and soft rescattering respectively. Motivated by a very active experimental form factor 
program at intermediate values of momentum transfers,  $Q^{2}\sim 5-15~\text{GeV}^{2}$, 
where an understanding in terms of only a hard rescattering mechanism cannot yet be expected, 
we investigate in this work the soft rescattering contribution using
 soft collinear effective theory (SCET). 
 Within such a description, the form factor is characterized, besides the hard scale $Q^2$, by a 
 hard-collinear scale $Q \Lambda$,  which arises due to the presence of soft spectators, 
 with virtuality $\Lambda^2$ ($\Lambda \sim 0.5$~GeV), such that   
 $Q^{2}\gg Q\Lambda\gg \Lambda^{2}$.  We show that in this case a two-step factorization can be successfully carried out  using  the SCET approach.
  In a first step (SCET$_I$), we perform the 
leading-order matching of the  QCD electromagnetic current onto the relevant SCET$_I$ operators  and 
perform a resummation of large logarithms using renormalization group equations. 
 We then discuss the further matching onto a SCET$_{II}$ framework, 
 and propose the  factorization formula (accurate to leading logarithmic approximation) for the Dirac form 
 factor, accounting for both hard and soft contributions.  
We also present a qualitative discussion of the phenomenological consequences 
 of this new framework.

\end{abstract}

\date{\today}

\maketitle

\section{Introduction}

The study of the nucleon form factors (FFs) is one of the central topics in 
hadronic physics ( 
for recent reviews see, {\it e.g.} , 
Refs.~\cite{HydeWright:2004gh,Arrington:2006zm,Perdrisat:2006hj}). 
Substantial progress has been achieved in this field over the 
past decade, mainly thanks to new experimental methods, using polarization observables, 
which allow for  precise measurements of the FFs.  The results for the proton FFs, obtained 
over the past few years at JLab \cite{Jones00, Punjabi:2005wq, Gayou02, Puckett:2010ac} 
up to a momentum transfer $Q^2 \simeq 8.5$~GeV$^2$, considerably
boosted our knowledge about the distribution of the electric charge inside the proton. 
A substantial program to extend the measurements of the nucleon FFs up to 
$Q^2 \simeq 15$~GeV$^2$ in the spacelike region will be performed in the near future at the 
JLab $12$~GeV upgrade.
In parallel, the PANDA Collaboration at GSI is planning to carry
out precise measurements of the proton FFs at large timelike
momentum transfers, up to around 20 GeV$^2$,  
using the annihilation process $p+\bar{p}\rightarrow e^{+} + e^{-}$
\cite{Sudol:2009vc}. These experiments will provide us with precious information
on the FF behaviors in the region of large momentum transfers.

On the theory side, an understanding of the nucleon FFs at large momentum transfers,
both spacelike and timelike, from the underlying QCD dynamics, still remains a challenge. 
At present, the FF behavior for moderate and large values of $Q^{2}$ is still not well understood and
an adequate description, allowing for quantitative predictions, is absent.

\begin{figure}[th]
\begin{center}
\includegraphics[height=1.0784in,width=2.0539in]{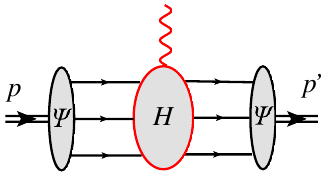}
\end{center}
\caption{Reduced diagram describing the hard scattering picture}%
\label{hs-reduced-diagram}%
\end{figure}

The leading power behavior of the FFs was studied a long time ago using
the QCD factorization approach ( see, e.g., \cite{Chernyak:1983ej, Lepage:1979zb} and
references therein). It was established that the dominant contribution can be
represented by a reduced diagram as shown in Fig.\ref{hs-reduced-diagram}.
In this figure, the hard blob describes the hard scattering of quarks
and gluons with virtualities of order $Q^{2}$. Such a hard subprocess can be
systematically computed in perturbative QCD (pQCD) order-by-order. The soft blobs, denoted by
$\mathbf{\ \Psi}$, describe the soft, nonperturbative subprocesses, and can
be parametrized in terms of universal matrix elements known as distribution
amplitudes (DAs). Such a picture suggests the well known factorization formula
for the Dirac FF%
\begin{equation}
F_{1}=\int dx_{i}~\int dy_{i}~\mathbf{\Psi}(x_{i})~\mathbf{H}(x_{i}%
,y_{i}|Q)~\mathbf{\ \Psi} (y_{i})\equiv\mathbf{\Psi}*\mathbf{H}*\mathbf{\Psi
},\label{F1:fact}%
\end{equation}
and predicts the scaling behavior
\begin{equation}
F_{1}\sim\frac{\Lambda^{4}}{Q^{4}}\times\left[  \ln Q/\Lambda\right]
^{\gamma},
\end{equation}
where the logarithmical corrections can be systematically computed order-by-order.

Unfortunately, for the Pauli FF $F_2$ this approach cannot provide such a systematic
picture and suggests only the power estimate%
\begin{equation}
F_{2}\sim\frac{\Lambda^{6}}{Q^{6}}.\label{F2:as}%
\end{equation}

Almost simultaneously, it was found that the picture described by
Fig.\ref{hs-reduced-diagram} is not complete. In Ref.~\cite{Dun1980} it was
demonstrated that the exchange of soft quarks between initial and final states may
also produce contribution of order $1/Q^{4}$ times logarithms. In
Refs.~\cite{Fadin1981,Fadin1982} all such contributions were computed with the
leading logarithmic accuracy at 2 and 3 loops. Using these results 
it was assumed~\cite{Fadin1982} that these ``nonrenormalization'' logarithms
probably can be resummed to all orders into an exponent similar to the well known
Sudakov logarithms \cite{Sudakov:1954sw}. However, this effect was ignored in
many later publications. In particular, in Ref.~\cite{Lepage:1979zb} it was
suggested that such contributions could be strongly suppressed due to those
Sudakov logarithms and therefore can be ignored at large values of $Q^{2}$.

At the same time, many phenomenological studies of the hard rescattering
picture support the conclusion that in the region of moderate $Q^{2}\simeq
5-10~$GeV$^{2}$ the factorization approach expressed by Eq.~(\ref{F1:fact}) cannot describe the
data properly ( see, for instance, \cite{Bolz:1994hb}). 
Moreover, existing data for the FF ratio $F_{2}/F_{1}$ measured up to
$Q^{2}=8.5~$GeV$^{2}$ \cite{Puckett:2010ac} also do not support the expectation
of Eq.~(\ref{F2:as}) which assumes that in the asymptotic region $Q^{2}F_{2}%
/F_{1}\sim const.$ Therefore, it was suggested that the so-called Feynman
mechanism~\cite{Feynman:1973xc}, associated with the scattering of the hard 
virtual photon off one active quark, dominates the nucleon FFs at moderate
values of $Q^{2}$. The other spectators remain soft and therefore very often
such scattering is associated with the soft overlap of the nucleon wave functions.

Such a picture is supported by different phenomenological approaches, such as 
QCD-motivated models for the hadronic wave functions \cite{Isgur:1984jm,
Isgur:1988iw, Isgur:1989cy}, QCD sum rules \cite{Ioffe:1982qb,
Nesterenko:1982gc} and light-cone sum rules \cite{Braun:2001tj, Braun:2006hz}.
The Sudakov suppression in this case is always assumed to be relatively
small. The aim of the present work is to develop a systematic approach for the
specific soft contribution described first in Ref.~\cite{Dun1980}, and 
formulate it through a factorization theorem. We apply the effective theory
approach, known as soft collinear effective theory (SCET), 
 in order to describe contributions from different regions of virtualities in the diagrams. 
 
 %%%PRDv2
 
The effective theory is a very convenient tool in this case because soft rescattering is characterized
by subprocesses which exhibit different scales~: 
a hard rescattering involving particles with momenta of order $Q^{2}$,
hard-collinear scattering processes with virtualities of order $\Lambda Q$, 
and soft nonperturbative modes with momenta of order $\Lambda^{2}$.
Therefore one has to perform a two-step matching procedure 
in order to perform full factorization of such a process.   

\begin{figure}[th]
\begin{center}
\includegraphics[height=1.5956in]{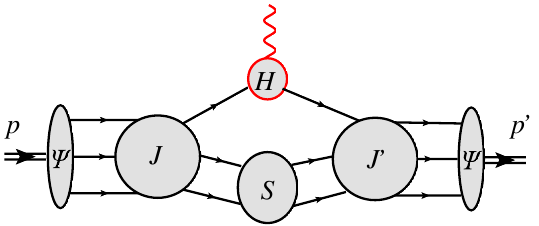}
\end{center}
\caption{Interpretation of the soft rescattering as a reduced diagram}%
\label{s-reduced-diagram}%
\end{figure}

Following this scheme we  obtain that the full description of  large-$Q^{2}$ behavior of the nucleon FF $F_{1}$  is given by the sum 
of two contributions associated with the soft  and hard  rescattering picture:
\begin{equation}
F_{1}\simeq F_{1}^{(s)}+F_{1}^{(h)}.%
\end{equation}
The hard rescattering part $F_{1}^{(h)}$ is well known and described by (\ref{F1:fact}). 
One can expect that the soft contribution can also be presented in a factorized form but with the more 
complicated structure reflecting the presence of different scales. Performing the leading logarithmic analysis of the leading power  contribution ($\sim 1/Q^{4}$) we 
demonstrate in this work that  the corresponding soft term can be presented in the following form,
\begin{eqnarray}
F_{1}^{(s)} \simeq  
H(Q)
\int Dy_{i}\mathbf{\Psi}(y_{i})
 \int_{0}^{\infty}d\omega_{1}d\omega_{2}~\mathbf{J}^{\prime}(y_{i},\omega_{i}Q)
%\nonumber\\ & & \times
 \int Dx_{i}\mathbf{\Psi}(x_{i}) 
\int_{0}^{\infty}d\nu_{1}d\nu_{2}~\mathbf{J}(x_{i},\nu_{i}Q)\boldsymbol{S}(\omega_{i},\nu_{i}),
\label{F1s:int}%
\end{eqnarray}
which can be interpreted in  terms of a reduced diagram as in Fig.\ref{s-reduced-diagram}. 
This result involves a hard coefficient function  $H$, and two hard-collinear jet functions $\mathbf{J}$   and $\mathbf{J}^\prime$   which can be computed in pQCD. 
They describe the subprocesses with hard momenta and hard collinear momenta respectively. 
The nonperturbative functions $\mathbf{\Psi}$ and $\boldsymbol{S}$ describe the scattering of collinear and soft modes.   
The convolution integrals in Eq.~(\ref{F1s:int})
are performed with respect to the collinear fractions $x_{i}$ and $y_{i}$, and 
with respect to the soft spectator fractions $\omega_{i}, \nu_{i}\sim \Lambda$.   

In the case of the Pauli FF $F_{2}$, 
we can  also perform a factorization of  the soft-overlap contribution but only partially, 
separating  the hard modes with momenta of order $Q^{2}$. 
The full factorization is problematic due to overlapping 
integration regions corresponding with soft and collinear contributions, which lead to well known 
end-point singularities in the convolution integrals. 
However, such a partial result can be used to carry out a phenomenological analysis of the FFs in the region of intermediate $Q^{2}$ values. 
Such a region corresponds to momentum transfers 
where $Q^{2}$ is large enough, allowing us to perform a power expansion, but where the second, 
 hard-collinear  scale $\sim \Lambda Q$ is still relatively small, so that one expects 
 the dominance of the leading power asymptotic term. 
Such a situation may indeed be relevant to interpret existing data and planned experiments. 

The specific feature of the factorization for the soft-overlap contribution is the presence of the Sudakov logarithms which can be ressummed
using the renormalization group in effective field theory. 
 It was suggested, see e.g. \cite{Radyushkin00} , that these logarithms  could play an important role in the timelike region
providing an enhancement of the timelike FFs compared to the spacelike region 
(the so-called $K$ factor). Within the factorization picture  such an enhancement  can
be clearly studied  in a model independent way.       
  
 %%%
 
Our paper is organized as follows. In Sec.~\ref{sec2} , we consider as
an example the analysis of the dominant regions for certain Feynman diagrams
and demonstrate the existence of the soft spectator contribution at leading
power (in the hard scale $Q$) for both the Dirac and Pauli FFs. In Sec.~\ref{sec3}, we
discuss the factorization scheme for such contributions, perform the leading-order 
matching between QCD and the SCET, and perform a resummation of   large logarithms. 
In Sec.~\ref{sec4}
%%%PRDv2 
%, we demonstrate how the results from Section~\ref{sec2} are reproduced in SCET and 
we discuss the SCET power counting in $1/Q$ and 
derive the factorization formula (\ref{F1s:int}).   
%%%
In Sec.~\ref{sec5}, 
we perform a first qualitative discussion of the phenomenological consequences following 
from our results. In Sec.~\ref{sec6}, we summarize our findings.

\section{Soft rescattering mechanism: examples}
\label{sec2}

In this section we consider specific examples of soft rescattering contributions. 
For the Dirac FF our analysis overlaps
with results of the work of Ref.~\cite{Dun1980}, whereas for the Pauli FF 
this is discussed here for the first time.

In our consideration we use the Breit frame
\begin{equation}
q=p^{\prime}-p=Q\left(  \frac{n}{2}-\frac{\bar{n}}{2}\right)
,~\ n=(1,0,0,-1),~\bar{n}=(1,0,0,1),~\ (n\bar{n})=2,~
\end{equation}
and define the external momenta as%
\begin{equation}
p=\mathcal{Q}\frac{\bar{n}}{2}+\frac{m_{N}^{2}}{\mathcal{Q}}\frac{n}%
{2},~\ \ p^{\prime}=\mathcal{Q}\frac{n}{2}+\frac{m_{N}^{2}}{\mathcal{Q}}%
\frac{\bar{n}}{2},~\ \ \mathcal{Q=}Q\frac{1}{2}\left[  1+\sqrt{1+\frac
{4m_{N}^{2}}{Q^{2}}}\right]  =Q+\mathcal{O}(m_{N}^{2}/Q^{2}),
\end{equation}%
\begin{equation}
~\ 2(pp^{\prime})=\mathcal{Q}^{2}+\frac{m_{N}^{4}}{\mathcal{Q}^{2}}\approx
Q^{2}, 
\end{equation}
where $m_{N}$ is the nucleon mass. 
For the incoming and outgoing collinear quarks we always imply%
\begin{equation}
p_{i}=x_{i}\mathcal{Q}\frac{\bar{n}}{2}+p_{\bot i}+\left(  x_{i}^{\prime
}~\frac{m_{N}^{2}}{\mathcal{Q}}\right)  \frac{n}{2},~\ \ ~p_{i}^{\prime}%
=y_{i}\mathcal{Q}\frac{n}{2}+p_{\bot i}^{\prime}+\left(  y_{i}^{\prime}%
~\frac{m_{N}^{2}}{\mathcal{Q}}\right)  \frac{\bar{n}}{2}%
,\ \ \ \label{quark mom}%
\end{equation}
with the transverse momenta%
\[
p_{\bot}^{2}\sim p_{\bot}^{\prime2}\sim\Lambda^{2},
\]
and where $x_i$ and $x^\prime_i$ denote fractions of the corresponding momentum component.  
%%PRDv2
In what follow we shall use the convenient notation   $\bar x_{i}=1-x_{i}$. 
%%%
We also use the following notation for scalar products%
\begin{equation}
(a\cdot n)\equiv a_{+}~,~(a\cdot\bar{n})\equiv a_{-}\ .
\end{equation}
and Dirac contractions
\begin{equation}
p_{\mu}\gamma^{\mu}\equiv
\setbox0=\hbox{$p$}\dimen0=\wd0\setbox1=\hbox{/}\dimen1=\wd1\ifdim\dimen0>\dimen1\rlap{\hbox to \dimen0{\hfil/\hfil}}p\else\rlap{\hbox to \dimen1{\hfil$p$\hfil}}/\fi\equiv
\hat{p}.
\end{equation}

%%%PRDv2
Nucleon FFs are defined as the matrix elements of the electromagnetic (e.m.) current
between the nucleon states:%
\begin{equation}
\left\langle p^{\prime}\right\vert J_{e.m.}^{\mu}(0)\left\vert p\right\rangle
=\bar{N}(p^{\prime})\left[  \gamma^{\mu}(F_{1}+F_{2})-\frac{(p+p^{\prime
})^{\mu}}{2m_{N}}~F_{2}\right]  N(p),\label{FF:def}%
\end{equation}
with nucleon spinors normalized  as $\bar N N=2m_{N}$. We also use a standard normalization for particle states:
\begin{equation}
\langle p',s' |p,s\rangle=(2\pi)^{3}\, 2E\, \delta_{ss'}\delta(\vec p-\vec p^{\, \prime}).
\label{norm}
\end{equation}
%%%
In what follows, we shall compute the Feynman diagrams which provide
contributions to the nucleon FFs. The component of interest for our 
calculations is the soft matrix element describing the overlap of the partonic
configurations with the hadron state. In the case of the FFs such
overlap is described by DAs. In the case of the
nucleon, the corresponding leading twist DAs can be defined as%
\begin{equation}
~4\left\langle 0\left\vert u_{\alpha}W^{i}[\lambda_{1}n]u_{\beta}W^{j}%
[\lambda_{2}n]d_{\sigma}W^{k}[\lambda_{3}n]\right\vert p\right\rangle
=\frac{\varepsilon^{ijk}}{3!}\int Dx_{i}~e^{-ip_{+}\left(  \sum x_{i}%
\lambda_{i}\right)  }\mathbf{\Psi}(x_{i}),\label{DA:def}%
\end{equation}
where
\begin{equation}
q_{\alpha}W[x]\equiv q_{\alpha}(x)\text{P}\exp\left\{  ig\int_{-\infty}%
^{0}dt~(n\cdot A)(x+tn)\right\}  ,
\end{equation}
and the measure reads$\ ~Dx_{i}=dx_{1}dx_{2}dx_{3}\delta(1-x_{1}-x_{2}%
-x_{3}).$ The function $\mathbf{\Psi}(x_{i})$ can be further decomposed as
$\ \ $
\begin{align}
~\ \mathbf{\Psi}(x_{i}) &  =V(x_{i})~p_{+}\left[  {\scriptstyle\frac{1}{2}%
}%
\setbox0=\hbox{$\bar n$}\dimen0=\wd0\setbox1=\hbox{/}\dimen1=\wd1\ifdim\dimen0>\dimen1\rlap{\hbox to \dimen0{\hfil/\hfil}}\bar
{n}\else\rlap{\hbox to \dimen1{\hfil$\bar n$\hfil}}/\fi~C\right]
_{\alpha\beta}\left[  \gamma_{5}N^{+}\right]  _{\sigma}+A(x_{i})~p_{+}\left[
{\scriptstyle\frac{1}{2}}%
\setbox0=\hbox{$\bar n$}\dimen0=\wd0\setbox1=\hbox{/}\dimen1=\wd1\ifdim\dimen0>\dimen1\rlap{\hbox to \dimen0{\hfil/\hfil}}\bar
{n}\else\rlap{\hbox to \dimen1{\hfil$\bar n$\hfil}}/\fi\gamma_{5}C\right]
_{\alpha\beta}\left[  N^{+}\right]  _{\sigma}\nonumber\\
&  ~\ \ \ \ \ \ \ \ \ \ \ \ \ \ \ \ \ \ \ \ \ \ \ \ \ \ \ \ +T(x_{i}%
)~p_{+}\left[  {\scriptstyle\frac{1}{2}}%
\setbox0=\hbox{$\bar n$}\dimen0=\wd0\setbox1=\hbox{/}\dimen1=\wd1\ifdim\dimen0>\dimen1\rlap{\hbox to
\dimen0{\hfil/\hfil}}\bar{n}\else\rlap{\hbox to \dimen1{\hfil$\bar
n$\hfil}}/\fi\gamma_{\bot}~C\right]  _{\alpha\beta}\left[  \gamma^{\bot}%
\gamma_{5}N^{+}\right]  _{\sigma},\label{Psi:def}%
\end{align}
The large component $N^{+}$ of the nucleon spinor is defined as
\begin{equation}
N^{+}=\frac
{\setbox0=\hbox{$\bar n$}\dimen0=\wd0\setbox1=\hbox{/}\dimen1=\wd1\ifdim\dimen0>\dimen1\rlap{\hbox to \dimen0{\hfil/\hfil}}\bar
{n}%
\else\rlap{\hbox to \dimen1{\hfil$\bar n$\hfil}}/\fi\setbox0=\hbox{$n$}\dimen0=\wd0\setbox1=\hbox{/}\dimen1=\wd1\ifdim\dimen0>\dimen1\rlap{\hbox to \dimen0{\hfil/\hfil}}n\else\rlap{\hbox to
\dimen1{\hfil$n$\hfil}}/\fi}{4}N,
\end{equation}
and $C$ is the charge conjugate matrix $C:~C^{-1}\gamma_{\mu}C=-\gamma_{\mu}^{T}.$
The nucleon DA $\mathbf{\Psi}(x_{i})$ is shown by the soft blobs in
Fig.\ref{hs-reduced-diagram}. \ For simplicity we restrict our consideration
to the proton state. In what follows we always assume that in pQCD diagrams the
first and second top lines correspond to $u$ quarks. \ Assuming projections on
the leading twist DAs, we can considerably simplify certain considerations
substituting instead of DAs on-shell quark spinors. Such substitution is
possible because at leading-order (LO)  power accuracy we can neglect the small components in
the external quark momenta:%
\begin{equation}
p\simeq Q\frac{\bar{n}}{2},~p_{i}\simeq x_{i}p,~\ p^{\prime}\simeq Q\frac
{n}{2},\ ~p_{i}^{\prime}\simeq y_{i}p^{\prime},\ \ \
\end{equation}
and assume that external collinear quarks are on-shell. This is possible
because the leading twist projectors $\left[  \Gamma_{X}^{u}~C\right]
_{\alpha\beta}\left[  \Gamma_{X}^{d}~N^{+}\right]  _{\sigma}$ in
(\ref{DA:def}) satisfy to following relations
\begin{equation}
\Gamma_{X}^{u}~\hat{p}=\hat{p}\Gamma_{X}^{u}=\hat{p}\Gamma_{X}^{d}N^{+}%
\simeq0,\label{tw3:in}%
\end{equation}
that are compatible with the free equation of motion  for quark spinors and allow us to use
on-shell spinors in the intermediate calculations. The contribution to the
physical amplitude can be obtained by resubstitution of the quark spinors by
the hadronic matrix element (\ref{DA:def}).

\subsection{Soft rescattering contribution for the Dirac FF $F_{1}$}

\begin{figure}[th]
\begin{center}
\includegraphics[
natheight=1.464100in,
natwidth=3.174700in,
height=1.4641in,
width=3.1747in
]{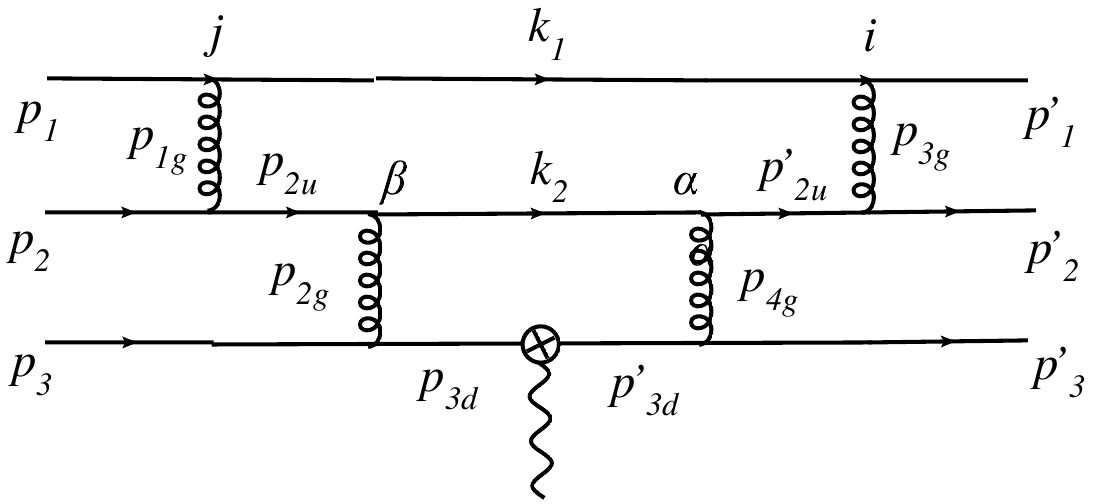}
\end{center}
\caption{The simplest diagram with soft exchanges.}%
\label{d-quark-2-loop}%
\end{figure}

Consider, following Ref.~\cite{Dun1980}, the diagram in Fig.\ref{d-quark-2-loop}. The
incoming and outgoing particles must have invariant mass $\sim\Lambda^{2}$ in
order to overlap with nucleon states. This is guaranteed by the momenta
in Eq.~(\ref{quark mom}). The interactions between the external quarks are soft and
described by DAs. For simplicity, the corresponding soft blobs are not shown in Fig.
\ref{d-quark-2-loop}. 
One can easily find that%
\begin{align}
p_{1g}=p_{1}-k_{1},~p_{2g}=p-p_{3}-k_{1}-k_{2},~\ p_{3g}=k_{1}-p_{1}^{\prime
},\ p_{4g}=k_{1}+k_{2}-p^{\prime}+p_{3}^{\prime},\\
p_{2u}=p-p_{3}-k_{1},~\ p_{3d}=p-k_{1}-k_{2},\ p_{2u}^{\prime}=p^{\prime
}-p_{3}^{\prime}-k_{1}, ~\ p_{3d}^{\prime}=p^{\prime}-k_{1}-k_{2}.
\end{align}
The analytical expression for the diagram of Fig.~\ref{d-quark-2-loop}, where the 
quark line with momenta $p_3$ and $p^\prime_3$ represents a $d$ quark, reads%
\begin{align}
{D_{\mu}}=~\mathcal{C}\int d^4 k_{1} d^4 k_{2}\frac{1}{\left[  k_{2}^{2}%
-m^{2}\right]  \left[  k_{1}^{2}-m^{2}\right]  } \frac{\bar{d}(p_{3}^{\prime
})~\gamma^{\alpha}\left( \hat p^{\prime}-\hat k_{1}-\hat k_{2}\right)
\gamma^{\mu}\left(  \hat p-\hat k_{1}-\hat k_{2}\right)  \gamma^{\beta
}~d(p_{3})~} {\left(  p^{\prime}-k_{1}-k_{2}\right)  ^{2}\left(  p-k_{1}%
-k_{2}\right)  ^{2} \left(  p_{1}-k_{1}\right)  ^{2}\left( k_{1}-p_{1}%
^{\prime}\right) ^{2} }\nonumber\\[0.5mm]
\times \frac{~\bar{u}(p_{1}^{\prime})\gamma^{i}\left( \hat k_{1}+m\right)  \gamma
^{j}u(p_{1})~ \bar{u}(p_{2}^{\prime})~\gamma^{i}(\hat p^{\prime}-\hat
p_{3}^{\prime}-\hat k_{1})\gamma^{\alpha}(\hat k_{2}+m)\gamma^{\beta}(\hat
p-\hat p_{3}-\hat k_{1})\gamma^{j}~u(p_{2})~~}{\left(  p-p_{3}-k_{1}\right)
^{2}\left(  p^{\prime}-p_{3}^{\prime}-k_{1}\right)  ^{2} \left(  k_{1}%
+k_{2}-p^{\prime}+p_{3}^{\prime}\right) ^{2}\left(  p-p_{3}-k_{1}-k_{2}\right)
^{2} },\label{Ds}%
\end{align}
where the numerical factor $\mathcal{C}$ accumulates all color factors and vertex
and propagator factors, and $m$ denotes the quark mass \footnote{For simplicity we
do not show explicitly the color indices. The quarks mass in written only in
the propagators where it can be relevant.}. We write on-shell quark spinors
instead of projectors on the nucleon DA as it was described above.

According to the factorization expressed by Eq.~(\ref{F1:fact}), one could expect that dominant
integration regions (providing contributions of order $\Lambda^{4}/Q^{4}$) can
be described as follows:%
\begin{equation}
\text{hard region: }k_{i}^{\mu}\sim Q,~~k_{i}^{2}\ \sim Q^{2},\label{hard}%
\end{equation}%
\begin{equation}
\text{collinear-}p\text{ region:~}k_{i}:(kn)\sim Q,~\ (k\bar{n})\sim
\Lambda^{2}/Q,~k_{\bot}\sim\Lambda\,,\text{\ }k_{i}^{2}\ \sim\Lambda
^{2},\label{coll-p}%
\end{equation}%
\begin{equation}
\text{collinear-}p^{\prime}\text{ region:~}k_{i}:(k\bar{n})\sim Q,~\ (kn)\sim
\Lambda^{2}/Q,~k_{\bot}\sim\Lambda\,,\text{\ }k_{i}^{2}\ \sim\Lambda
^{2},\label{coll-p'}%
\end{equation}
Then factorization formula (\ref{F1:fact}) implies that the general structure of
any 2-loop diagram can be interpreted as
\begin{eqnarray}
{D}  &  =&\mathbf{\Psi}\ast T^{(2)}\ast\mathbf{\Psi}+\mathbf{\Psi}^{(1)}\ast
T^{(1)}\ast\mathbf{\Psi}+\mathbf{\Psi}\ast T^{(1)}\ast\mathbf{\Psi}^{(1)}
\nonumber \\
&  +&\mathbf{\Psi}^{(1)}\ast T^{(0)}\ast\mathbf{\Psi}^{(1)}
+\mathbf{\Psi}^{(11)}\ast T^{(0)}\ast\mathbf{\Psi}+\mathbf{\Psi}\ast T^{(0)}\ast\mathbf{\Psi}^{(11)}
\nonumber \\ 
&+&\mathbf{\Psi}^{(2)}\ast T^{(0)}\ast\mathbf{\Psi}+\mathbf{\Psi}\ast T^{(0)}\ast\mathbf{\Psi}^{(2)},
\end{eqnarray}
where%
\begin{equation}
\mathbf{\Psi}^{(i)}={\cal V}^{(i)}\ast\mathbf{\Psi}, \quad \mathbf{\Psi}^{(11)}={\cal V}^{(1)}\ast {\cal V}^{(1)} \ast \mathbf{\Psi},
\end{equation}
denotes the convolution of the collinear evolution kernel ${\cal V}^{(i)}$ of order $i~$with DA
%%%%
$\mathbf{\Psi}$. Such contributions related with the collinear regions
(\ref{coll-p}) and (\ref{coll-p'}). The hard kernels $T^{(0,1,2)}$ denote the
contributions to the hard coefficient function in LO, next-to-leading order and next-to-next-to-leading order order, respectively.

However, this description is not the full answer at the leading-order accuracy
in $1/Q$. There is one more region, which cannot be interpreted in the form
of the reduced diagram in Fig. \ref{hs-reduced-diagram}, and is defined as
the \textit{soft} region:%
\begin{equation}
k_{i}^{\mu}\sim\Lambda,~\ k_{i}^{2}\ \sim\Lambda^{2}.\label{soft}%
\end{equation}
Let us compare the power contribution from this region with the contribution
from the hard region (\ref{hard}). The latter provides%
\begin{equation}
{D}^{(h)}_{\mu_{\bot}}\sim Q^{8}\frac{\text{Num}}{\text{Den}}\sim Q^{8}%
\frac{Q^{6}}{\left[  Q^{2}\right]  ^{10}}~\bar{\xi}_{1}^{\prime}\Gamma_{1}%
\xi_{1}~\bar{\xi}_{2}^{\prime}\Gamma_{2}\xi_{2}~\bar{\xi}_{3}^{\prime}%
\gamma_{\bot}^{\mu}\xi_{3}\sim\frac{1}{Q^{6}}~\bar{\xi}_{1}^{\prime}\Gamma
_{1}\xi_{1}~\bar{\xi}_{2}^{\prime}\Gamma_{2}\xi_{2}~\bar{\xi}_{3}^{\prime
}\gamma_{\bot}^{\mu}\xi_{3},\label{Dh}%
\end{equation}
where $\Gamma_{i}$ denote some scaleless Dirac structures, and  the factor $\sim
Q^{8}$ arises from the measure. The term with $\gamma_{\bot}^{\mu}$ in
(\ref{Dh})~reflects the requirements of one transverse index $\sim\gamma
_{\bot}^{\mu} $. In order to arrive at the formula (\ref{Dh}) we also used
the decomposition of the quark spinors onto large and small components:%
\begin{equation}
\bar{q}(p_{i}^{\prime})=\bar{\xi}_{i}^{\prime}+\bar{\eta}_{i}^{\prime}%
,~\ \bar{\xi}_{i}^{\prime}=\bar{q}(p_{i}^{\prime})\frac{
\setbox0=\hbox{$\bar n$} \dimen0=\wd0 \setbox1=\hbox{/} \dimen1=\wd1
\ifdim\dimen0>\dimen1 \rlap{\hbox to \dimen0{\hfil/\hfil}} \bar n
\else \rlap{\hbox to
\dimen1{\hfil$\bar n$\hfil}} / \fi \setbox0=\hbox{$n$} \dimen0=\wd0
\setbox1=\hbox{/} \dimen1=\wd1 \ifdim\dimen0>\dimen1 \rlap{\hbox to
\dimen0{\hfil/\hfil}} n \else \rlap{\hbox to \dimen1{\hfil$n$\hfil}} /
\fi }{4},~\bar{\eta}_{i}^{\prime}=\bar{q}(p_{i}^{\prime})\frac{
\setbox0=\hbox{$n$} \dimen0=\wd0 \setbox1=\hbox{/} \dimen1=\wd1
\ifdim\dimen0>\dimen1 \rlap{\hbox to \dimen0{\hfil/\hfil}} n
\else \rlap{\hbox to
\dimen1{\hfil$n$\hfil}} / \fi \setbox0=\hbox{$\bar n$} \dimen0=\wd0
\setbox1=\hbox{/} \dimen1=\wd1 \ifdim\dimen0>\dimen1 \rlap{\hbox to
\dimen0{\hfil/\hfil}} \bar n \else \rlap{\hbox to \dimen1{\hfil$\bar
n$\hfil}} / \fi }{4},\ \label{xi-out}%
\end{equation}%
\begin{equation}
q(p_{i})=\xi_{i}+\eta_{i},~\ \xi_{i}=\frac{ \setbox0=\hbox{$\bar n$}
\dimen0=\wd0 \setbox1=\hbox{/} \dimen1=\wd1 \ifdim\dimen0>\dimen1
\rlap{\hbox to
\dimen0{\hfil/\hfil}} \bar n \else \rlap{\hbox to \dimen1{\hfil$\bar
n$\hfil}} / \fi \setbox0=\hbox{$n$} \dimen0=\wd0 \setbox1=\hbox{/}
\dimen1=\wd1 \ifdim\dimen0>\dimen1 \rlap{\hbox to \dimen0{\hfil/\hfil}} n
\else \rlap{\hbox to \dimen1{\hfil$n$\hfil}} / \fi }{4}q(p_{i}),~\eta
_{i}=\frac{ \setbox0=\hbox{$n$} \dimen0=\wd0 \setbox1=\hbox{/} \dimen1=\wd1
\ifdim\dimen0>\dimen1 \rlap{\hbox to \dimen0{\hfil/\hfil}} n
\else \rlap{\hbox to
\dimen1{\hfil$n$\hfil}} / \fi \setbox0=\hbox{$\bar n$} \dimen0=\wd0
\setbox1=\hbox{/} \dimen1=\wd1 \ifdim\dimen0>\dimen1 \rlap{\hbox to
\dimen0{\hfil/\hfil}} \bar n \else \rlap{\hbox to \dimen1{\hfil$\bar
n$\hfil}} / \fi }{4}q(p_{i}).\label{xi-in}%
\end{equation}
One can easily obtain that the small component $\eta$ is suppressed relative to
the large component $\xi$ as%
\begin{equation}
~\ \eta\sim\xi/Q.\label{xi-eta}. %
\end{equation}

Consider now the contribution from the soft region. Neglecting in the denominator
(\ref{Ds}) by small terms one obtains:%
\begin{align}
\text{Den}  &  \simeq\left[  k_{2}^{2}-m^{2}\right]  \left[  k_{1}^{2}%
-m^{2}\right]  \left\{  y_{1}\bar{y}_{3}^{2}\left[  -2p^{\prime}\cdot
(k_{1}+k_{2})\right]  ^{2}\left[  -2(k_{1}\cdot p^{\prime})\right]
^{2}\right\} \nonumber\\
&  \times\left\{  x_{1}\bar{x}_{3}^{2}\left[  -2p\cdot(k_{1}+k_{2})\right]
^{2}\left[  -2\left(  k_{1}\cdot p\right)  \right]  ^{2}\right\}  \sim
\Lambda^{2}\Lambda^{2}\left(  \Lambda Q\right)  ^{8},
\end{align}
%%%PRDv2
where recall, $\bar x_{i}=1-x_{i}$.
%%%%
Therefore
\begin{equation}
{D}^{(s)}\sim\Lambda^{8}\frac{\text{Num}}{\text{Den}}\sim\Lambda^{8}%
\frac{\text{Num}}{\Lambda^{2}\Lambda^{2}\left(  \Lambda Q\right)  ^{8}%
}.\label{Ds:1}%
\end{equation}
From Eq.(\ref{Ds:1}) we see that the numerator must contain the soft scale at
least in the power $\sim\Lambda^{4}$ or higher. \ 

We next compute the largest terms in the numerator. Neglecting the small momenta
and the small spinor components in the $d$-quark line we obtain:%
\begin{align}
\bar{d}(p_{3}^{\prime})~\gamma^{\alpha}\left( \hat p^{\prime}-\hat k_{1}-\hat
k_{2}\right)  \gamma^{\mu}\left( \hat p-\hat k_{1}-\hat k_{2}\right)  \gamma
^{\beta}~d(p_{3})  &  \simeq\bar{d}(p_{3}^{\prime})~\gamma^{\alpha}\hat
p^{\prime}\gamma^{\mu}\hat p\gamma^{\beta}~d(p_{3})\nonumber\\
&  \simeq4p^{\prime\alpha}p^{\beta}~\bar{\xi}_{3}^{\prime}\gamma_{\bot}^{\mu
}\xi_{3}.
\end{align}
Then the second $u$-quark line can be rewritten as
\begin{equation}
p^{\prime\alpha}p^{\beta}~\bar{u}(p_{2}^{\prime})~\gamma^{i}(\hat p^{\prime
}-\hat p_{3}^{\prime}-\hat k_{1})\gamma^{\alpha}(\hat k_{2}+m)\gamma^{\beta
}(\hat p-\hat p_{3}-\hat k_{1})\gamma^{j}~u(p_{2})\simeq\bar{\xi}_{2}^{\prime
}\gamma^{i}\hat k_{1}\hat p^{\prime}(\hat k_{2}+m)\hat p\hat k_{1}\gamma
^{j}~\xi_{2}.
\end{equation}
%%%%PRDv2
The product of the $u$-quark  lines yields
%%%%%
\begin{align}
~\bar{\xi}_{1}^{\prime}\gamma^{i}\left( \hat k_{1}+m\right)  \gamma^{j}\xi_{1}
&  \bar{\xi}_{2}^{\prime}\gamma^{i}\hat k_{1}\hat p^{\prime}(\hat k_{2}+m)\hat
p\hat k_{1}\gamma^{j}~\xi_{2}\nonumber\\
\simeq &  2(p^{\prime}\cdot k_{1})2(p\cdot k_{1})~\ \bar{\xi}_{1}^{\prime
}\gamma^{i}\left(  \hat k_{1}+m\right)  \gamma^{j}\xi_{1}~\bar{\xi}%
_{2}^{\prime}\gamma^{i}(\hat k_{2}+m)\gamma^{j}~\xi_{2}.
\end{align}
Therefore we obtain%
\begin{align}
\text{Num}  &  = 2(p^{\prime}\cdot k_{1})2(p\cdot k_{1})~\bar{\xi}%
_{1}^{\prime}\gamma^{i}\left( \hat k_{1}+m\right)  \gamma^{j}\xi_{1}~\bar{\xi
}_{2}^{\prime}\gamma^{i}(\hat k_{2}+m)\gamma^{j}~\xi_{2}~\bar{\xi}_{3}%
^{\prime}\gamma_{\bot}^{\mu}\xi_{3} \nonumber \\
&  \sim Q^{2}\Lambda^{4}~\bar{\xi}_{1}^{\prime}\Gamma_{1}\xi_{1}~\bar{\xi}%
_{2}^{\prime}\Gamma_{2}\xi_{2}~\bar{\xi}_{3}^{\prime}\gamma_{\bot}^{\mu}%
\xi_{3}.
\end{align}
Substituting this into (\ref{Ds:1}) yields%
\begin{equation}
{D}^{(s)}_{\mu_{\bot}}\sim\Lambda^{8}\frac{Q^{2}\Lambda^{4}~}{\Lambda
^{2}\Lambda^{2}\left(  \Lambda Q\right)  ^{8}}~\bar{\xi}_{1}^{\prime}%
\Gamma_{1}\xi_{1}~\bar{\xi}_{2}^{\prime}\Gamma_{2}\xi_{2}~\bar{\xi}%
_{3}^{\prime}\gamma_{\bot}^{\mu}\xi_{3}\sim\frac{1}{Q^{6}}~\bar{\xi}%
_{1}^{\prime}\xi_{1}~\bar{\xi}_{2}^{\prime}\xi_{2}~\bar{\xi}_{3}^{\prime
}\gamma_{\bot}^{\mu}\xi_{3},
\end{equation}
i.e. we obtain the same power of $Q$ as for the hard region in (\ref{Dh}).
Therefore we established that the soft region is the additional relevant
region which is not accounted for in the factorization formula (\ref{F1:fact}). In
Ref.~\cite{Fadin1981} all diagrams with the soft spectator quarks have been
computed with the leading logarithmic accuracy. Their sum does not cancel
providing some nontrivial answer. Hence we can avoid consideration of such a possibility.

Consider now  the whole expression for the soft contribution:
\begin{eqnarray}
{D}_{\mu_{\bot}}^{(s)}  &=& 4\mathcal{C}\int d^4 k_{1} d^4 k_{2}~\left[  \gamma_{\bot
}^{\mu}~\right]  _{\alpha_{3}\beta_{3}}~\ \left[  \frac{\left(  \hat{k}%
_{1}+m\right)  _{\alpha_{1}\beta_{1}}~(\hat{k}_{2}+m)_{\alpha_{2}\beta_{2}}%
}{\left[  k_{2}^{2}-m^{2}\right]  \left[  k_{1}^{2}-m^{2}\right]  }\right]
\nonumber\\
& \times & \left[  \frac{~\left[  \bar{\xi}_{1}^{\prime}\gamma^{i}\right]  _{\alpha
_{1}}\left[  \bar{\xi}_{2}^{\prime}\gamma^{i}\right]  _{\alpha_{2}}~\left[
\bar{\xi}_{3}^{\prime}~\right]  _{\alpha_{3}}\ ~}{y_{1}\bar{y}_{3}^{2}\left[
-2p^{\prime}\cdot(k_{1}+k_{2})\right]  ^{2}\left[  -2(k_{1}\cdot p^{\prime
})\right]  ~}\right]  \left[  \frac{\ ~\left[  \ \gamma^{j}\xi_{1}\right]
_{\beta_{1}}~\left[  \gamma^{j}~\xi_{2}\right]  _{\beta_{2}}~\left[  \xi
_{3}\right]  _{\beta_{3}}}{~x_{1}\bar{x}_{3}^{2}~\left[  -2p\cdot(k_{1}%
+k_{2})\right]  ^{2}\left[  -2\left(  k_{1}\cdot p\right)  \right]  }\right]
.\label{Ds:2}%
\end{eqnarray}
Each expression in the square brackets  describes some subprocess involving
the particles with appropriate virtualities and momenta. We consider them
term by term. The factor
\begin{equation}
\left[  \frac{\left(  \hat{k}_{1}+m\right)  _{\alpha_{1}\beta_{1}}~(\hat
{k}_{2}+m)_{\alpha_{2}\beta_{2}}}{\left[  k_{2}^{2}-m^{2}\right]  \left[
k_{1}^{2}-m^{2}\right]  }\right]  ,
\end{equation}
describes the propagation of the soft spectator quarks and includes only soft
particles with $k_{i}^{2}\sim\Lambda^{2}$. This term can be  associated with
the soft part of the diagram. The factor%
\begin{equation}
\left[  \frac{~\left[  \bar{\xi}_{1}^{\prime}\gamma^{i}\right]  _{\alpha_{1}%
}\left[  \bar{\xi}_{2}^{\prime}\gamma^{i}\right]  _{\alpha_{2}}~\left[
\bar{\xi}_{3}^{\prime}~\right]  _{\alpha_{3}}\ ~}{y_{1}\bar{y}_{3}^{2}\left[
-2p^{\prime}\cdot(k_{1}+k_{2})\right]  ^{2}\left[  -2(k_{1}\cdot p^{\prime
})\right]  ~}\right]  ,
\end{equation}
describes the transition of two soft spectator quarks and one active quark into
three collinear quarks. It is described by the subdiagram with the two-gluon
exchange. As one can see, all propagators  have virtualities of order
$k_{i}\cdot p^{\prime}\sim Q\Lambda$ \ and all involved momenta have a large
component $\sim Q$ along the $p^{\prime}$ direction. 

In a the similar way one can describe the second subprocess given by%
\begin{equation}
\left[  \frac{\ ~\left[  \ \gamma^{j}\xi_{1}\right]  _{\beta_{1}}~\left[
\gamma^{j}~\xi_{2}\right]  _{\beta_{2}}~\left[  \xi_{3}\right]  _{\beta_{3}}%
}{~x_{1}\bar{x}_{3}^{2}~\left[  -2p\cdot(k_{1}+k_{2})\right]  ^{2}\left[
-2\left(  k_{1}\cdot p\right)  \right]  }\right]  .
\end{equation}
The difference from the previous case is only in the involved momenta. They
have large components $\sim Q$ along the $p$ direction.

The simple vertex  factor $\left[  \gamma_{\bot}^{\mu}~\right]  _{\alpha
_{3}\beta_{3}}$ can be associated with the hard scattering vertex of the
subprocess $\gamma^{\ast}+d(p)\rightarrow d(p^{\prime})$. It is clear that
this subprocess in general involves particles with large momenta of order
$Q^{2}$.   

Taking into account the different virtualities of the particles:~$\Lambda
^{2}\ll Q\Lambda\ll Q^{2}$ one can try to factorize the whole result
of Eq.~(\ref{Ds:2}) in accordance with the described subprocesses.
In order to do this we introduce the Sudakov decomposition
\begin{equation}
\left(  k_{i}\cdot n\right)  =k_{i}^{+},~\ \left(  k_{i}\cdot\bar{n}\right)
=k_{i}^{-},~dk_{i}=\frac{1}{2}dk_{i}^{+}dk_{i}^{-}dk_{\bot}~
\end{equation}
and rewrite Eq.~(\ref{Ds:2}) as%
\begin{align}
{D}_{\mu_{\bot}}^{(s)}  & =~\left[  \gamma_{\bot}^{\mu}~\right]  _{\alpha
_{3}\beta_{3}}\int dk_{1}^{\pm}dk_{2}^{\pm}~~\ \left[  \mathcal{C}\int
dk_{12\bot}\frac{\left(  \hat{k}_{1}+m\right)  _{\alpha_{1}\beta_{1}}~\left(\hat
{k}_{2}+m \right)_{\alpha_{2}\beta_{2}}}{\left[  k_{1}^{2}-m^{2}\right]  \left[
k_{2}^{2}-m^{2}\right]  }\right]  \nonumber\\
& \times \left[  \frac{1}{Q^{3}}\frac{~\left[  \bar{\xi}_{1}^{\prime}\gamma
^{i}\right]  _{\alpha_{1}}\left[  \bar{\xi}_{2}^{\prime}\gamma^{i}\right]
_{\alpha_{2}}~\left[  \bar{\xi}_{3}^{\prime}~\right]  _{\alpha_{3}}\ ~}%
{y_{1}\bar{y}_{3}^{2}~(k_{1}^{+}+k_{2}^{+})^{2}\left[  -k_{1}^{+}\right]
~}\right]  \left[  \frac{1}{Q^{3}}\frac{\ ~\left[  \ \gamma^{j}\xi_{1}\right]
_{\beta_{1}}~\left[  \gamma^{j}~\xi_{2}\right]  _{\beta_{2}}~\left[  \xi
_{3}\right]  _{\beta_{3}}}{~x_{1}\bar{x}_{3}^{2}~(k_{1}^{-}+k_{2}^{-}%
)^{2}\left[  -k_{1}^{-}\right]  }\right]  .\label{Ds:3}%
\end{align}
This equation  almost represents the required form. To make it more obvious
Eq.~(\ref{Ds:3}) can be rewritten as:
\begin{equation}
{D}_{\mu_{\bot}}^{(s)}=\left[  \gamma_{\bot}^{\mu}\right]  _{\alpha_{3}%
\beta_{3}}~\int d\omega_{1,2}~~J_{(\alpha)}^{\prime}\left(  y_{i},\omega
_{i}\right)  \int d\nu_{1,2}~~J_{(\beta)}\left(  x_{i},\nu_{i}\right)
~~\mathbf{S}_{(\alpha\beta)}\left(  \omega_{i},\nu_{i}\right)  .~\label{fact}%
\end{equation}
where we introduced
\begin{align}
~\mathbf{S}_{(\alpha\beta)}\left(  \omega_{i},\nu_{i}\right)   &
=\mathcal{C}\int dk_{1}^{\pm}dk_{2}^{\pm}~dk_{12\bot}\delta(\omega_{1}%
-k_{1}^{+})\delta(\omega_{2}-k_{2}^{+})\delta(\nu_{1}-k_{1}^{-})\delta(\nu
_{2}-k_{2}^{-})\nonumber\\
&  \times\frac{\left(  \hat k_{1}+m\right)  _{\alpha_{1}\beta_{1}}~}{\left[
k_{1}^{2}-m^{2}\right]  }\frac{\left( \hat k_{2}+m \right)_{\alpha_{2}\beta_{2}}}{\left[
k_{2}^{2}-m^{2}\right]  }.\label{S:def}%
\end{align}
The two functions $~J_{(\alpha)}^{\prime}$ and $~J_{(\beta)},$ which we will refer to as 
\textit{jet functions, } read%
\begin{align}
~J_{(\alpha)}^{\prime}\left(  y_{i},\omega_{i}\right)   &  =\frac{1}{Q^{3}%
}\frac{1}{y_{1}\bar{y}_{3}^{2}}\frac{1}{(\omega_{1}+\omega_{2})^{2}\left[
-\omega_{1}\right]  }\left[  ~\bar{\xi}_{3}^{\prime}\right]  _{\alpha_{3}%
}~\left[  ~\bar{\xi}_{2}^{\prime}\gamma_{\bot}^{i}\right]  _{\alpha_{2}%
}\left[  \bar{\xi}_{1}^{\prime}\gamma_{\bot}^{i}\right]  _{\alpha_{1}%
},\label{Jalf}\\
~J_{(\beta)}\left(  x_{i},\nu_{i}\right)   &  =\frac{1}{Q^{3}}\frac{1}%
{x_{1}\bar{x}_{3}^{2}}~\frac{1}{\left[  \nu_{1}+\nu_{2}\right]  ^{2}\left[
-\nu_{1}\right]  }~\left[  \xi_{3}\right]  _{\beta_{3}}~\left[  \gamma_{\bot
}^{j}\xi_{2}\right]  _{\beta_{2}}~\left[  \gamma_{\bot}^{j}\xi_{1}\right]
_{\beta_{1}},\label{Jbet}%
\end{align}
where the index in brackets denotes multi-index: $(\alpha)\equiv\{\alpha
_{1},\alpha_{2},\alpha_{3}\}$. These functions describe the scattering of the
%%%PRDv2
particles with \textit{hard-collinear} virtualities:
%%%
\begin{equation}
p_{i}^{2}\sim Q\Lambda.
\end{equation}
Such fluctuations appear in the case of scattering  collinear and soft particles
and in our case they have momenta components which scale as
\begin{equation}
\text{in }J\text{-function}:(p_{i}^{+}\sim Q,~\ p_{i}^{-}\sim\Lambda
,~~p_{i\bot}\sim\sqrt{Q\Lambda}),\label{hc-in}%
\end{equation}%
\begin{equation}
\text{in }J^{\prime}\text{-function}:(p_{i}^{\prime+}\sim\Lambda
,~\ p_{i}^{\prime-}\sim Q,~~p_{i\bot}^{\prime}\sim\sqrt{Q\Lambda}).\label{hc-out}%
\end{equation}
Such modes are often refereed to as \textit{hard-collinear} particles.

The \textit{soft correlation function }$\mathbf{S}\left(  \omega_{i},\nu
_{i}\right)  $ defined in (\ref{S:def}) describes the contribution of
the subdiagram with the soft momenta and low virtualities. In this particular
case this is the simple product of the two soft propagators. Taking into
account that the jet functions do not depend on the transverse momenta the soft
part can be represented as a light-cone correlation function (CF):%
\begin{align}
~\mathbf{S}_{(\mathbf{\alpha\beta})}\left(  \omega_{i},\nu_{i}\right)   &
=\mathcal{C}\int\frac{d\lambda_{1}}{2\pi}e^{i\omega_{1}\lambda_{1}}\int
\frac{d\lambda_{2}}{2\pi}e^{i\omega_{2}\lambda_{2}}\int\frac{d\eta_{1}}{2\pi
}e^{-i\nu_{1}\eta_{1}}\int\frac{d\eta_{2}}{2\pi}~e^{-i\nu_{2}\eta_{2}}~ \nonumber \\
& ~~\ \ \ \ \ \ \ \ \ \ \  \ \ \ \ \ \times \left\langle 0\left\vert
q_{\alpha_{1}}\left(  \lambda_{1}n\right)  ~\bar{q}_{\beta_{1}}\left(
\eta_{1}\bar{n}\right)  \left\vert 0\right\rangle \left\langle 0\right\vert
q_{\alpha_{2}}\left(  \lambda_{2}n\right)  ~\bar{q}_{\beta_{2}}\left(
\eta_{2}\bar{n}\right)  \right\vert 0\right\rangle .\label{SGG}%
\end{align}

In pQCD, the leading-order CF factorizes into the product of two propagators
: $\left\langle 0\left\vert ...\right\vert 0\right\rangle ~\left\langle
0\left\vert ...\right\vert 0\right\rangle $. But it is clear that this is
specific for the perturbative result. In the general case one can expect \ general
matrix element $\left\langle 0\left\vert ...\right\vert 0\right\rangle $. Let
us note that such  a CF is a vacuum loop for the transverse momentum and
simultaneously a 4-point CF function from the point of view of longitudinal
subspace. Therefore integration over transverse components is UV-divergent.
Computing the convolution integrals $\int d\omega_{1,2}~\int d\nu_{1,2}$ and
integrating over the soft quark momenta, we reproduce the factorization breaking logarithmic contribution 
computed in   Ref.~\cite{Dun1980}.
In our example we have only the leading-order, simple contribution from the hard
subprocess: tree-level scattering of the transverse hard photon on the
hard-collinear quark. The corresponding amplitude $\sim\gamma_{\bot}^{\mu}$ can be
associated with the hard coefficient function.

The answer (\ref{fact}) can be interpreted in terms of a reduced diagram as in
Fig.\ref{s-reduced-diagram}. 
We observe that in this case the scattering process contains soft
spectators and involves two large scales: the hard scale $Q^{2}$ and hard-collinear
scale of order $\Lambda Q$. The presence of the soft spectators, allows us to
associate the contribution from the soft region with the Feynman mechanism
\cite{Feynman:1973xc}. In the following, we shall refer to it, for simplicity, as the soft
rescattering mechanism.

\subsection{Soft rescattering contribution for the Pauli FF $F_{2}$}

A calculation of the helicity flip FF $F_{2}$ carried out in the hard rescattering
picture cannot provide a well defined result because the convolution integral
(\ref{F1:fact}) is divergent. This divergence can be understood as an
indication that the definition of relevant regions according to
Fig.\ref{hs-reduced-diagram} is not complete. However, such a calculation allows us
to define the power behavior (\ref{F2:as}). As one can observe, \ $F_{2}$ is
suppressed as $1/Q^{2}$ compared to $F_{1}$. This is a consequence of the helicity
flip which requires us to involve one unit of orbital quark momentum that leads to
suppression of order $\Lambda/Q$ [one more factor $Q$ arises from the
kinematical prefactor $(p+p^{\prime})^{\mu}$ in the FF definition, see
Eq.(\ref{FF:def})]. 

\begin{figure}[th]
\begin{center}
\includegraphics[
height=1.4961in,
width=3.0787in
]{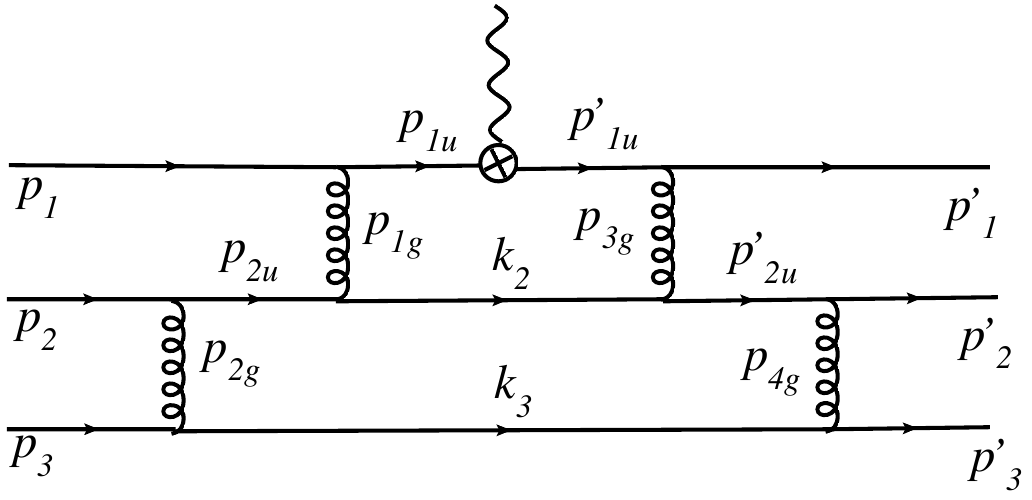}
\end{center}
\caption{Two-loop diagram for the helicity flip FF}%
\label{2-loop-graphF2}%
\end{figure}

Consider now the diagram as shown in Fig.\ref{2-loop-graphF2}, where 
we calculate the contribution where the hard photon couples to a $u$ quark. 
Following the same definitions of the external momenta as before, 
the internal momenta read%
\begin{align}
p_{1u}=p-k_{2}-k_{3},~\ p_{2u}=\bar{x}_{1}p-k_{3},~\ p_{1u}^{\prime}%
=p^{\prime}-k_{2}-k_{3},\ p_{2u}^{\prime}=\bar{y}_{1}p-k_{3}, \nonumber \\
p_{1g}=k_{2}+k_{3}-\bar{x}_{1}p,~\ \ p_{2g}=k_{3}-p_{3},~ p_{3g}=\bar{y}%
_{1}p^{\prime}-k_{2}-k_{3},~\ p_{4g}=p_{3}^{\prime}-k_{3}.~\
\end{align}
and the analytical expression for the diagram reads:%
\begin{align}
D^{\mu}  &  =\mathcal{C}\int\frac{dk_{2}dk_{3}}{\left[  k_{2}^{2}
-m^{2}\right]  \left[  k_{3}^{2}-m^{2}\right]  }\frac{\bar{\xi}_{1}^{\prime
}\gamma^{\alpha}(\hat p^{\prime}-\hat k_{2}-\hat k_{3})\gamma^{\mu}(\hat
p-\hat k_{2}-\hat k_{3})\gamma^{\beta}\xi_{1}}{\left(  p^{\prime}-k_{2}
-k_{3}\right)  ^{2}\left(  p-k_{2}-k_{3}\right)  ^{2}}\nonumber\\
&  \times \frac{\bar{\xi}_{2}^{\prime}~\gamma^{i}\left(  \bar{y}
_{1}\hat p^{\prime}-\hat k_{3}\right)  \gamma^{\alpha}\left(  \hat{k}
_{2}+m\right)  \gamma^{\beta}\left(  \bar{x}_{1}\hat p-\hat k_{3}\right)
\gamma^{j}\xi_{2} \,
 \bar{\eta}_{3}^{\prime}\gamma^{i} \left(\hat{k}_{3}+m \right)\gamma^{j}\xi_{3}} 
{\left(  \bar{y}_{1} p-k_{3}\right)  ^{2}\left(  \bar{x}
_{1}p-k_{3}\right)  ^{2}\left(  k_{3}-p_{3}\right)  ^{2}\left(  k_{3}
-p_{3}^{\prime}\right)  ^{2} 
\left(  k_{2}+k_{3}-\bar{x}_{1}p\right)
^{2}\left(  \bar{y}_{1}p^{\prime}-k_{2}-k_{3}\right)  ^{2}}
.\label{DmuF2}%
\end{align}
Let us add few comments to this formula. Following conventions, we assume that the first
and second spinor lines correspond to $u$ quarks and we substitute instead of
spinors their large components $\bar{\xi}_{1,2}^{\prime}$ and $\xi_{1,2}$ as
defined in (\ref{xi-out}) and (\ref{xi-in}). However, we cannot perform such a
substitution for all external quarks as we did in the case of the Dirac FF $F_{1}$. In order to
obtain a nontrivial helicity flip amplitude, we need to project the \textit{in} or
\textit{out} collinear partonic state on the higher twist (twist-4) DAs. The
projections on twist-4 DAs are well known and can be written in the same form
as for the twist-3 case \cite{Braun:2000kw} . Contrary to the twist-3 case,
the twist-4 projections do not satisfy the full set of relations (\ref{tw3:in})
because twist-4 operator includes one small component of the collinear quark
field:%
\begin{equation}
\text{twist-3 DA}\sim\left\langle 0\right\vert \xi~\xi~\xi~\left\vert
p\right\rangle ,~~\ \ \text{twist-4 DA}\sim\left\langle 0\right\vert \xi
~\xi~\eta~\left\vert p\right\rangle .
\end{equation}
For instance, one obtains the following projector (in general, there are 9 twist-4
projections \cite{Braun:2000kw})
\begin{equation}
\text{twist-4: }\Gamma_{V_{2}^{\prime}}^{u}\otimes\Gamma_{V_{2}^{\prime}}%
^{d}\sim~ \setbox0=\hbox{$n$} \dimen0=\wd0 \setbox1=\hbox{/} \dimen1=\wd1
\ifdim\dimen0>\dimen1 \rlap{\hbox to \dimen0{\hfil/\hfil}} n
\else \rlap{\hbox to \dimen1{\hfil$n$\hfil}} / \fi C\otimes\gamma_{5}
\setbox0=\hbox{$\bar{n}$} \dimen0=\wd0 \setbox1=\hbox{/} \dimen1=\wd1
\ifdim\dimen0>\dimen1 \rlap{\hbox to \dimen0{\hfil/\hfil}} \bar{n}
\else \rlap{\hbox to
\dimen1{\hfil$\bar{n}$\hfil}} / \fi ,
\end{equation}
where the $u$ quarks projected on large components but $d$ quark on the small
component. Therefore in order to obtain such configuration one has to
substitute instead of a $d$-quark spinor its small projection (\ref{xi-out}):
\begin{equation}
\bar{d}(p_{3}^{\prime})\rightarrow\bar{\eta}_{3}^{\prime}=\bar{d}%
(p_{3}^{\prime})\frac{ \setbox0=\hbox{$n$} \dimen0=\wd0 \setbox1=\hbox{/}
\dimen1=\wd1 \ifdim\dimen0>\dimen1 \rlap{\hbox to \dimen0{\hfil/\hfil}} n
\else \rlap{\hbox to \dimen1{\hfil$n$\hfil}} / \fi \setbox0=\hbox{$\bar
n$} \dimen0=\wd0 \setbox1=\hbox{/} \dimen1=\wd1 \ifdim\dimen0>\dimen1
\rlap{\hbox to \dimen0{\hfil/\hfil}} \bar n \else \rlap{\hbox to
\dimen1{\hfil$\bar n$\hfil}} / \fi }{4},~\text{with}\ \ \bar{\eta}_{3}%
^{\prime} \setbox0=\hbox{$n$} \dimen0=\wd0 \setbox1=\hbox{/} \dimen1=\wd1
\ifdim\dimen0>\dimen1 \rlap{\hbox to \dimen0{\hfil/\hfil}} n
\else \rlap{\hbox to \dimen1{\hfil$n$\hfil}} / \fi \neq0.
\end{equation}
We take into account this particular case in the expression (\ref{DmuF2}) and
do not consider the other configurations (with the small $u$-quark components)
for the sake of simplicity.

Consider first the contribution from the hard region, Eq.~(\ref{hard}). In order to
project the index $\mu$ in Eq.~(\ref{hard}) onto  the longitudinal subspace, we perform a contraction
\begin{equation}
\bar{D}_{\Vert}^{(h)}=(n+\bar{n})^{\mu}~\bar{D}_{\mu}^{(h)}.
\end{equation}
Simple dimensional counting provides%
\begin{equation}
\bar{D}^{(h)}\sim\frac{1}{Q^{6}}~\bar{\xi}_{1}^{\prime}\Gamma_{1}\xi_{1}%
~\bar{\xi}_{2}^{\prime}\Gamma_{2}\xi_{2}~\bar{\eta}_{3}^{\prime}\Gamma_{3}%
\xi_{3}\sim~(p+p^{\prime})\cdot(n+\bar{n})\frac{1}{Q^{8}}~\bar{\xi}%
_{1}^{\prime}\Gamma_{1}\xi_{1}~\bar{\xi}_{2}^{\prime}\Gamma_{2}\xi_{2}~\left[
Q\bar{\eta}_{3}^{\prime}\right]  \Gamma_{3}\xi_{3},
\end{equation}
where we took into account the kinematical factor $~(p+p^{\prime})\cdot(n+\bar
{n})\sim Q$ and the fact that the small component $\eta_{3}^{\prime}$ is
suppressed, according to Eq.~(\ref{xi-eta}). Then we observe that the hard part of $F_{2}$ is
suppressed compared to $F_{1}~$(\ref{Dh}) as expected.

Consider now the soft region, expressed by Eq.~(\ref{soft}). In the denominator $D_{\Vert}%
^{(s)}$ we obtain:%
\begin{align}
\text{Den}  &  \simeq\left[  k_{2}^{2}-m^{2}\right]  \left[  k_{3}^{2}%
-m^{2}\right] \nonumber\\
&\times  Q^{2}\bar{y}_{1}\left[  (k_{2}+k_{3})\cdot n\right]  ^{2}~Q^{2}\bar{y}%
_{1}y_{3}\left[  k_{3}\cdot n\right]  ^{2}\ \ Q^{2}\bar{x}_{1}\left[
(k_{2}+k_{3})\cdot\bar{n}\right]  ^{2}~Q^{2}\bar{x}_{1}x_{3}\left[  k_{3}%
\cdot\bar{n}\right]  ^{2}.\label{DenDsII}%
\end{align}
In the numerator, the first spinor line gives~:%
\begin{eqnarray}
&&\bar{\xi}_{1}^{\prime}\gamma^{\alpha}(\hat p^{\prime}-\hat k_{2}-\hat k_{3})(
\setbox0=\hbox{$n$} \dimen0=\wd0 \setbox1=\hbox{/} \dimen1=\wd1
\ifdim\dimen0>\dimen1 \rlap{\hbox to \dimen0{\hfil/\hfil}} n
\else \rlap{\hbox to \dimen1{\hfil$n$\hfil}} / \fi + \setbox0=\hbox{$\bar n$}
\dimen0=\wd0 \setbox1=\hbox{/} \dimen1=\wd1 \ifdim\dimen0>\dimen1
\rlap{\hbox to \dimen0{\hfil/\hfil}} \bar n \else \rlap{\hbox to
\dimen1{\hfil$\bar n$\hfil}} / \fi )(\hat p-\hat k_{2}-\hat k_{3}%
)\gamma^{\beta}\xi_{1} \nonumber \\
&&\simeq-2(k_{2}+k_{3})\cdot n~2p^{\beta}\bar{\xi}_{1}^{\prime}\gamma_{\bot
}^{\alpha}\xi_{1}-2(k_{2}+k_{3})\cdot\bar{n}~2p^{\prime\alpha}~\bar{\xi}%
_{1}^{\prime}\gamma_{\bot}^{\beta}~\xi_{1}
\label{1st2ndL}%
\end{eqnarray}
Combining the contribution $\sim p^{\beta}\gamma_{\bot}^{\alpha}$ with the second
and third lines we obtain:
\begin{eqnarray}
&& p^{\beta}~\bar{\xi}_{2}^{\prime}~\gamma^{i}\left(  \bar{y}_{1}\hat p^{\prime
}-\hat k_{3}\right)  \gamma^{\alpha}\left(  \hat{k}_{2}+m\right)
\gamma^{\beta}\left(  \bar{x}_{1}\hat p-\hat k_{3}\right)  \gamma^{j}\xi
_{2}~\bar{\eta}_{3}^{\prime}\gamma^{i}(\hat{k}_{3}+m)\gamma^{j}\xi_{3} \nonumber \\
&&=-4\bar{y}_{1}Q(k_{3}\cdot\bar{n})~\bar{\xi}_{2}^{\prime}\gamma_{\bot}%
^{\alpha}\left(  \hat{k}_{2}+m\right)  \gamma^{j}\xi_{2}~\bar{\eta}%
_{3}^{\prime}~\hat{p}^{\prime}~(\hat{k}_{3}+m)\gamma^{j}\xi_{3}.
\end{eqnarray}
The same combination of the second term $\sim p^{\prime\alpha}\gamma_{\bot}%
^{\beta}$ in Eq.~(\ref{1st2ndL}) provides trivial results:%
\begin{eqnarray}
&& p^{\prime\alpha}~\bar{\xi}_{2}^{\prime}~\gamma^{i}\left(  \bar{y}_{1}%
p^{\prime}-\hat k_{3}\right)  \gamma^{\alpha}\left(  \hat{k}_{2}+m\right)
\gamma^{\beta}\left(  \bar{x}_{1}\hat p-\hat k_{3}\right)  \gamma^{j}\xi
_{2}~\bar{\eta}_{3}^{\prime}\gamma^{i}(\hat{k}_{3}+m)\gamma^{j}\xi_{3} \nonumber \\
&&=\bar{x}_{1}~\bar{\xi}_{2}^{\prime}~\gamma^{i}\left(  -\hat k_{3}\right)
p^{\prime}\left(  \hat{k}_{2}+m\right)  \gamma_{\bot}^{\beta}~\hat{p}%
~\gamma^{j}\xi_{2}~\bar{\eta}_{3}^{\prime}\gamma^{i}(\hat{k}_{3}+m)\gamma
^{j}\xi_{3} \nonumber \\
&&=2\bar{x}_{1}~\bar{\xi}_{2}^{\prime}~\gamma^{i}\left(  -\hat k_{3}\right)
p^{\prime}\left(  \hat{k}_{2}+m\right)  \gamma_{\bot}^{\beta}~\xi_{2}%
~\bar{\eta}_{3}^{\prime}\gamma^{i}(\hat{k}_{3}+m)~\underbrace{\hat{p}\ \xi
_{3}}=0.
\end{eqnarray}
Therefore, we can write%
\begin{equation}
\text{Num}\simeq16~Q~\bar{y}_{1}(k_{3}\cdot\bar{n})(k_{2}+k_{3})\cdot
n~~\bar{\xi}_{1}^{\prime}\gamma_{\bot}^{\alpha}\xi_{1}~\bar{\xi}_{2}^{\prime
}\gamma_{\bot}^{\alpha}\left(  \hat{k}_{2}+m\right)  \gamma^{j}\xi_{2}%
~~\bar{\eta}_{3}^{\prime}\hat{p}^{\prime}(\hat{k}_{3}+m)\gamma^{j}\xi
_{3}.\label{NumDsII}%
\end{equation}
Combining Eqs.~(\ref{DenDsII}) and (\ref{NumDsII}) yields%
\begin{align}
D_{\Vert}^{(s)}=\frac{(p+p^{\prime})\cdot(n+\bar{n})}{Q^{7}} &  \int
\frac{8\mathcal{C~}dk_{2}dk_{3}}{\left[  k_{2}^{2}-m^{2}\right]  \left[
k_{3}^{2}-m^{2}\right] }\nonumber\\
& \times\frac{~\bar{\xi}_{1}^{\prime}\gamma_{\bot}^{\alpha}\xi_{1}~\bar{\xi
}_{2}^{\prime}\gamma_{\bot}^{\alpha}\left(  \hat{k}_{2}+m\right)  \gamma
^{j}\xi_{2}~~\bar{\eta}_{3}^{\prime}~\hat{n}~(\hat{k}_{3}+m)\gamma^{j}\xi_{3}%
}{\bar{x}_{1}^{2}x_{3}~\bar{y}_{1}y_{3}\left[  (k_{2}+k_{3})\cdot n\right]
\left[  (k_{2}+k_{3})\cdot\bar{n}\right]  ^{2}~\left[  k_{3}\cdot n\right]
^{2}\left[  k_{3}\cdot\bar{n}\right]  }.\label{Dss:F2}%
\end{align}
By simple power counting, we obtain%
\begin{eqnarray}
D_{\Vert}^{(s)} &  \sim & \frac{(p+p^{\prime})\cdot(n+\bar{n})}{Q^{7}}%
~\Lambda^{8}\frac{\Lambda^{2}}{\Lambda^{10}}~\bar{\xi}_{1}^{\prime}\Gamma
_{1}\xi_{1}~\bar{\xi}_{2}^{\prime}\Gamma_{2}\xi_{2}~\bar{\eta}_{3}^{\prime
}\Gamma_{3}\xi_{3} \nonumber \\
& \sim & \frac{(p+p^{\prime})\cdot(n+\bar{n})}{Q^{8}}~~\bar{\xi}_{1}^{\prime
}\Gamma_{1}\xi_{1}~\bar{\xi}_{2}^{\prime}\Gamma_{2}\xi_{2}~\left[  Q\bar{\eta
}_{3}^{\prime}\right]  \Gamma_{3}\xi_{3}.
\end{eqnarray}
One notices that we obtain the same power counting as for the hard region. Therefore we can
conclude that the soft rescattering is also relevant for the helicity flip
case and is not suppressed compared to the hard rescattering mechanism.

Let us perform an interpretation of Eq.~(\ref{Dss:F2}) in terms of
hard, jet and soft functions introduced in the previous section.
\begin{align}
\label{F2:J'andJet} & D_{\Vert}^{(s)} =\underset{\text{hard CF}%
}{\mathcal{~}\underbrace{\frac{(p+p^{\prime})\cdot(n+\bar{n})~}{m}~\frac
{m}{Q^{2}}~}} & \nonumber\\
&  \times~\underset{J^{\prime}\text{-function}}{\underbrace{\int d\omega
_{2}d\omega_{3}~\frac{1}{Q^{2}}\frac{\left[  \bar{\xi}_{1}^{\prime}%
\gamma_{\bot}^{\alpha}\right]  _{\alpha_{1}}\left[  \bar{\xi}_{2}^{\prime
}\gamma_{\bot}^{\alpha}\right]  _{\alpha_{2}}~\left[  \bar{\eta}_{3}^{\prime
}~\hat{n}\right]  _{\alpha_{3}}}{~\bar{y}_{1}y_{3}(\omega_{2}+\omega
_{3})\omega_{3}^{2}}}}~~\underset{J\text{-function, same as in }F_{1},\text{
see Eq.(\ref{Jbet})}}{\underbrace{\int d\nu_{2}~d\nu_{3}\frac{1}{Q^{3}}%
\frac{\left[  \xi_{1}\right]  _{\beta_{1}}\left[  \gamma^{j}\xi_{2}\right]
_{\beta_{2}}\left[  \gamma^{j}\xi_{3}\right]  _{\beta_{3}}}{\bar{x}_{1}%
^{2}x_{3}~(\nu_{2}+\nu_{3})^{2}~\nu_{3}}}} & \\
&  \times\underset{\text{soft corr. f.}~\mathbf{S}\left[  \omega_{i},\nu_{i}\right]
}{\underbrace{\mathcal{C}\int dk_{2}^{\pm}dk_{3}^{\pm}dk_{23\bot}\frac{ (\hat
k_{2}+m) _{\alpha_{2}\beta_{2}}} {\left[  k_{2}^{2}-m^{2}\right]  }\frac{
(\hat k_{3}+m)_{\alpha_{3}\beta_{3}}} {\left[  k_{3}^{2}-m^{2}\right]  }
\delta(\omega_{2}-k_{2}^{+})\delta(\omega_{3}-k_{3}^{+})\delta(\nu_{2}%
-k_{2}^{-})\delta(\nu_{3}-k_{3}^{-})}} & \nonumber
\end{align}
We observe that the soft part and one jet function (twist-3 projection) are
the same as for $F_{1}$, but the outgoing jet function (twist-4 projection) is
different. We may expect that the soft rescattering for $F_{2}$ can also be
described in terms of a reduced diagram as in Fig.~\ref{hs-reduced-diagram}.

\subsection{QCD factorization for the soft rescattering picture}

The specific feature of the soft rescattering is the presence of two
subprocesses related to the two hard scales: a hard subprocess with typical scale
of order $Q^{2}$ and a hard-collinear subprocess with typical scale of order $\Lambda
Q$. Therefore description of such processes could be carried out in two
steps: first, one integrates over hard fluctuations so that the remaining degrees
of freedom describe hard-collinear and soft processes. From the previous analysis
we may conclude that such degrees of freedom include hard-collinear
(\ref{hc-in},\ref{hc-out}), collinear:%
\begin{eqnarray}
p_{c}  &  \sim & \left( p_{c}^{+}\sim Q,~p_{\bot}\sim\Lambda,~p_{c}^{-}\sim\Lambda
^{2}/Q \right),~~p_{c}^{2}\sim\Lambda^{2}, \nonumber \\
 p_{c}^{\prime}  &  \sim & \left( p_{c}^{\prime+}\sim\Lambda^{2}/Q,~p_{\bot}
\sim\Lambda,~p_{c}^{^{\prime}-}\sim Q \right),~\ p_{c}^{\prime2}\sim\Lambda^{2},
\end{eqnarray}
and soft,
\begin{equation}
p_{s}^{\mu}\sim\Lambda,~\ p_{s}^{2}\sim\Lambda^{2}.
\end{equation}
Therefore one needs the effective theory describing the dynamics of such a system.
Such effective theory, known as SCET, was
built already for the description of heavy quark decays and some other hadronic
reactions. Therefore we can apply it also for description of the soft
rescattering mechanism.

If $Q$ is large enough and $\Lambda Q\gg\Lambda^{2}$ one can further use
perturbation theory and factorize the hard-collinear fluctuations, leaving at the
end only collinear and soft modes which describe soft QCD dynamics.
Technically, such two-step factorization is described as matching of full QCD
onto the soft collinear effective theory at the scale $\mu=Q$ (SCET$_{I}$), which
is equivalent to calculating the hard coefficient functions in front of an
operator constructed from SCET$_{I}$ fields described above. The second step
is the matching of SCET$_{I}$ at the scale $\mu=\sqrt{\Lambda Q}$ to
SCET$_{II}$, which again corresponds to the pQCD calculation of hard-collinear
coefficient functions (which are usually called jet functions) in front of
operators constructed only from the collinear and soft fields. In the next
section, we perform the matching of QCD to the SCET$_{I}$ effective theory.

\section{Matching QCD to SCET$_{I}$ and resummation of leading logarithms}
\label{sec3}

\subsection{Soft Collinear Effective Theory}

In this section we briefly describe the main ingredients of SCET
\cite{Bauer:2000ew, Bauer2000, Bauer:2001ct, Bauer2001, BenCh, BenFeld03}. The
effective Lagrangian can be obtained from QCD Lagrangian by integrating over hard
fluctuations and performing a systematical expansion with respect to the small
dimensionless parameter $\lambda$ related to the large scale $Q$. \ We define
$\lambda\sim\sqrt{\Lambda/Q,}$where $\Lambda$ is the typical hadronic scale of
the order of a few hundred MeV. In general, the \ physical amplitude describing
a hard exclusive reaction can be defined in a convenient reference frame, for
instance the Breit frame. Then external particles usually are hard or collinear.
The fast moving hadron consists of energetic partons carrying collinear
momentum:%
\begin{equation}
p_{c}^{\mu}=(p_{c}\cdot n)\frac{\bar{n}^{\mu}}{2}+p_{\bot c}^{\mu}+(p_{c}%
\cdot\bar{n})\frac{n^{\mu}}{2}\equiv(p_{c}^{+},p_{\bot c},p_{c}^{-}%
),~\ p_{c}^{2}\sim\lambda^{4}Q^{2}.\label{mom}%
\end{equation}
The individual momentum components have the following scaling behavior:%
\[
p_{c}^{\mu}\sim Q(1,\lambda^{2},\lambda^{4}),
\]
as required by (\ref{mom}). However, as we could see in the example above, the relevant regions
could involve fluctuations with different momenta. We classify the different
regions following the terminology suggested in Refs.~\cite{BenF, Hill:2002vw}:
\textit{hard }$~p_{h}\sim Q(1,1,1),~\ $ \textit{semihard} $~p_{sh}\sim
Q(\lambda,\lambda,\lambda),$ \textit{hard-collinear} $p_{hc}\sim
Q(1,\lambda,\lambda^{2})$ or $p_{hc}^{\prime}\sim Q(\lambda^{2},\lambda,1),$
\textit{collinear} $p_{c}\sim Q(1,\lambda^{2},\lambda^{4})$ or $p_{c}^{\prime
}\sim Q(\lambda^{4},\lambda^{2},1)$ and \textit{soft} $p_{s}\sim Q(\lambda
^{2},\lambda^{2},\lambda^{2})$.

The large (small) components $\xi_{hc}^{\prime}$ of the quark fields describing
particles with momentum $p_{hc}^{\prime}$ have been introduced through
a decomposition of exact collinear quark fields $\psi_{hc}^{\prime}$:
\begin{equation}
\xi_{hc}^{\prime}(x)=\frac{ \setbox0=\hbox{$n$} \dimen0=\wd0 \setbox1=\hbox{/}
\dimen1=\wd1 \ifdim\dimen0>\dimen1 \rlap{\hbox to
\dimen0{\hfil/\hfil}} n \else \rlap{\hbox to \dimen1{\hfil$n$\hfil}} /
\fi \setbox0=\hbox{$\bar n$} \dimen0=\wd0 \setbox1=\hbox{/} \dimen1=\wd1
\ifdim\dimen0>\dimen1 \rlap{\hbox to \dimen0{\hfil/\hfil}} \bar n
\else \rlap{\hbox to \dimen1{\hfil$\bar n$\hfil}} / \fi }{4}\psi_{hc}^{\prime
},~\eta_{hc}^{\prime}(x)=\frac{ \setbox0=\hbox{$\bar n$} \dimen0=\wd0
\setbox1=\hbox{/} \dimen1=\wd1 \ifdim\dimen0>\dimen1 \rlap{\hbox to
\dimen0{\hfil/\hfil}} \bar n \else \rlap{\hbox to \dimen1{\hfil$\bar
n$\hfil}} / \fi \setbox0=\hbox{$n$} \dimen0=\wd0 \setbox1=\hbox{/}
\dimen1=\wd1 \ifdim\dimen0>\dimen1 \rlap{\hbox to \dimen0{\hfil/\hfil}} n
\else \rlap{\hbox to \dimen1{\hfil$n$\hfil}} / \fi }{4}\psi_{hc}^{\prime},
\end{equation}
with $\hat{n}~\xi_{hc}^{\prime}=0.$ The small components $~\eta_{hc}^{\prime}
$ are suppressed \ with respect to those of $\xi_{hc}^{\prime}$ by a factor
$\lambda^{2}\sim\Lambda/Q$ and are integrated out when constructing the effective Lagrangian.

Such definitions set the following scaling relations for the corresponding
effective fields:%
\begin{equation}
\xi_{hc}^{\prime}\sim\lambda,~\ \bar{n}\cdot A_{hc}^{\prime}\sim1,~A_{\bot
hc}^{\prime}\sim\lambda~\ ,n\cdot A_{hc}^{\prime}\sim\lambda^{2}%
,\label{hc:count}%
\end{equation}%
\begin{equation}
\xi_{c}^{\prime}\sim\lambda^{2},~\ \bar{n}\cdot A_{c}^{\prime}\sim1,~A_{\bot
c}^{\prime}\sim\lambda^{2}~\ ,n\cdot A_{c}^{\prime}\sim\lambda^{4}%
,\label{col:count}%
\end{equation}%
\begin{equation}
A_s^{\mu}\sim\lambda^{2},~\ q \sim\lambda^{3}, 
\label{soft:count}
\end{equation}
with $A_{hc}^{\prime \, \mu}$, $A_{c}^{\prime \, \mu}$, $A_{s}^{\prime \, \mu}$ denoting 
the gauge fields in the SCET, and $q$ the soft quark field. 

After integration over hard modes we reduce full QCD to the SCET$_{I}$ which
describes the interaction of particles with hard-collinear and soft momenta.
This theory still includes the particles with large virtuality of order
$\Lambda Q\gg\Lambda^{2}$ if $Q$ is large enough. Therefore, if possible, one
can perform a matching of the SCET$_{I}$ to the effective theory which contains only
collinear and soft particles (SCET$_{II}$). In present paper we consider in
detail the matching of QCD to SCET$_{I}$ and resummation of Sudakov
logarithms which arises due to the evolution of the SCET operators.

The effective action describing the interaction of the hard-collinear and soft
particles can be written as an expansion with respect to $\lambda~$\cite{BenCh,
BenFeld03} \footnote{There are two different technical formulations of SCET
developed in \cite{Bauer2000, Bauer2001, BenCh, BenFeld03}. In the present paper
we follow the technique suggested Beneke et al. in Ref.~\cite{BenCh}.}:
\begin{equation}
\mathcal{L}_{\text{SCET}_{I}}=\mathcal{L}_{\xi}^{(0)}+\mathcal{L}_{\xi}%
^{(1)}+\mathcal{L}_{q\xi}^{(1)}+\mathcal{O}(\lambda^{2})+\mathcal{L}%
_{\text{YM}}+\mathcal{L}_{s},\label{LSCET}%
\end{equation}
where%
\begin{equation}
\mathcal{L}_{\xi}^{(0)}=\int d^{4}x~\bar{\xi}_{hc}^{\prime}(x)\left(  in\cdot
D+g~n\cdot A_{s}(x_{-})+i\hat{D}_{\bot}~\frac{1}{i\bar{n}\cdot D}~i\hat
{D}_{\bot}\right)  \frac{\bar{n}}{2}\xi_{hc}^{\prime}(x),\label{L0xi}%
\end{equation}%
\begin{equation}
\mathcal{L}_{q\xi}^{(1)}=\int d^{4}x~\bar{\xi}_{hc}^{\prime}(x)i\hat{D}_{\bot
}W^{\prime}(x)q(x_{-})+\bar{q}(x_{-})W^{\prime\dag}(x)i\hat{D}_{\bot}\xi
_{hc}^{\prime}(x).\label{L1qxi}%
\end{equation}
where $x_{-}\equiv{\scriptstyle \frac12}(x\cdot\bar{n})n$. The fields $\{\xi
_{hc}^{\prime},$ $A_{hc}^{\prime}\}$ and $\{q,A_{s}\}$ describe hard-collinear
($(p_{hc}^{\prime}\cdot\bar{n})\sim Q$) and soft fields respectively,
whereas the covariant derivative reads$~iD=i\partial+gA_{hc}^{\prime}$. ~The Wilson lines
are defined as
\begin{equation}
W^{\prime}(x)=\text{P}\exp\left\{  ig\int_{-\infty}^{0}ds~\bar{n}\cdot
A_{hc}^{\prime}(x+s\bar{n})\right\}  .
\end{equation}

We do not write explicitly the gluon part $\mathcal{L}_{\text{YM}}$ because
we will not use it in this work. $\mathcal{L}_{s}$ is the usual QCD Lagrangian with the soft
fields. In Eqs.~(\ref{L0xi}) and (\ref{L1qxi}) we provide only the
contributions which will be relevant for our discussion.

The expressions in Eqs.~(\ref{L0xi}) and (\ref{L1qxi}) have definite (homogeneous)
scaling in $\lambda$ which is indicated by the number in the superscript brackets. In order to
achieve this the arguments of the soft fields are expanded with respect to
the small components of the position arguments\footnote{In momentum space such
"multipole " expansion corresponds to the expansion with respect to small
momentum components.}. From Eq.~(\ref{L0xi}) one can see that the soft-gluon
fields couple to the hard collinear fields only via the longitudinal component
$n\cdot A_{s}$. Using the field redefinition
\begin{equation}
\xi_{hc}^{\prime}(x)\rightarrow S_{n}(x_{-})\xi_{hc}^{\prime(0)}%
(x),~\ A_{hc}^{\prime}(x)\rightarrow S_{n}(x_{-})A_{hc}^{\prime(0)}%
(x)S_{n}^{\dag}(x_{-}),\label{dec-tr}%
\end{equation}
with the soft Wilson line%
\begin{equation}
S_{n}(x)=\text{P}\exp\left\{  ig\int_{-\infty}^{0}ds~n\cdot A_{s}%
(x+sn)\right\}  ,\label{Ws}%
\end{equation}
we can eliminate the soft field from the leading-order Lagrangian
(\ref{L0xi}). However the soft Wilson lines $S_{n}$ remain in the external
operators with soft fields in order to ensure the gauge invariance.

Obviously, all the above results also hold for the second collinear region with
momentum $(p_{hc}\cdot n)\sim Q$, by merely interchanging light-cone
vectors $n\leftrightarrow\bar{n}$ and substituting corresponding hard collinear
fields $\{\xi_{hc},$ $A_{hc}\}$.

%%%PRDv2
The formulation of SCET described above can be extended by introducing the so-called soft-collinear or messenger modes as 
discussed in \cite{Becher:2003qh}. However, 
 such particles have virtualities  which are much smaller then the typical hadronic scale  $p^{2}_{sc}\ll \Lambda^{2}$.
This situation was investigated in detail in several papers ( see e.g. \cite{Beneke:2003pa, Bauer:2003td, Manohar:2005az}). It was shown that the 
existence of such modes depends on the precise form of the IR regularization used in massless pQCD. 
Therefore it was suggested that in the processes with  real hadrons, where all nonperturbative 
effects have typical scales of order $\Lambda$, 
such low-mass degrees of freedom cannot appear because 
they are clearly an artefact of perturbation theory.  
Therefore we do not include them in the present considerations. 
   
%%%%

\subsection{Construction of the operator basis  and leading order coefficient functions}

%%%PRDv2
In this section we briefly  describe  the matching of QCD to the relevant leading-order
operators in the SCET$_{I}$.    The leading-order matching the e.m. current onto the SCET operators
 has been already introduced and studied earlier. In Ref.~\cite{Manohar:2003vb} it was used for a description of DIS at large
$x\rightarrow1$ and in Ref.~\cite{Becher:2007ty} for the description of Drell-Yan
production.  The matching onto subleading operators was also discussed in \cite{Chay:2005rz}.   
For the convenience of the reader we repeat here the main steps of these calculations in order to introduce required notations.  
%%%

In order to obtain the allowed SCET operators we take
into account the restrictions imposed by the SCET counting rules, gauge invariance,
and invariance under the reparametrization transformations. Explicit
construction of such operators can be performed in the same way as it was done
for heavy-to-light transitions in the works of Refs.~\cite{Chay:2002vy, Pirjol2002, Bauer2003,
Beneke2004, HillB2004}. The building blocks, invariant under collinear gauge
transformations are well known and read
\begin{align}
\left\{  \left(  \bar{\xi}_{hc}^{\prime}W^{\prime}\right)  ,~\left(  W^{\dag
}\xi_{hc}\right)  \right\}   &  \sim\lambda,\ \label{xi-jet}\\
\left\{  \left[  W^{\prime\dag}iD_{\mu}W^{\prime}\right]  ,~\ \left[  W^{\dag
}iD_{\mu}W\right]  \right\}   &  \equiv\left\{  \mathcal{A}_{\mu}^{\prime
},\mathcal{A}_{\mu}\right\}  \sim\lambda.\label{A-jet}%
\end{align}
In the terms with $[...]$, the derivative is only applied inside the brackets.

For the LO operator one can easily construct the expression which consists of two quark jets:%
\begin{equation}
O(s_{1},s_{2})=\left[  \bar{\xi}_{hc}^{\prime}W^{\prime}\right]  (s_{1}\bar
{n})\otimes\left[  W^{\dag}\xi_{hc}\right]  (s_{2}n ) \label{O2}%
\end{equation}
where we do not explicitly write the color and spinor indices for each jet and
the symbol $\otimes$ is used to stress that their indices are not contracted.
From the previous discussion it is clear that such an operator is relevant for
the Dirac FF $F_{1}$. \ For the case of Pauli FF $F_{2}$ we need the subleading
operator involving the transverse gluon field as in Eq.~(\ref{A-jet}). In this case it is
useful to take into account the constraints imposed by reparametrization
transformations. The details were already discussed in the literature, 
and we refer to Refs.~\cite{BenCh, Chay:2002vy, Pirjol2002, HillB2004} for these details. 
%%%PRDv2
%Taking account of the subleading corrections it is convenient
%to rewrite Eq.~(\ref{O2}) as%
%\begin{equation}
%O(s_{1},s_{2})= \left[  \left[  \bar{\xi}_{hc}^{\prime}W^{\prime}\right]
%(s_{2}\bar{n})\left(  1-\frac{i\overleftarrow{ \setbox0=\hbox{$\partial$}
%\dimen0=\wd0 \setbox1=\hbox{/} \dimen1=\wd1 \ifdim\dimen0>\dimen1
%\rlap{\hbox to
%\dimen0{\hfil/\hfil}} \partial\else \rlap{\hbox to
%\dimen1{\hfil$\partial$\hfil}} / \fi }_{\bot}}{i(\bar{n}\cdot\overleftarrow
%{\partial})}\frac{ \setbox0=\hbox{$\bar n$} \dimen0=\wd0 \setbox1=\hbox{/}
%\dimen1=\wd1 \ifdim\dimen0>\dimen1 \rlap{\hbox to
%\dimen0{\hfil/\hfil}} \bar n \else \rlap{\hbox to \dimen1{\hfil$\bar
%n$\hfil}} / \fi }{2}\right)  \right]  ~\otimes~~\left[  \left(  1-\frac{
%\setbox0=\hbox{$n$} \dimen0=\wd0 \setbox1=\hbox{/} \dimen1=\wd1
%\ifdim\dimen0>\dimen1 \rlap{\hbox to \dimen0{\hfil/\hfil}} n
%\else \rlap{\hbox to
%\dimen1{\hfil$n$\hfil}} / \fi }{2} \frac{i \setbox0=\hbox{$\partial$}
%\dimen0=\wd0 \setbox1=\hbox{/} \dimen1=\wd1 \ifdim\dimen0>\dimen1
%\rlap{\hbox to \dimen0{\hfil/\hfil}} \partial\else \rlap{\hbox to
%\dimen1{\hfil$\partial$\hfil}} / \fi _{\bot} } {i \left( n\cdot\partial
%\right)  } \right)  \left[  W^{\dag}\xi_{hc} \right]  (s_{1}n) \right]
%.\label{O2:der}%
%\end{equation}
The relevant for our consideration subleading operators can be
written as%
\begin{equation}
O_{\bar{n}}(s_{1},s_{2},s_{3})=\left[  \bar{\xi}_{hc}^{\prime}W^{\prime
}\right]  (s_{2}\bar{n}) \; \setbox0=\hbox{$\mathcal{A}$} \dimen0=\wd0
\setbox1=\hbox{/} \dimen1=\wd1 \ifdim\dimen0>\dimen1 \rlap{\hbox to
\dimen0{\hfil/\hfil}} \mathcal{A} \else \rlap{\hbox to
\dimen1{\hfil$\mathcal{A}$\hfil}} / \fi _{\bot}^{\prime}(s_{3}\bar{n}%
)~\otimes\left[  W^{\dag}\xi_{hc}\right]  (s_{1}n),\label{O3:out}%
\end{equation}%
\begin{equation}
O_{n}(s_{1},s_{2},s_{3})=\left[  \bar{\xi}_{hc}^{\prime}W^{\prime}\right]
(s_{2}\bar{n})\otimes\setbox0=\hbox{$\mathcal{A}$} \dimen0=\wd0
\setbox1=\hbox{/} \dimen1=\wd1 \ifdim\dimen0>\dimen1 \rlap{\hbox to
\dimen0{\hfil/\hfil}} \mathcal{A} \else \rlap{\hbox to
\dimen1{\hfil$\mathcal{A}$\hfil}} / \fi _{\bot}(s_{3}n)\left[  W^{\dag}%
\xi_{hc}\right]  (s_{1}n).\label{O3:in}%
\end{equation}
So that for matching of vector current \ we can write%
\begin{align}
\bar{q}(0)\gamma^{\mu}q(0)  &  = \int d\hat{s}_{1}d\hat{s}_{2}%
~~\text{tr}\left[  \tilde{C}^{\mu}(\hat{s}_{1},\hat{s}_{2})~O^{q}(s_{1}%
,s_{2})\right] \nonumber\\
&  +\int d\hat{s}_{1}d\hat{s}_{2}d\hat{s}_{3}~~\text{tr}\left[  ~\tilde
{C}_{\bar{n}}^{\mu}(\hat{s}_{1},\hat{s}_{2},\hat{s}_{3})~O_{\bar{n}}^{q}%
(s_{1},s_{2},s_{3})+~\tilde{C}_{n}^{\mu}(\hat{s}_{1},\hat{s}_{2},\hat{s}%
_{3})~O_{n}^{q}(s_{1},s_{2},s_{3})~\right] \label{Jem:match}%
\end{align}
where $\hat{s}_{i}\equiv s_{i}Q.$ The coefficient functions $\tilde{C}$ are
defined as matrices in the spinor and color indices and the trace has to be
understood in a sense of contractions of all the  indices between coefficient
functions and operators, for instance:
\begin{equation}
~\text{tr}\left[  C^{\mu}(\hat{s}_{1},\hat{s}_{2},Q/\mu)~O(s_{1}%
,s_{2})\right]  =\left[  ~\tilde{C}^{\mu}(\hat{s}_{1},\hat{s}_{2})\right]
_{\alpha\beta}~\left[  \bar{\xi}_{hc~\alpha}^{\prime}W^{\prime}\right]
(s_{1}\bar{n})\left[  W^{\dag}\xi_{hc~\beta}\right]  (s_{2}n).
\end{equation}

%%%PRDv2

 The further details of our
calculations are presented in  Appendix  A.  
It is convenient to pass in momentum space where the final result can be presented in the compact form:
%%%
\begin{equation}
\bar{q}(0)\gamma^{\mu}q(0)=C_{A}(Q,\mu)~O_{A}^{\mu_{\bot}}~-\frac{\left(  n^{\mu
}+\bar{n}^{\mu}\right)  }{Q}\int_{0}^{1}d\tau~C_{B}(\tau,Q,\mu)~O_{B}%
[\tau]+~...\label{Jem:LLexp}%
\end{equation}
where the scalar coefficient functions $C_{A,B}$ include all relevant
contributions with large logarithms, and the operators are defined as%
\begin{align}
~O_{A}^{\mu_{\bot}}  &  =\left(  \bar{\xi}_{hc}^{\prime}W^{\prime}\right)
(0)~\gamma_{\bot}^{\mu}~\left(  W^{\dag}\xi_{hc}\right)  (0)\equiv\left(
\bar{\xi}^{\prime}W^{\prime}\right)  \gamma_{\bot}^{\mu}\left(  W^{\dag}\xi\right)
~,\label{OA}\\
O_{B}[\tau]  &  =\left(  \bar{\xi}_{hc}^{\prime}W^{\prime}\right)
(0)\ \int\frac{d\hat{s}}{2\pi}~\left[  ~e^{-is(P^{\prime}\cdot\bar{n})~\tau}
\setbox0=\hbox{$\mathcal{A}$} \dimen0=\wd0 \setbox1=\hbox{/} \dimen1=\wd1
\ifdim\dimen0>\dimen1 \rlap{\hbox to \dimen0{\hfil/\hfil}} \mathcal{A}
\else \rlap{\hbox to \dimen1{\hfil$\mathcal{A}$\hfil}} / \fi _{\bot}^{\prime
}(s\bar{n})+e^{is(P\cdot n)~\tau} \setbox0=\hbox{$\mathcal{A}$} \dimen0=\wd0
\setbox1=\hbox{/} \dimen1=\wd1 \ifdim\dimen0>\dimen1 \rlap{\hbox to
\dimen0{\hfil/\hfil}} \mathcal{A} \else \rlap{\hbox to
\dimen1{\hfil$\mathcal{A}$\hfil}} / \fi _{\bot}(sn)\right]  \left(  W^{\dag
}\xi_{hc}\right)  (0)\nonumber\\
& \equiv\left(  \bar{\xi}^{\prime}W^{\prime}\right)  
\left[
\setbox0=\hbox{$\mathcal{A}$} \dimen0=\wd0 \setbox1=\hbox{/} \dimen1=\wd1
\ifdim\dimen0>\dimen1 \rlap{\hbox to \dimen0{\hfil/\hfil}} \mathcal{A}
\else \rlap{\hbox
to \dimen1{\hfil$\mathcal{A}$\hfil}} / \fi _{\bot}^{\prime}(\tau)+
\setbox0=\hbox{$\mathcal{A}$} \dimen0=\wd0 \setbox1=\hbox{/} \dimen1=\wd1
\ifdim\dimen0>\dimen1 \rlap{\hbox to \dimen0{\hfil/\hfil}} \mathcal{A}
\else \rlap{\hbox
to \dimen1{\hfil$\mathcal{A}$\hfil}} / \fi _{\bot}(\tau)\right]  \left(
W^{\dag}\xi \right).
 \label{OB}
\end{align}
From the tree-level calculations it follows that
\begin{equation}
C_{A}(Q,\mu=Q)=1+O(\alpha_{S}),~\ C_{B}(\tau,Q,\mu=Q)=1+O(\alpha
_{S})~.\ \label{CABtree}%
\end{equation}
Note that the SCET operators depend also on the renormalization scale $\mu$ that
was ignored for simplicity.

%Eqs.~(\ref{Jem:LLexp}) and (\ref{CABtree}) can be considered as the main
%technical result of this subsection\footnote{After publication of the first  version of paper we have been informed that 
%the similar results were obtained earlier
%%%

\subsection{Resummation of large logarithms}

The next important step is the resummation of the large logarithms or,
equivalently, the solution for the evolution of the SCET$_{I}$ operators. As
we described above, we expect that the scale for the remaining hard-collinear
subprocesses is of order $\Lambda Q$. Therefore it is natural to set the 
factorization scale $\mu^{2}$ to be of order $\Lambda Q$. However, we then obtain
in pQCD large logarithms $\ln Q^{2}/\mu^{2}$ which must be resummed to all
orders. Such resummation can be easily performed with the help of
the renormalization group (RG) and has been carried out for many
applications. We therefore only briefly describe the main steps and
provide the final results.

%%%PRDv2
%We start our discussion from the LO operator, Eq.~(\ref{OA}),  and want to determine its scale dependence. 
%Using standard technique, we introduce the renormalized and bare
%operators and write%
%\begin{equation}
%\left[  \left(  \bar{\xi}^{\prime}W^{\prime}\right)  \gamma_{\bot}^{\mu
%}\left(  W^{\dag}\xi\right)  \right]  ^{\text{bare}}=Z~\left[  \left(
%\bar{\xi}^{\prime}W^{\prime}\right)  \gamma_{\bot}^{\mu}\left(  W^{\dag}%
%\xi\right)  \right]  _{R},
%\end{equation}
%where we assume the absence of  operator mixing. A perturbative expansion
%of the renormalization constant yields
%\begin{equation}
%Z=1+\frac{1}{\varepsilon}~\left(  \frac{\alpha_{s}}{4\pi}Z_{LO}^{(1)}+\left(
%\frac{\alpha_{s}}{4\pi}\right)  ^{2}Z_{NLO}^{(1)}+~...\right)  +~O\left(
%\varepsilon^{-2}\right)  ~\ .
%\end{equation}
%and the corresponding anomalous dimension reads:
%\begin{equation}
%\tilde{\gamma}\left(  \alpha_{s}\right)  =\mu\frac{d}{d\mu}\ln Z=-2\alpha
%_{s}\frac{\partial}{\partial\alpha_{s}}Z^{(1)}.\label{gamZ}%
%\end{equation}
%The RG equation for the renormalized operator $O_{R}^{\mu}(\mu)$ can be written as
%\begin{equation}
%\frac{d}{d\ln\mu}O_{R}^{\mu}(\mu)=-~\tilde{\gamma}\left(  \alpha_{s}\right)
%~O_{R}^{\mu}(\mu).
%\end{equation}
%It can easily be translated to a RG equation for the corresponding coefficient
%function:%

We start our discussion from the coefficient function $C_{A}$ in front of the LO operator, Eq.~(\ref{OA}).   The corresponding RG equation reads
\begin{equation}
\frac{d}{d\ln\mu}~C_{A}(Q,\mu)=C_{A}(Q,\mu)~\tilde{\gamma}\left(  \alpha
_{s}\right)  ,~\ \ C_{A}(Q,\mu=Q)=1+\mathcal{O}(\alpha_{s}),\label{RGE1}%
\end{equation}
The anomalous dimension $\tilde{\gamma}\left(  \alpha_{s}\right)$  is defined by  renormalization of the operator $O_{A}$ ( see, e.g.,  \cite{Manohar:2003vb}). 
%
%$\tilde{\gamma}\left(  \alpha_{s}\right) $ :
%\begin{equation}
%\tilde{\gamma}\left(  \alpha_{s}\right)  =-\frac{\alpha_{s}(\mu)}{4\pi}%
%~C_{F}\left(  6+4\ln\left[  \frac{\mu^{2}}{Q^{2}}\right]  \right).
%\label{gammaO}%
%\end{equation}
%
%In order to find the explicit expression for the anomalous dimension
%$\tilde{\gamma}\left(  \alpha_{s}\right)  $ one has to compute the diagrams
%shown in Fig.\ref{scet-nlo-diagrams} . 
%\begin{figure}[th]
%\begin{center}
%\includegraphics[
%natheight=0.470500in,
%natwidth=5.076400in,
%height=0.4705in,
%width=5.0764in
%]{SCET-NLO-diagrams.pdf}
%\end{center}
%\caption{One-loop diagrams required for renormalization of the leading order
%operator. The soft gluon line is indicated by index $s$. }%
%\label{scet-nlo-diagrams}%
%\end{figure}Using dimension regularization ($\overline{MS}$ scheme) we
%confirm the result derived before in Ref.~\cite{Manohar:2003vb}):
%\begin{equation}
%\tilde{\gamma}\left(  \alpha_{s}\right)  =-\frac{\alpha_{s}(\mu)}{4\pi}%
%~C_{F}\left(  6+4\ln\left[  \frac{\mu^{2}}{Q^{2}}\right]  \right)
%.\label{gammaO}%
%\end{equation}
%The non-local contribution $\sim\ln Q$ arises, as usually, due to the simultaneous
%presence of the soft and collinear modes. In performing the above calculation we used
%the transverse momentum as intermediate IR regularization. 
It is well known that to all orders the anomalous dimension $\tilde{\gamma}$  can be represented as%
\begin{equation}
\tilde{\gamma}\left(  \alpha_{s}\right)  =-~\Gamma_{\text{cusp}}(\alpha
_{s})\ln\frac{\mu^{2}}{Q^{2}}~+\gamma\left(  \alpha_{s}\right), 
\label{gam:str}%
\end{equation}
where the coefficient in front of the logarithm in Eq.~(\ref{gam:str}) is known as the
universal cusp anomalous dimension, and controls the leading Sudakov double
logarithms. Such specific term is usual when Sudakov logarithms appear 
for the quantity under consideration. The single-logarithmic evolution is controlled by the
$\gamma\left(  \alpha_{s}\right)  $. 
%Using Eq.~(\ref{gam:str}), the RG equation (\ref{RGE1}) reads
%\begin{equation}
%\mu\frac{d}{d\mu}~C_{A}(Q,\mu)=\left[  -~\Gamma_{\text{cusp}}(\alpha_{s}%
%)\ln\frac{\mu^{2}}{Q^{2}}~+\gamma\left(  \alpha_{s}\right)  \right]
%~C_{A}(Q,\mu),
%\label{RGE2}
%\end{equation}
%with%

The solution of Eq.~(\ref{RGE1}) provides a systematic resummation of large logarithms
in pQCD.  
In order to find $C_{A}$ in the next-to-leading logarithmic (NLL) approximation,
\begin{equation}
C_{A}^{\text{NLL}}\sim\exp\left\{  \sum a_{n}\alpha_{s}^{n}\ln^{n+1}%
+b_{n}~\alpha_{s}^{n}\ln^{n}\right\}, 
\label{LLstruktur}%
\end{equation}
one needs to know the 2-loop cusp anomalous dimension $\Gamma_{0,1}$ and the
leading-order term $\gamma_{1}$:
\begin{equation}
\Gamma_{\text{cusp}}(\alpha_{s})=\frac{\alpha_{s}(\mu)}{4\pi}\Gamma
_{0}+\left(  \frac{\alpha_{s}(\mu)}{4\pi}\right)  ^{2}\Gamma_{1}+~...~,~~
\gamma(\alpha_{s})=\frac{\alpha_{s}(\mu)}{4\pi}\gamma_{1}+~...~.
\end{equation}
where \cite{ Korch2, Manohar:2003vb}
\begin{equation}
\Gamma_{0}=4C_{F},~ \Gamma_{1}=4C_{F}\left[  \left(  \frac{67}{9}-\frac{\pi^{2}}{3}\right)
C_{A}-\frac{10}{9}n_{f}~\right] ,~ \gamma_{1}=-6C_{F}.
\end{equation}

The explicit NLL solution reads%
\begin{equation}
C_{A}^{\text{NLL}}(Q,\mu)=e^{-S(Q,\mu_{h},\mu)}~U_{A}(\mu_{h},\mu)\left[
1+\mathcal{O}(\alpha_{s}(\mu_{h}))~\right]  ,\label{CALL}%
\end{equation}
where
\begin{align}
S(Q,\mu_{h},\mu)  &  =-\frac{\Gamma_{0}}{\beta_{0}}\ln r\ln\frac{\mu_{h}}%
{Q}+\frac{\Gamma_{0}}{2\beta_{0}^{2}}\left[  \frac{4\pi}{\alpha_{s}(\mu_{h}%
)}\left(  \ln r-1+\frac{1}{r}\right)  -\frac{\beta_{1}}{2\beta_{0}}\ln
^{2}r\right. \nonumber\\
&  \left.  +\left(  \frac{\Gamma_{1}}{\Gamma_{0}}-\frac{\beta_{1}}{\beta_{0}%
}\right)  \left[  r-1-\ln r\right]  ~\right]  ,\label{S}%
\end{align}%
\begin{equation}
U_{A}(\mu_{h},\mu)=r^{-\frac{\gamma_{1}}{2\beta_{0}}}.\label{UA}%
\end{equation}
with $r=\alpha_{s}(\mu)/\alpha_{s}(\mu_{h})>1$ and $\beta-$function
coefficients%
\begin{equation}
\beta_{0}=\frac{11}{3}C_{A}-\frac{2}{3}n_{f},~\ \beta_{1}=\frac{34}{3}%
C_{A}^{2}-\left(  \frac{10}{3}C_{A}+2C_{F}\right)  n_{f}.\label{bet}%
\end{equation}
In Eq.(\ref{CALL}) we assume that evolution is running from the initial scale
$\mu_{h}$ (which should be of order $Q$) \ to scale $\mu$~of order
$(Q\Lambda)^{1/2}$.

A similar technique can  also be used for the subleading operator of Eq.~(\ref{OB}).
Notice that in this case our calculation also provides the practical check for
the existence of the convolution integral in Eq. (\ref{Jem:LLexp}). If it does not
exist then our suggestion about the factorization must be reconsidered.

\begin{figure}[th]
\begin{center}
\includegraphics[
natheight=0.949600in,
natwidth=4.342200in,
height=0.9496in,
width=4.3422in
]{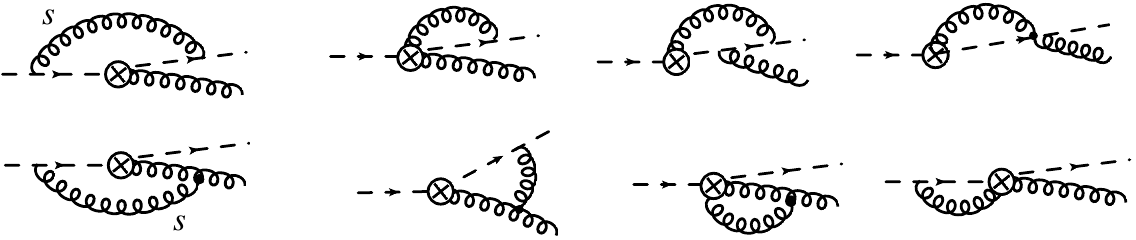}
\end{center}
\caption{One-loop diagrams required for renormalization of the three-particle
SCET$_{I}$ operator. As before, soft gluon lines indicated by an index $s$. The
wave function renormalization diagram is not shown for simplicity. }%
\label{scet-operator-renorm-diagrams}%
\end{figure}

In order to find the anomalous dimension one has to compute the diagrams shown
in Fig.\ref{scet-operator-renorm-diagrams}. 
The operator RG equation reads 
\begin{equation*}
\mu\frac{d}{d\mu}O_{R}[\tau]=-\int d\tau^{\prime}\gamma\lbrack\tau
,\tau^{\prime}]~O_{R}[\tau^{\prime}], 
\end{equation*}
with the evolution kernel $\gamma\lbrack\tau,\tau^{\prime}]$:%
\begin{equation}
\gamma\lbrack\tau,\tau^{\prime}]=\left\{ -~\delta(\tau-\tau^{\prime})\Gamma_{%
\text{cusp}}[\alpha_{s}]\ln\left( \frac{{\mu}^{2}}{Q^{2}}\right) +\frac{%
\alpha_{s}}{\pi}V[\tau,\tau^{\prime}]\right\} ,   \label{gam-t}
\end{equation}
where (c.f. \cite{Chay:2005rz})
\begin{align}
V[\tau,\tau^{\prime}] = &   ~-\delta(\tau-\tau^{\prime})\left(
C_{F}\left[  \frac{5}{2}-\ln\bar{\tau}\right]  +\frac{C_{A}}{2}\ln\frac
{\bar{\tau}}{\tau}\right)  -\frac{C_{A}}{2}\left[  \frac{\theta(\tau
<\tau^{\prime})}{\left(  \tau^{\prime}-\tau\right)  }+\frac{\theta
(\tau^{\prime}<\tau)}{\left(  \tau-\tau^{\prime}\right)  }\right]
_{+}   \notag \\
&  +\left(  C_{F}-\frac{C_{A}}{2}\right)  ~\left[  \frac{\tau^{\prime}}%
{\bar{\tau}^{\prime}}~\theta(\tau^{\prime}<\bar{\tau})+\theta(\bar{\tau}%
<\tau^{\prime})\frac{\bar{\tau}}{\tau}\right]  +~C_{F}~\bar{\tau}
 \notag  \\
&   -\frac{C_{A}}{2}\left[  \theta(\tau<\tau^{\prime})\frac{\bar{\tau}%
}{\tau^{\prime}}\left(  \frac{\tau^{\prime}}{\bar{\tau}}-\frac{3}{2}\right)
+\theta(\tau^{\prime}<\tau)\frac{\bar{\tau}}{\bar{\tau}^{\prime}}\left(
\frac{3}{2}-\frac{\bar{\tau}^{\prime}}{\tau}-\frac{1}{\bar \tau}\right)  \right]  ,   
\label{V-t}
\end{align}
where the prescription $[...]_{+}$ is defined for the
symmetrical kernel $f(\tau,\tau^{\prime})=f(\tau^{\prime},\tau)$ as%
\begin{equation*}
\left[ f(\tau,\tau^{\prime})\right] _{+}=\int\tau^{\prime}f(\tau
,\tau^{\prime})\left[ \phi(\tau^{\prime})-\phi(\tau)\right] . 
\end{equation*}
Computing the convolution integral with the LO $C_{B}[\tau]=1$ yields the well
defined expression: 
\begin{equation}
\int_{0}^{1}d\tau^{\prime}V[\tau^{\prime},\tau]=-\left[
2C_{F}-\frac{3}{8}C_{A}\right]=-\gamma_{B} , 
\label{eigenf}
\end{equation}
which does not depend on $\tau$. Hence we can conclude that the leading
logarithmic convolution integral in Eq.~(\ref{Jem:LLexp}) is also well defined.

\begin{figure}[th]
\begin{center}
\includegraphics[
height=1.727in,
width=4.7686in
]{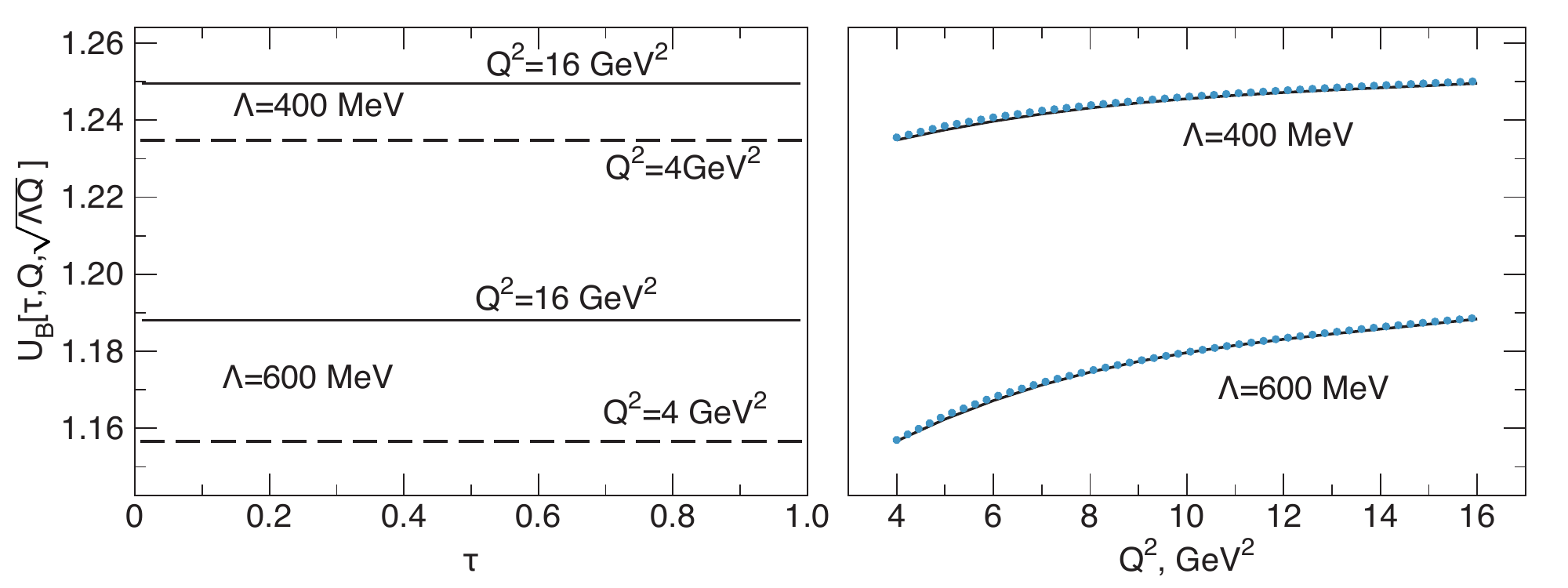}
\end{center}
\caption{Left panel: numerical evaluation of the evolution kernel $U_{B}[\tau
,Q,\sqrt{Q\Lambda}]$ as a function of $\tau$ at fixed $Q^2$. Right panel:
$U_{B}[\tau,Q,\sqrt{Q\Lambda}]$ as a function of $Q$ at fixed $\tau=0.5$ for
two different values of $\Lambda$ (solid curves). The approximate solution
$U^{\text{app} }_{B} \left[ \mu_{h},\mu\right] $ of Eq.~(\ref{Uapp}) is shown by dotted (blue)
curves. }
\label{ukernel}
\end{figure}

The corresponding RG equation for the coefficient function reads:%
\begin{align}
\mu\frac{d}{d\mu}C_{B}(\tau,Q;\mu)=- 
%&
\Gamma_{\text{cusp}}[\alpha_{s}]\ln\left(  \frac{{\mu}^{2}}{Q^{2}}\right)
C_{B}(\tau,Q;\mu)
%\nonumber \\  & 
+\frac{\alpha_{s}}{\pi}\int_{0}^{1}d\tau^{\prime}~V[\tau^{\prime},\tau
]C_{B}(\tau^{\prime},Q;\mu)~.\label{CB-RGe}%
\end{align}
Similar equations have been studied already in heavy-light decays (see, e.g.,
Refs.~\cite{HillB2004, Beneke:2005gs}). The NLL solution of this equation can be
written as
\begin{equation}
C_{B}^{\text{NLL}}(\tau,Q;\mu)=e^{-S(Q;\mu_{h},\mu)}\int_{0}^{1}d\tau^{\prime
}U\left[  \tau,\tau^{\prime};\mu_{h},\mu\right]  C_{B}^{(0)}(\tau^{\prime
},Q;\mu_{h}),\label{C2LL}%
\end{equation}
where the evolution kernel satisfies the integro-differential equation%
\begin{equation}
\mu\frac{d}{d\mu}U\left[  \tau,\tau^{\prime};\mu_{h},\mu\right]  =\frac
{\alpha_{s}}{\pi}\int_{0}^{1}d\tau^{\prime\prime}~V[\tau^{\prime\prime}%
,\tau]U\left[  \tau^{\prime\prime},\tau^{\prime};\mu_{h},\mu\right]
\label{Ueq}%
\end{equation}
with initial condition $U\left[  \tau,\tau^{\prime};\mu_{h},\mu_{h}\right]
=\delta(\tau-\tau^{\prime}).~$Recall that, in order to sum the large logarithms the
initial scale $\mu_{h}$ should be of order $Q$, and the evolution ends at
$\mu$ of order $(\Lambda Q)^{1/2}$. The Sudakov factor $S(Q;\mu_{h},\mu)$ is the
same as in Eq.~(\ref{S}). Taking into account that at NLL approximation the
initial condition $C_{B}^{(0)}(\tau^{\prime},Q;\mu_{h})$ is given by
the tree-level expression Eq.~(\ref{CABtree}), one can perform 
the integration over $\tau^{\prime}$ in Eq.(\ref{C2LL}), yielding~:
\begin{equation}
C_{B}^{\text{NLL}}(\tau,Q,\mu)=e^{-S(Q;\mu_{h},\mu)}U_{B}\left[  \tau~;\mu
_{h},\mu\right]  ,\label{CBLL}%
\end{equation}
with%
\begin{equation}
\mu\frac{d}{d\mu}U_{B}\left[  \tau;\mu_{h},\mu\right]  =\frac{\alpha_{s}}{\pi
}\int_{0}^{1}d\tau^{\prime}~V[\tau^{\prime},\tau]U_{B}\left[  \tau^{\prime
};\mu_{h},\mu\right] \label{U:RGE}%
\end{equation}
and $U_{B}\left[  \tau;\mu_{h},\mu_{h}\right]  =1$. The solution of this equation
can be found numerically. We have found that to a very good accuracy the
approximate solution can be written as
\begin{equation}
U_{B}\left[  \tau;\mu_{h},\mu\right] \approx U^{\text{app}}_{B}\left[  \mu
_{h},\mu\right] = \left( \frac{\ln[\mu_{h}/\Lambda^{(n_{f})}] }{\ln
[\mu/\Lambda^{(n_{f})}]}\right) ^{2\gamma_{B}/\beta_{0}}\label{Uapp}%
\end{equation}
with effective anomalous dimension $\gamma_{B}$ defined in Eq.(\ref{eigenf})
and with soft scale $\Lambda^{(n_{f})}$ used for calculating the running coupling
$\alpha_{s}$. To avoid confusion let us note that the soft scale $\Lambda$
which we used to define the hard-collinear scale $\sim Q\Lambda$ is different
$\Lambda\neq\Lambda^{(n_{f})}$. This difference provides the slow dependence
on $Q$ in the approximate solution of Eq.~(\ref{Uapp}). In Fig.~\ref{ukernel}
we show $U_{B}\left[  \tau~;Q,\sqrt{Q\Lambda}\right]  $ computed
for different values of $Q$ and $\Lambda$ and 
compare it with the approximate solution
$U^{\text{app}}_{B}$. We obtained that for all considered cases to a very good
accuracy the kernel does not depend on the momentum fraction $\tau$ and evolves 
quite slowly with respect to $Q$ according to (\ref{Uapp}). At the end let us
note that the similar approximate solution for the single-logarithmic
evolution was also found for the heavy-light current in Ref.~\cite{Beneke:2005gs}.

The obtained results already lead to some qualitative features
when applying this formalism to the proton FFs, as will be discussed in 
the next section.

\section{ QCD factorization at leading order using the SCET approach}
\label{sec4}

In this section, we consider the matching on SCET$_{II}$ and discuss the
factorization formula for the soft rescattering mechanism. 
We perform an analysis of the dominant regions using the
methods of the effective theory. 
%%%PRDv2
We restrict our consideration only to the  terms
relevant at leading logarithmic approximation both at SCET$_{I}$ and SCET$_{II}$  levels. 
%%%
The general, all order analysis is much more complicated and goes
beyond our present considerations. However, using the results obtained above, 
we suggest a leading
order factorization formula (i.e. restricted by leading logarithms) for the Dirac FF $F_{1}$ which includes soft and hard
rescattering contributions.

In this section we would like to demonstrate that the soft rescattering
contribution can be estimated in SCET using the counting rules (\ref{hc:count}%
-\ref{soft:count}) without direct calculation of the diagrams as we did
before. Such counting is an important ingredient of a factorization proof and
can be considered as a quite general argument in support of the nontrivial soft
rescattering contribution.

In addition to the field relations, we also need the counting of the energetic
(collinear) hadronic state. It reads%
\begin{equation}
\left\vert p_{c}\right\rangle \sim\lambda^{-2},
\end{equation}
and follows from the conventional normalization (\ref{norm}).

Let us start from the well known hard rescattering picture. From existing
results one can easily obtain:
\begin{equation}
\left\langle p^{\prime}\left\vert J_{\bot}^{\mu}\right\vert p\right\rangle
\overset{Q\rightarrow\infty}{\sim}\bar{N}^{\prime}\gamma_{\bot}^{\mu}%
N~F_{1}\sim\bar{N}^{\prime}\gamma_{\bot}^{\mu}N\frac{f_{N}^{~2}}{Q^{4}}%
\sim~\bar{N}^{\prime}\gamma_{\bot}^{\mu}N~\lambda^{8},\label{F1HR}%
\end{equation}
and%
\begin{equation}
\left\langle p^{\prime},\lambda^{\prime}\left\vert ~J_{\Vert}^{\mu}\right\vert
p,\lambda\right\rangle =\frac{(p+p^{\prime})^{\mu}}{m_{N}}~\bar{N}^{\prime
}\hat{1}N~F_{2}(Q^{2})\overset{Q\rightarrow\infty}{\sim}\frac{Q}{m_{N}}%
~\frac{f_{N}^{2}~m_{N}^{2}}{Q^{6}}~\bar{N}^{\prime}\hat{1}N~\sim\bar
{N}^{\prime}\hat{1}N~\lambda^{10}.
\end{equation}
where the nonperturbative scale is presented by the nucleon mass $m_{N}\sim\Lambda$
and the overall normalization constant for the nucleon distribution
amplitude$f_{N}\sim\Lambda^{2}$.

The same counting in SCET is obtained directly from the dimensional analysis 
of the leading operators constructed from the collinear and soft fields, which
represent the main degrees of freedom of SCET$_{II}$. However, difficulties
arise due to nonlocal contributions with inverse powers of $1/\Lambda$
momenta if one computes time-ordered products involving 
SCET$_{I}$ fields. We shall follow the strategy suggested in
Refs.~\cite{Bauer:2002aj, Bauer:2003mga}. Then matching onto SCET$_{II}$,
\begin{equation}
\text{SCET}_{I}\left[  p_{hc}^{2}\sim Q\Lambda,~k_{s}^{2}\sim\Lambda
^{2}\right]  \overset{\mu^{2}\sim Q\Lambda}{\longrightarrow}\text{SCET}%
_{II}\left[  p_{c}^{2}\sim\Lambda^{2},~k_{s}^{2}\sim\Lambda^{2}\right]
\end{equation}
can be performed in two steps: decoupling the soft fields from the hard
collinear modes using field redefinitions expressed by Eq.~(\ref{dec-tr}), 
and subsequently matching hard-collinear modes to collinear ones%
\begin{equation}
\{\xi_{hc}^{\prime(0)},~A_{hc}^{\prime(0)},\xi_{hc}^{(0)},~A_{hc}%
^{(0)}\}\rightarrow\{\xi_{c}^{\prime},~A_{c}^{\prime},\xi_{c},~A_{c}\},
\end{equation}
lowering the off-shellness of the external hard-collinear fields. Notice that
the last step changes the power counting of the fields from Eq.~(\ref{hc:count})
to Eq.~(\ref{col:count}).

In the case of the hard rescattering we perform the matching of QCD directly
onto SCET$_{II}$. Therefore, the power counting is simple because it does not
involve the intermediate effective theory.

Matching for the Dirac FF $F_{1}$\ involves the six-quark operator constructed
only from the collinear fields $\xi_{c}^{\prime}$, $\xi_{c}$ and Wilson lines
with longitudinal collinear gluons $\bar{n}\cdot A^{\prime}$ and $n\cdot A$
respectively. It is the product of two twist-3 \ 3-quark operators which
define the leading twist nucleon DA (\ref{DA:def}). Then using
(\ref{col:count}) one obtains
\begin{equation}
\left\langle p^{\prime}\left\vert J_{\bot}^{\mu}\right\vert p\right\rangle
^{(h)}\sim\left\langle p^{\prime}\left\vert \bar{\xi}_{c}^{\prime}\bar{\xi
}_{c}^{\prime}\bar{\xi}_{c}^{\prime}|0\left\rangle \ast C_{\bot}^{\mu}%
(Q)\ast\right\langle 0|\xi_{c}\xi_{c}\xi_{c}\right\vert p\right\rangle
\sim\lambda^{8}~\bar{N}^{\prime}\gamma_{\bot}^{\mu}N.\label{Jbot}%
\end{equation}
For the helicity flip FF $F_{2}$, the matching involves
the product of twist-3 and twist-4 operators, as we discussed in 
Sec.~\ref{sec3}. Schematically this situation can be described by substituting $\xi
_{c}\rightarrow\eta_{c}\sim\xi_{c}/Q$. Then%
\begin{align}
\left\langle p^{\prime}\left\vert J_{\Vert}^{\mu}\right\vert p\right\rangle
^{(h)}\sim\left\langle p^{\prime}\left\vert \bar{\xi}_{c}^{\prime}\bar{\xi
}_{c}^{\prime}\bar{\xi}_{c}^{\prime}|0\left\rangle \ast C_{\Vert}^{\mu}%
(Q)\ast\right\langle 0|\eta_{c}\xi_{c}\xi_{c}\right\vert p\right\rangle  &
\sim(n+\bar{n})^{\mu}~\lambda^{10}~\bar{N}^{\prime}\hat{1}N\\
& \sim\lambda^{12}~\frac{(p+p^{\prime})^{\mu}}{m}\bar{N}^{\prime}\hat
{1}N~,\label{JL:HS}%
\end{align}
and we obtain that $F_{2}$ is suppressed as $1/Q^{2}$ relative to $F_{1}$ as
it should be.

In order to estimate the soft rescattering contribution one has to perform a more
complicated analysis with the two-step matching: QCD$\rightarrow$SCET$_{I}$
$\rightarrow$SCET$_{II}$. \ The matching onto SCET$_{I}$ has been done in
Section~\ref{sec4} and for the electromagnetic current at leading order it yields the formula
Eq.~(\ref{Jem:LLexp}). \ In matching onto SCET$_{II}$ we need at least six
collinear quarks in order to have overlap with the \textit{in} and
\textit{out} nucleon states. Next, guided by the perturbative QCD calculations from Sec.~\ref{sec2}
we need higher order vertices $\mathcal{L}_{\xi q}^{(n)}$ in order to describe
the soft spectators in the intermediate state.

%%%PRDv2
\subsection{Leading-order SCET analysis for Dirac FF $F_{1}$}
Let us start with a discussion for the FF $F_{1}$. To leading order in $1/Q$  we can restrict our  consideration by the 
first term in Eq.~(\ref{Jem:LLexp}). Therefore
our task is to compute the time-ordered product  which contributes to the matrix element
\begin{equation}
\left\langle p^{\prime}\left\vert J_{\bot}^{\mu}~(0)\right\vert p\right\rangle
^{(s)}\simeq \left\langle p^{\prime}\right\vert ~T\left\{  C_{A}~O_{A}^{\mu_{\bot}}~e^{i\mathcal{L}^{(\bar{n})}_{\text{SCET}_{I}}
+i\mathcal{L}^{(n)}_{\text{SCET}_{I}}+i\mathcal{L}_{s}}\right\}
~\left\vert p\right\rangle .\label{FFme}%
\end{equation}
The calculations amount to integrating out hard-collinear modes and, if possible, to deriving the expression
for the vector current $J_{\bot}^{\mu}$ in terms of SCET$_{II}$ collinear and soft fields which can be schematically written as 
\begin{eqnarray}
&& T \left \{
  C_{A}O_{A}^{ {\mu}_{\bot} } 
~e^{i\mathcal{L}^{(\bar{n})}_{\text{SCET}_{I}}+i\mathcal{L}^{(n)}_{\text{SCET}_{I}}+i\mathcal{L}_{s}} 
\right \}
\nonumber \\
&& \simeq C_{A} \text{\bf Tr}\left[  \gamma^{\mu}_{\bot} \,
T\left\{ \mathbf{O}_{out} \left(  \varphi_{c}^{\prime} \right) ~e^{i\mathcal{L}_{c}^{(\bar{n})} }  \right\} 
\ast \mathbf{J}^{\prime} \ast
T\left\{  \mathbf{S} \left(  \varphi_{s} \right) ~e^{i\mathcal{L}_{s} }  \right\} 
\ast \mathbf{J} \ast 
 T\left\{ \mathbf{O}_{in} \left(  \varphi_{c} \right) ~e^{i\mathcal{L}_{c}^{({n})} }   \right\}
\,  
\right],
\label{f-f}
\end{eqnarray}
where $\mathbf J$ and $\mathbf{J}^{\prime}$ are jet functions, 
$\mathcal{L}_{c}^{(n)}$ denotes the collinear Lagrangian,  {\bf Tr} denotes contractions over the Dirac and color indices which are not shown explicitly,
and where we used  the notation $\varphi_{c,s}\equiv\{\xi_{c,s},A_{c,s}\}$. We also 
assumed that the collinear operators $\mathbf{O}_{in,out}$  have nontrivial overlap with nucleon states.
The Dirac matrix $\gamma^{\mu}_{\bot}$ is associated  with  the vertex of the $O_{A}$ operator. 
It is clear that the existence of the factorized  representation (\ref{f-f}) is equivalent to establishing the  factorization theorem. 
Guided by our QCD analysis, carried out in Sec.~II,  
we demonstrate below that at leading order in $1/Q$ such a contribution definitely exists. 
 For simplicity, we  restrict our consideration to a leading-order analysis in $\alpha_{s}$, 
 and consider it as a first step towards a complete proof.

 Obviously, the time-ordered
product in left hand side of (\ref{f-f}) can be represented as the product of two:
\begin{equation}
T\left\{  ...\right\}  =T\left(  \bar{\xi}_{hc}^{\prime}W^{\prime
}(0)~e^{i\mathcal{L}^{(\bar{n})}_{\text{SCET}_{I}} }~\right)  \gamma_{\bot}^{\mu}~T\left(
W\xi_{hc}(0)~ e^{i\mathcal{L}^{(n)}_{\text{SCET}_{I}} } \right) \equiv T_{out} ~\gamma_{\bot}^{\mu} ~T_{in}
\label{Tprod}%
\end{equation}
where we ``freeze'' the soft fields, i.e.,  consider them as external. As  calculations of the each
of the $T$ products are almost identical, we only consider one of them. The result of the integration over 
hard-collinear modes can be schematically written as
\begin{equation}
 T_{out} =T\left(  \bar{\xi}_{hc}^{\prime}W^{\prime}~e^{i\mathcal{L}^{(\bar{n})}_{\text{SCET}_{I}}%
}~\right)  \ \simeq\mathbf{~~}\bar{\xi}_{c}^{\prime}\bar{\xi}_{c}^{\prime}%
\bar{\xi}_{c}^{\prime}\ast\mathbf{J}^{\prime}\ast qq~,\ \ \ \label{Jdef}%
\end{equation}
where the last equation shows the desired structure in terms of collinear
and soft fields. Combining such results  for $T_{in}$ and $T_{out}$  we obtain desired representation (\ref{f-f}).

\begin{figure}[th]
\begin{center}
\includegraphics[
natheight=0.531900in,
natwidth=6.103000in,
height=0.5319in,
width=6.103in
]{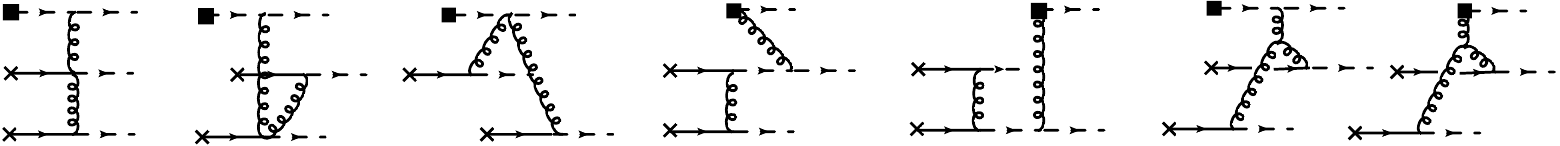}
\end{center}
\caption{ Leading-order SCET diagrams required for the calculation of jet
functions. The inner dashed and curly lines denote hard-collinear quarks and
gluons, external dashed lines correspond to collinear quarks, fermion lines 
with crosses denote soft quarks. Black squares denote the vertex of
the SCET$_{I}$ operator. }%
\label{jet-f-diagrams}%
\end{figure}

Let us now consider in details the calculation of the right-hand side of Eq.(\ref{Jdef}).  
The relevant $T$ product is of order $\lambda^{3}$ and
to leading order in $\alpha_{s}$ reads
\begin{equation}
T_{out}\equiv T_{out}^{(3)}\simeq\int d^{4}x_{1}\int d^{4}x_{2}\int d^{4}%
x_{3}~T\left(  \bar{\xi}_{hc}^{\prime}W^{\prime}(0),\mathcal{L}_{\xi q}%
^{(1)}(x_{1}),\mathcal{L}_{\xi q}^{(1)}(x_{2}),\mathcal{L}_{\xi}^{(0)}%
(x_{3})\right),  
\label{Axi}%
\end{equation}
where $\mathcal{L}_{\xi q}^{(1)}$ is the leading-order soft-collinear
contribution in Eq.~(\ref{L1qxi}). We did not find  the other possibilities to
obtain the leading in the $1/Q$ result. Time-ordered products with insertions of
other higher order contributions $\mathcal{L}_{\xi q}^{(n)}$ with $n\geq 2$  from the collinear or soft-collinear sectors
can provide only suppressed  operators in SCET$_{II}$ and therefore
can be excluded from the consideration. Performing a decoupling of the soft
field with the help of Eq.~(\ref{dec-tr}) we obtain%
\begin{equation}
\bar{\xi}_{hc}^{\prime}W^{\prime}(0)\rightarrow\bar{\xi}_{hc}^{\prime
(0)}W^{\prime}(0)S_{n}^{\dag}(0),~\mathcal{L}_{\xi q}^{(1)}(x)\rightarrow
\bar{\xi}_{hc}^{\prime(0)}~i\hat{D}_{\bot}W~\left[  S_{n}^{\dag}(x_{-}%
)q(x_{-})\right]  .\ \
\end{equation}
The eikonal factors $S_{n}^{\dag}$ ensure the gauge invariance of the soft
sector described by the soft quark fields. Subsequently, we compute the
contractions of the hard-collinear fields which can be conveniently presented
by Feynman graphs. The leading-order contribution to $T_{out}^{(3)}$ is given by the set of diagrams shown in Fig.\ref{jet-f-diagrams}.
Note that the last two diagrams with the three-gluon vertex have
zero color factor and therefore do not contribute. This is in full agreement
with the similar observation made in Ref.~\cite{Fadin1981}. The remaining diagrams
have a similar topology and the corresponding power counting can be easily
established.  The contractions of the hard-collinear fields yield%
\begin{equation}
\int d^{4}x_{1}~\left\langle A_{hc\bot}^{\alpha}(x_{1})A_{hc\bot}^{\beta
}(x_{2})\right\rangle \sim\int d^{4}x_{2}~\left\langle (\bar{n}\cdot
A_{hc})(x_{2})(n\cdot A_{hc})(x_{3})\right\rangle \sim\lambda^{-2},\label{AA}%
\end{equation}%
\begin{equation}
\int d^{4}x_{3}\left\langle ~\bar{\xi}_{hc}(x_{1})\xi_{hc}(x_{3})\right\rangle
\sim\lambda^{-2},\label{xi-xi}%
\end{equation}
i.e., all hard-collinear contractions cost $\lambda^{-2}$, which results from 
the hard-collinear propagators in momentum space.  
Remember  that  we assume that external hard-collinear particles  are matched onto collinear  ones.  Therefore taking into account the
external collinear and soft fields we obtain%
\begin{equation}
T_{out}^{(3)}\sim\underset{\text{h-coll contractions}}{\underbrace
{\lambda^{-2}\lambda^{-2}\lambda^{-2}}\ }\times\underset{\text{2 soft fields}%
}{\underbrace{\lambda^{3}\lambda^{3}}}\times\underset{\text{3 coll fields}%
}{\underbrace{\lambda^{2}\lambda^{2}\lambda^{2}}}~\sim\lambda^{6}%
.\label{T3:count}%
\end{equation}
The same counting is also relevant  for the second time-ordered product in 
Eq.~(\ref{Tprod}). Therefore the order of the total contribution in SCET$_{II}$ now reads
\begin{equation}
\left\langle p^{\prime}\left\vert ~J_{\bot}^{\mu}\right\vert p\right\rangle
_{(s)}\sim T_{out}^{(3)}\times T_{in}^{(3)}\times\lambda^{-4}%
~~\bar{N}^{\prime}\gamma_{\bot}^{\mu}N\sim\lambda^{8}~\bar{N}^{\prime}%
\gamma_{\bot}^{\mu}N.
\end{equation}
We observe that the soft rescattering contribution has the same power suppression
as the hard one [ Eq.~(\ref{Jbot})].

Let us briefly discuss the general structure of the soft rescattering contribution. 
It is clear from the above consideration that the leading-order jet functions can be computed from the diagrams 
in Fig. \ref{jet-f-diagrams}.  
The details of their calculations and explicit expressions will be presented in
a different publication \cite{inprep}.  From the QCD calculation, we noticed that
at tree level the transverse momentum is completely defined by external
soft and collinear fields and therefore it scales as $k_{\bot}\sim
Q \lambda^{2}$. Such counting ratio remains true for the hard-collinear
lines  inside diagrams due to the momentum conservation. Therefore the
transverse components in the hard-collinear propagators (for tree
diagrams only!) can be neglected. Consequently, the arguments of the external collinear and
soft fields are local in transverse space \footnote{We choose $x=0$ in the
(\ref{FFme}) that correspond to $x_{\bot}=0$.} and depend only on the
relevant light-cone components. In SCET the same properties follow from the
multipole expansion of the fields with respect to the small parameter $\lambda$.
Therefore computing the diagrams in Fig.~\ref{jet-f-diagrams} and passing to momentum space one obtains%
\begin{equation}
T_{out}^{(3)}\simeq\int Dy_{i}\boldsymbol{O}_{out}(y_{i})\, \int d\omega
_{1,2}~\mathbf{J}^{\prime}\left(  y_{i},\omega_{i}Q \right)\, \boldsymbol{S}%
_{n}(\omega_{i}),
\end{equation}
with the following collinear and soft operators:%
\begin{eqnarray}
\boldsymbol{O}_{out}(y_{i})   =4~\prod_{i=1}^{3}\int\frac{d\hat{z}_{i}^{+}}{2\pi} 
&& e^{-\frac{i}{2}\left(  {P}^{\prime}\cdot\bar{n}\right)  \left(  y_{1}%
z_{1}^{+}+y_{2}z_{2}^{+}+y_{3}z_{3}^{+}\right)  } \nonumber \\
&& \times \varepsilon^{ijk}~\bar{\xi}_{c}^{\prime}W_{c}^{\prime i}%
({{\scriptstyle \frac12}}z_{1}^{+}~\bar{n})~\bar{\xi}_{c}^{\prime}%
W_{c}^{\prime j}({{\scriptstyle \frac12}}z_{2}^{+}~\bar{n})~\bar{\xi}%
_{c}^{\prime}W_{c}^{\prime k}({{\scriptstyle \frac12}}z_{3}^{+}~\bar{n}),
\end{eqnarray}
and%
\begin{eqnarray}
\boldsymbol{S}_{n}(\omega_{i})  =\varepsilon^{i^{\prime}j^{\prime}k^{\prime
}}\int\frac{dz_{1,2}^{-}}{2\pi}e^{\frac{i}{2}(\omega_{1}z_{1}^{-}+\omega
_{2}z_{2}^{-})} \left[  S_{n}^{\dag}(0)\right]  ^{k^{\prime}l}\left[  S_{n}^{\dag
}\frac{\Dbn \Dn}{4}q({{\scriptstyle \frac12}}z_{1}^{-}n)\right]  ^{j^{\prime}%
}\left[  S_{n}^{\dag}\frac{\Dbn \Dn}{4}q({{\scriptstyle \frac12}}z_{2}%
^{-}n)\right]  ^{k^{\prime}}
\label{Snbold}
\end{eqnarray}
where we do not show for simplicity the  spinor indices, ${P}^{\prime}$
denotes the total collinear momentum operator, $z_{i}^{-}\equiv(z_{i}\cdot
\bar{n})$, $z_{i}^{+}\equiv(z_{i}\cdot n)$, and $d\hat{z}_{i}^{+}\equiv d {z}_{i}^{+}Q$. The structure for 
$T_{in}^{(3)}$ can be obtained in an analogous way.  Combining these results we obtain operator expression with  structure (\ref{f-f})
which schematically can be written as
\begin{eqnarray}
T\{  C_{A}O_{A}^{ {\mu}_{\bot} }  \}
\simeq  C_{A}\, \text{\bf Tr}\left[\,  \gamma^{\mu}_{\bot} 
\mathbf{O}_{out}(y_{i}) \ast \mathbf{J}^{\prime}\left[  y_{i},\omega_{i}Q \right]\ast
\left\{  \mathbf{S}_{\bar n}~ \mathbf{S}_{n}   \right\} \ast
\mathbf{J} [ x_{i},\nu_{i}Q ]  \ast   \mathbf{O}_{in} (x_{i})\, 
\right].
\label{f-f2}
\end{eqnarray}

\begin{figure}[th]
\begin{center}
\includegraphics[
height=3cm,
]{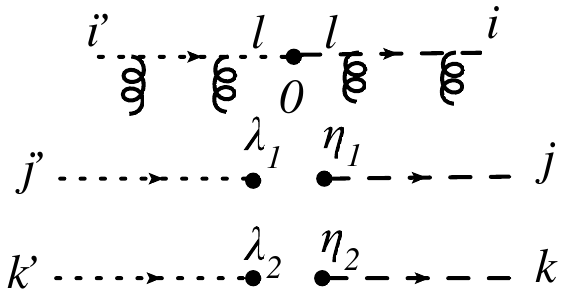}
\end{center}
\caption{ Graphical representation of the operator of Eq.~(\ref{opOS}) entering 
the soft correlation function. 
Dashed and dotted lines denote the different light-cone directions. 
}
\label{osstructure}%
\end{figure}

Substituting these results into the matrix element
of Eq.~(\ref{FFme}) and taking account of the decoupling of the collinear and soft
modes (see, e.g., \cite{Bauer2001, Bauer:2003mga}) one obtains three matrix elements: two with the collinear fields and  the soft
correlation function. The collinear matrix elements can be easily converted
into DAs (\ref{DA:def}):%
\begin{equation}
\int Dy_{i} \left\langle p^{\prime}\right\vert \boldsymbol{O}_{out}(y_{i})\left\vert
0\right\rangle ~\mathbf{J}^{\prime}(y_{i},\omega_{i}Q)=\int Dy_{i}%
~~\mathbf{\Psi}^{\prime}(y_{i})~\mathbf{J}^{\prime}(y_{i},\omega_{i}Q).
\label{Psi-J}%
\end{equation}
Rewriting the initial matrix element $\left\langle 0\right\vert
~\xi_{c}\xi_{c}\xi_{c}\left\vert p\right\rangle \sim\mathbf{\Psi}$ in the same way and
combining all contributions, we obtain the factorization formula for the soft
rescattering contribution:%
\begin{eqnarray}
F_{1}^{(s)}(Q^{2})   &\simeq&  C_{A}(Q,\mu_{I})
\int Dy_{i}\mathbf{\Psi}^{\prime}(y_{i},\mu_{II})
 \int_{0}^{\infty}d\omega_{1}d\omega_{2}~\mathbf{J}^{\prime}(y_{i},\omega_{i}Q,\mu_{I},\mu_{II})
\nonumber\\
& & \times \int Dx_{i}\mathbf{\Psi}(x_{i},\mu_{II}) 
\int_{0}^{\infty}d\nu_{1}d\nu_{2}~\mathbf{J}(x_{i},\nu_{i}Q,\mu_{I},\mu_{II})\boldsymbol{S}(\omega_{i},\nu_{i};\mu_{II}),
\label{F1s}%
\end{eqnarray}
where the soft correlation function is defined as 
\begin{equation}
~\boldsymbol{S}(\omega_{i},\nu_{i};\mu_{II})=\int\frac{d\eta_{1}}{2\pi}%
\int\frac{d\eta_{2}}{2\pi}~e^{-i\eta_{1}\nu_{1}-i\eta_{2}\nu_{2}}\int
\frac{d\lambda_{1}}{2\pi}\int\frac{d\lambda_{2}}{2\pi}e^{i\lambda_{1}%
\omega_{1}+i\lambda_{2}\omega_{2}}\left\langle 0\right\vert \mathbf{O}%
_{S}(\eta_{i},\lambda_{i})\left\vert 0\right\rangle ,\label{Sdef}%
\end{equation}
with the operator%
\begin{eqnarray}
\mathbf{O}_{S}(\eta_{i},\lambda_{i}) &  = & \varepsilon^{i^{\prime}j^{\prime
}k^{\prime}}\left[  S_{n}^{\dag}(0)\right]  ^{i^{\prime}l}\left[  S_{n}^{\dag
}\frac{\Dbn \Dn}{4}q(\lambda_{1}n)\right]  _{\sigma}^{j^{\prime}}\left[
S_{n}^{\dag}\frac{\Dbn \Dn}{4}q(\lambda_{2}n)\right]  _{\rho}^{k^{\prime}} \nonumber \\
&  \times & \varepsilon^{ijk}\left[  S_{\bar{n}}(0)\right]  ^{li}\left[  \bar
{q}\frac{\Dbn \Dn}{4}S_{\bar{n}}(\eta_{1}\bar{n})\right]  _{\alpha}^{j}\left[
\bar{q}\frac{\Dbn \Dn}{4}S_{\bar{n}}(\eta_{2}\bar{n})\right]  _{\beta}^{k},
\label{opOS}
\end{eqnarray}
which is shown graphically in Fig.~\ref{osstructure}. 
In the last equation we assume $\bar{q}S_{\bar{n}}(\eta_{2}\bar{n})=\bar
{q}(\eta_{2}\bar{n})S_{\bar{n}}(\eta_{2}\bar{n})$, and the color and Dirac indices
are shown explicitly. Furthermore in Eq.~(\ref{F1s}), $C_{A}$ denotes the hard coefficient
function which has been computed in the leading-order approximation (\ref{CABtree}).%

In Eq.~(\ref{F1s}) we show explicitly two factorization scales $\mu_{I}$ and
$\mu_{II}$ . The total contribution, as usually, does not depend on these
auxiliary quantities. The first scale $\mu_{I}$ arises at the matching QCD to
SCET$_{I}$. The evolution equations at leading logarithmic approximation with
respect to $\mu_{I}$ were discussed above. In practical applications it is
convenient to fix this scale at the value $\mu_{I}^{2}\simeq\Lambda Q$. Then
the large logarithms $\ln[Q^{2}/\mu_{I}^{2}$] can be resummed solving RG
equations. The second scale $\mu_{II}$ appears when one performs the second
reduction to SCET$_{II}$. Usually, the corresponding coefficient functions ( =
jet functions $J^{\prime}$ and $J$~) are computed at $\mu_{II}^2\simeq$ $\Lambda
Q$ and then the scale is fixed to be $\mu_{II}^2\simeq$ $\Lambda^{2}$. \ Again,
arising large logarithms $\ln[Q\Lambda/\mu_{II}^{2}]$ must be resummed with
the help of evolution equations for nucleon DAs $\mathbf{\Psi}(x_{i},\mu
_{II})$ and CF $\boldsymbol{S}(\omega_{i},\nu_{i},\mu_{II})$. The evolution of
the DAs is well studied in the literature (see, e.g. Refs.~\cite{Chernyak:1983ej,
Lepage:1979zb} and \cite{Braun:1999te} for recent progress ) but the corresponding
equation for $\boldsymbol{S}(\omega_{i},\nu_{i},\mu_{II})$ is new and has not
been derived before. Such a calculation must be done because it will provide an
important check of the factorization formula (\ref{F1s}) at leading
logarithmic accuracy. A derivation of the jet functions and evolution kernel for
$\boldsymbol{S}(\omega_{i},\nu_{i},\mu)$ will be presented in a separate
publication \cite{inprep}. In the Appendix B to this paper we demonstrate how the 
perturbative QCD result of Eq.~(\ref{Jalf}) is reproduced from the 
corresponding SCET diagram in Fig.\ref{jet-f-diagrams}. \ 

It turns out that the product of the nucleon DA and jet function has the same
Dirac and color structure as $~\mathbf{\Psi}^{\prime}(y_{i})$ (\ref{Psi:def}):
\begin{align}
~\mathbf{J}(x_{i},\omega_{i})~\mathbf{\Psi}(x_{i}) &  =~\Omega_{V}%
(x_{i},\omega_{i})~~p_{+}\left[  {\scriptstyle\frac{1}{2}}%
\setbox0=\hbox{$\bar n$}\dimen0=\wd0\setbox1=\hbox{/}\dimen1=\wd1\ifdim\dimen0>\dimen1\rlap{\hbox to
\dimen0{\hfil/\hfil}}\bar{n}\else\rlap{\hbox to \dimen1{\hfil$\bar
n$\hfil}}/\fi~C\right]  _{\alpha\beta}\left[  \gamma_{5}N^{+}\right]
_{\sigma}\nonumber\\
&  +~\Omega_{A}(x_{i},\omega_{i})~~p_{+}\left[  {\scriptstyle\frac{1}{2}%
}%
\setbox0=\hbox{$\bar n$}\dimen0=\wd0\setbox1=\hbox{/}\dimen1=\wd1\ifdim\dimen0>\dimen1\rlap{\hbox to
\dimen0{\hfil/\hfil}}\bar{n}\else\rlap{\hbox to \dimen1{\hfil$\bar
n$\hfil}}/\fi\gamma_{5}C\right]  _{\alpha\beta}\left[  N^{+}\right]  _{\sigma
}\label{Psi-Jet}\\
&  \ +\Omega_{T}(x_{i},\omega_{i})~p_{+}\left[  {\scriptstyle\frac{1}{2}%
}%
\setbox0=\hbox{$\bar n$}\dimen0=\wd0\setbox1=\hbox{/}\dimen1=\wd1\ifdim\dimen0>\dimen1\rlap{\hbox to
\dimen0{\hfil/\hfil}}\bar{n}\else\rlap{\hbox to \dimen1{\hfil$\bar
n$\hfil}}/\fi\gamma_{\bot}C\right]  _{\alpha\beta}\left[  \gamma^{\bot}%
\gamma_{5}N^{+}\right]  _{\sigma}.\nonumber
\end{align}
where the coefficients$~$\ $\Omega_{X}(x_{i},\omega_{i})$ are linear
combinations of the nucleon DAs (\ref{Psi:def}) and hard-collinear jet functions%
\begin{equation}
~\Omega_{X}(x_{i},\omega_{i})=J_{XV}(x_{i},\omega_{i})V(x_{i})+J_{XA}(x_{i},\omega_{i})A(x_{i})+J_{XT}(x_{i},\omega_{i})T(x_{i}).
\end{equation}
Therefore, the jet function can be interpreted as a hard-collinear component of the
three-quark nucleon wave function describing the transition of the three
collinear quark state into configuration with one hard-collinear and two soft
quarks. Correspondingly, the CF $\boldsymbol{S}(\omega_{i},\nu_{i},\mu_{II})$
(\ref{Sdef}) describes the propagation of the soft diquark state in the
background of the soft-gluon field created by a fast moving active quark: i.e.,
it describes the soft overlap of the nucleon wave function. Therefore we
expect that the soft rescattering picture can also be associated with the 
well known mechanism, suggested by Feynman a long time ago
\cite{Feynman:1973xc}.

In the factorization formula of Eq.~(\ref{F1s}) we restricted the fractions
$\omega_{i}$ and $\nu_{i}$ to be defined on the real semiaxis, assuming that
 all the functions in (\ref{F1s}) are real functions. This allows us to avoid the
poles in the propagators of the tree diagrams in Fig.~\ref{jet-f-diagrams} and
ensures that the jet functions are real. Recall, that the reality of the nucleon form
factors is guaranteed by the time reversal invariance of QCD.

The other interesting observation which follows already from the QCD
computation (\ref{Jalf}) is the absence of
the end-points singularities in the convolution integrals of DAs with jet
functions in Eq.~(\ref{Psi-Jet}).  Let  us assume that the convolution integrals with respect to 
the soft fractions $\omega_{i}, \nu_{i}$ in (\ref{F1s}) are
also well defined.  Then this allows us to suggest that hard rescattering
and the soft rescattering mechanisms provide additive contributions to the
total FF $F_{1}$ at least to leading logarithmic accuracy:%
\begin{equation}
F_{1}\simeq F_{1}^{(s)}+F_{1}^{(h)},\label{F1:sum}%
\end{equation}
with the well known expression for the hard rescattering part: $F_{1}%
^{(h)}=\mathbf{\Psi}\ast\mathbf{H}\ast\mathbf{\Psi}$. Recall that the
convolution integrals for $F_{1}^{(h)}$ at leading logarithmic accuracy are
also well defined. Hence we may conclude that there is no double counting
in this case. The formula of Eq.~(\ref{F1:sum}), together with the result of Eq.~(\ref{F1s}), 
is our suggestion for the full factorization formula for the Dirac FF
at large $Q^{2}$. We would like to emphasize that the obtained
results have been derived at leading order and only partially verified at the
leading logarithmic approximation. A discussion of an all order factorization
proof for Eqs.~(\ref{F1s}) and (\ref{F1:sum}) requires more detailed analysis and
goes beyond this publication.

\subsection{Leading order SCET analysis for Pauli FF $F_{2}$}
In contrast to $F_{1}$, the description of the
Pauli FF $F_{2}$ is more complicated. First, it is well known that hard gluon
exchange can produce large logarithms,
\begin{equation}
F_{2}^{(h)}=\mathbf{\Psi}\ast\mathbf{H}\ast\mathbf{\Psi}\sim\alpha_{s}^{2}
\ln^{2}Q^{2}/\mu^{2},
\label{F2h}
\end{equation}
which arise due to the end-point singularities in the convolution integrals
\cite{Belitsky:2002kj}. This is perhaps, an indication that the hard and soft
rescattering mechanisms overlap. Therefore, in order to find the correct
description for $F_{2}$ one has to formulate a clear recipe for how to avoid double
counting in the calculation of soft and hard rescattering contributions. Such a
problem for $F_{2}$, probably, arises already at the level of matching QCD to
SCET$_{I}$. However, our analysis from the previous section does not show any
problems with the SCET$_{I}$ convolution integrals for the coefficient
function $C_{B}$ [see, e.g., Eq.~(\ref{eigenf})]. Moreover, resummation of the
leading Sudakov logarithms can be carried out exactly because the problematic
logarithms (\ref{F2h}) admix only at the next-to-leading accuracy. The
structure of the logarithms beyond the leading order is an important subject
which remains to be established for a full proof of the factorization theorem. 
As a first step in this direction, we 
demonstrate here that the SCET counting rules confirm the power suppression of the
soft rescattering contribution in $F_{2}$ obtained from the QCD calculation in Sec.~\ref{sec2}.

\begin{figure}[th]
\begin{center}
\includegraphics[
natheight=1.051600in,
natwidth=1.422600in,
height=1.0516in
]{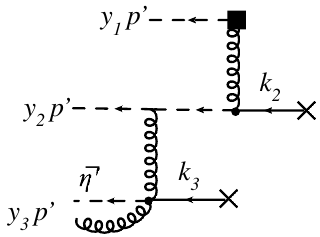}
\end{center}
\caption{ SCET diagram for matching the quark-gluon hard-collinear state onto
three collinear quarks for Pauli FF. The small component $\bar{\eta}%
_{c}^{\prime}$ is shown as a quark gluon state $\bar{\xi}_{c}^{\prime}A_{c\bot}%
$. }%
\label{scet-pauliff-jpr-diagram}%
\end{figure}

For this purpose, the relevant part of the SCET$_{I}$ vector current 
follows from Eq.~(\ref{Jem:LLexp}) as~:
\begin{equation}
J_{\Vert}^{\mu}~(0)=-\frac{1}{Q}\left(  n^{\mu}+\bar{n}^{\mu}\right)  \int
_{0}^{1}d\tau~~\left(  \bar{\xi}_{hc}^{\prime}W^{\prime}\right) \left[
\Dslash{\mathcal{A}}_{\bot}^{\prime}(\tau)+\Dslash{\mathcal{A}}_{\bot}(\tau)\right]  \left(
W^{\dag}\xi_{hc}\right) ~,
\label{JVert}
\end{equation}
with $\Dslash{\mathcal{A}}_{\bot}^{\prime}(\tau)+\Dslash{\mathcal{A}}_{\bot}(\tau)$ as given in
Eq.~(\ref{OB}). For the qualitative discussion we consider the first term with
$\Dslash{\mathcal{A}}_{\bot}^{\prime}$ only. Following the similar arguments as before
we arrive at an analysis of time-ordered products:%
\begin{align}
& \ \ \ \ \ \ \ \ \ \ T\left\{  J_{\Vert}^{\mu}(0)~e^{i\mathcal{L}^{(\bar{n})}+i\mathcal{L}^{(n)}}\right\}  =
\frac{\bar n^{\mu}}{(\bar n\cdot \partial )}~T_{out}^{(4)}~T_{in}^{(3)}+\dots ,
\label{158}
\\
T_{out}^{(4)}=  & T\left\{  \left(  \bar{\xi}_{hc}^{\prime}W^{\prime
}\right) (0) \Dslash{\mathcal{A}}_{\bot}^{\prime}(0) e^{i\mathcal{L}^{(\bar{n})}}\right\}  ,
~ T_{in}^{(3)}= T\left\{  \left(  W^{\dag}\xi_{hc}\right)  (0)e^{i\mathcal{L}^{(n)}} \right\}. 
\label{159}
\end{align}
To obtain Eq.(\ref{158}) we converted (\ref{JVert}) into position space and for simplicity wrote explicitly only the term $\sim \bar n^{\mu}$.
The second term $T_{in}^{(3)}$ in Eq.(\ref{159}) is the same as the one appearing in $F_{1}$. 
Hence we only have to consider the new term $T_{out}^{(4)}$. 
Consider the following contribution:%
\begin{equation}
T_{out}^{(4)}\simeq T\left(  \bar{\xi}_{c}^{\prime}(0)
\Dslash{\hat{A}}_{hc\bot}^{\prime}(0)\int dx_{1}\mathcal{L}_{\xi q}^{(1)}(x_{1})\int
dx_{2}\mathcal{L}_{\xi\xi}^{(0)}(x_{2})\int dx_{3}\mathcal{L}_{\xi q}
^{(1)}(x_{3})\right) \label{T4-1}%
\end{equation}
\begin{equation}
{\textstyle \simeq T\left(  \bar{\xi}_{c}^{\prime}(0)
\Dslash{\hat{A}}_{hc\bot}^{\prime}(0), \int dx_{1}~\bar{\xi}_{hc}^{\prime}A_{hc\bot}q(x_{1}),\int
dx_{2}~\bar{\xi}_{c}^{\prime}\ \left(  n\cdot A_{hc}\right)  \xi_{hc}%
(x_{2}),\int dx_{3}~\bar{\eta}_{c}^{\prime}\left(  \bar{n}\cdot A_{hc}\right)
q(x_{3})\right) , }
\label{T4-2}%
\end{equation}
where we substituted the small component
\begin{equation}
\bar{\eta}_{c}^{\prime}=\bar{\xi}_{c}^{\prime}i
\setbox0=\hbox{$\overleftarrow D$} \dimen0=\wd0 \setbox1=\hbox{/} \dimen1=\wd1
\ifdim\dimen0>\dimen1 \rlap{\hbox to \dimen0{\hfil/\hfil}} \overleftarrow D
\else \rlap{\hbox to \dimen1{\hfil$\overleftarrow D$\hfil}} / \fi _{\bot
c}(i\bar{n}\cdot\overleftarrow{D}_{c})^{-1}\frac{ \setbox0=\hbox{$\bar n$}
\dimen0=\wd0 \setbox1=\hbox{/} \dimen1=\wd1 \ifdim\dimen0>\dimen1
\rlap{\hbox to
\dimen0{\hfil/\hfil}} \bar n \else \rlap{\hbox to \dimen1{\hfil$\bar
n$\hfil}} / \fi }{2} \simeq\bar{\xi}_{c}^{\prime}~g\hat{A}_{c\bot}(i\bar
{n}\cdot\overleftarrow{\partial})^{-1}\frac{ \setbox0=\hbox{$\bar n$}
\dimen0=\wd0 \setbox1=\hbox{/} \dimen1=\wd1 \ifdim\dimen0>\dimen1
\rlap{\hbox to
\dimen0{\hfil/\hfil}} \bar n \else \rlap{\hbox to \dimen1{\hfil$\bar
n$\hfil}} / \fi }{2}.\label{eta-c}%
\end{equation}
The presence of the small component $\bar{\eta}_{c}^{\prime}$ can be explained
by interaction with the longitudinal photon. The outgoing collinear state must
have one collinear transverse gluon or transverse derivative in order to
satisfy conservation of the orbital momentum. Contracting the hard collinear
fields in Eq.~(\ref{T4-2}) one obtains the diagram as in
Fig.~\ref{scet-pauliff-jpr-diagram}. 
Using Eqs.~(\ref{AA}) and (\ref{xi-xi}) one easily obtains
\begin{equation}
T_{out}^{(4)}\sim
\underset{\text{2 soft fileds}}{ \underbrace{
  \lambda^{3}\lambda^{3}  }}\times
\underset{\text{~3 coll fields}}{\underbrace{
\lambda^{2}\lambda^{2}\lambda^{4}  }}~\times
\underset{ \text{h-coll contractions} }{ \underbrace{
\lambda^{-2}\lambda^{-2}\lambda^{-2}
}}
 \sim\lambda^{8}.
\end{equation}
Then the total contribution reads%
\begin{align}
\left\langle p^{\prime}\left\vert J_{\Vert}^{\mu}\right\vert p\right\rangle
&  \sim T_{out}^{(4)}\times T_{in}^{(3)}\times\lambda^{-4}%
~(n+\bar{n})^{\mu}~\bar{N}~\hat{1}N\ \\
& \sim\lambda^{6}~\lambda^{8}~\lambda^{-4}~(n+\bar{n})^{\mu}~\bar{N}~\hat{1}N
\sim\lambda^{10}~(n+\bar{n})^{\mu}~\bar{N}~\hat{1}N,
\end{align}
i.e., we obtained the same result as in the case of the hard rescattering mechanism
(\ref{JL:HS}). In the Appendix we demonstrate that the diagram in
Fig.~\ref{scet-pauliff-jpr-diagram} reproduces the QCD expression for $J^{\prime}$
from Eq.~(\ref{F2:J'andJet}). However, contrary to the Dirac FF, the
convolution integral with respect to the collinear momentum fraction $\Psi
^{\prime}\ast\mathbf{J}^{\prime}$ is not defined due to the end-point
divergencies. Therefore we assume that the matching onto SCET$_{II}$ for the
Pauli FF $F_{2}$ cannot provide a well defined expression. As we discussed
above, even matching onto SCET$_{I}$, one is  faced with the mixing problem between
hard and soft rescattering contributions.

\section{Phenomenological application to the nucleon FFs}
\label{sec5}

In order to perform a first phenomenological analysis we introduce SCET$_{I}$ form
factors defined as the following nucleon matrix elements,
\begin{equation}
\left\langle p^{\prime}\left\vert \left(  \bar{\xi}'_{q}W^{\prime}\right)
\gamma_{\bot}^{\mu}\left(  W^{\dag}\xi_{q}\right)  ~\right\vert p\right\rangle_{\text{SCET}_{I}}
=
\bar{N}(p^{\prime})
\nbn \gamma_{\bot}^{\mu} \nbn
N(p)~f_{1}^{q}(Q_{I},\mu)\equiv\bar{N}_{+}^{\prime
}~\gamma_{\bot}^{\mu}N_{+}~f_{1}^{q}(Q_{I},\mu),
\label{fq-1}%
\end{equation}
and%
%\begin{equation}
%\frac{(n^{\mu}+\bar{n}^{\mu})}{Q}\left\langle p^{\prime}\left\vert O_{B}%
%[\tau]~\right\vert p\right\rangle _{\text{SCET}_{I}}=\frac{(p+p^{\prime}%
%)^{\mu}}{2m_{N}}~\bar{N}_{+}^{\prime}~\hat{1}~N_{+}(p)~f_{2}^{q}(\tau
%;Q_{I},\mu),\label{fq-2}%
%\end{equation}
%new in v9
\begin{equation}
\left\langle p^{\prime}\left\vert O_{B}%
[\tau]~\right\vert p\right\rangle _{\text{SCET}_{I}}=%
~\bar{N}_{+}^{\prime}~\hat{1}~N_{+}(p)~
\frac{m_{N}}{2}f_{2}^{q}(\tau;Q_{I},\mu),\label{fq-2}%
\end{equation}
with the operator $O_{B}[\tau]$ defined in Eq.~(\ref{OB}). We also used the notation
$Q_{I}\sim\sqrt{Q\Lambda}$ in order to stress that the defined quantities do not
depend on the large scale $Q^{2}$. We indicate explicitly in  the right-hand side of 
Eqs.~(\ref{fq-1},\ref{fq-2}) the renormalization scale dependence. 
\begin{figure}[th]
\begin{center}
\includegraphics[
height=2.5cm
]{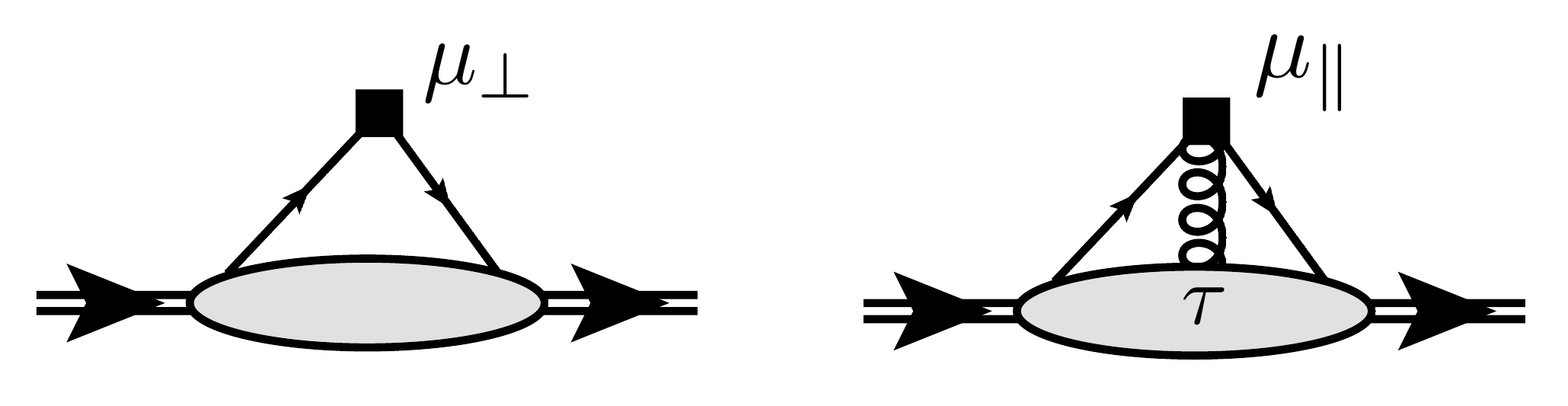}
\end{center}
\caption{Graphical representation of the SCET$_I$ FFs $f_1$ (left panel) and 
$f_2$ (right panel). In $f_2$, $\tau$ corresponds with the gluon momentum  fraction.}%
\label{scet1ff}%
\end{figure}
Taking the
nucleon matrix element from both sides of Eq.~(\ref{Jem:LLexp}), we obtain%
\begin{equation}
F_{1}^{(s)}(Q^{2})=C_{A}(Q,\mu_{I})\sum_{q}e_{q}~f_{1}^{q}(Q_{I},\mu
)=C_{A}(Q,\mu)~f_{1}(Q_{I},\mu),
\label{F1scet1}%
\end{equation}%
%new in v9
\begin{equation}
F_{2}^{(s)}(Q^{2})=\frac{m^{2}_{N}}{Q^{2}} \int_{0}^{1}d\tau~C_{B}(\tau,Q,\mu)\sum_{q}e_{q}~f_{2}%
^{q}(\tau;Q_{I},\mu)=\frac{m^{2}_{N}}{Q^{2}} \int_{0}^{1}d\tau~C_{B}(\tau,Q,\mu)~f_{2}(\tau;Q_{I}%
,\mu).\label{F2scet1}%
\end{equation}
Eqs.(\ref{F1scet1}) and (\ref{F2scet1}) are presented in Fig.\ref{scet1ff}  in graphical form.

Using NLL approximation for the coefficient functions (\ref{CALL}) and
(\ref{CBLL}) these results can be represented as
%new in v9
\begin{align}
F_{1}^{(s)}(Q^{2})  &  \simeq e^{-S(Q,\mu_{h},\mu_{I})}U_{A}~(\mu_{h},\mu
_{I})~f_{1}(Q_{I},\mu_{I}),~\ \ \label{F1LL}\\
F_{2}^{(s)}(Q^{2})  &  \simeq \frac{m^{2}_{N}}{Q^{2}} e^{-S(Q;\mu_{h},\mu_{I})}\int_{0}^{1}d\tau U_{B}
\left[  \tau~;\mu_{h},\mu_{I}\right]  \ f_{2}(\tau;Q_{I},\mu_{I}%
),~\ \label{F2LL}%
\end{align}
where the scale $\mu_{I}\simeq Q_{I}\sim\sqrt{\Lambda Q}$. From the right-hand side of 
Eqs.~(\ref{F1LL},\ref{F2LL}) one can see that the SCET$_{1}$ FFs $f_{1,2}$ depend now
only on the hard-collinear scales. All dependence from the large scale of order
$Q^{2}$ is factorized into Sudakov factors $e^{-S(Q,\mu_{h},\mu_{I})}$. This is
the main feature of the Feynman mechanism. We could expect that the hard
scattering contribution in $F_{1}$ provides corrections of order $\alpha
_{s}^{n+2}\ln^{n}Q^{2}/\mu^{2}$ which are suppressed relative to the contributions computed in
Eq.~(\ref{F1LL}), and therefore can be neglected:
\begin{equation}
F_{1}(Q^{2})\simeq F_{1}^{(s)}(Q^{2}).
\end{equation}

In the case of the Pauli FF $F_{2}$, the situation is more delicate due to a possible
overlap of hard and soft rescattering terms. From the calculations of the
hard scattering contribution \cite{Belitsky:2002kj}, one obtains, due to the
end-point singularities, contribution of order $\alpha_{s}^{2}\ln^{2}Q^{2}%
/\mu^{2}$. Such logarithms are of the same accuracy as next-to-leading Sudakov
or single logarithms in Eq.~(\ref{F2LL}). Therefore, Eq.~(\ref{F2LL}) is exact
only at the level of leading Sudakov logarithms. Beyond this accuracy one has to
perform a more accurate analysis in order to avoid double counting. 
For a first numerical estimate, we 
shall neglect the hard scattering contribution in $F_{2}$ assuming
\begin{equation}
F_{2}(Q^{2})\simeq F_{2}^{(s)}(Q^{2}).
\end{equation}
Such an approximation, perhaps, may work if $Q^{2}$ is not very large, of order a
few GeV$^2$, and one may expect that the dominant contribution is provided by the soft
spectator contributions of Eqs.~(\ref{F1LL}) and (\ref{F2LL}).

\begin{figure}[th]
\begin{center}
\includegraphics[
height=7cm
]{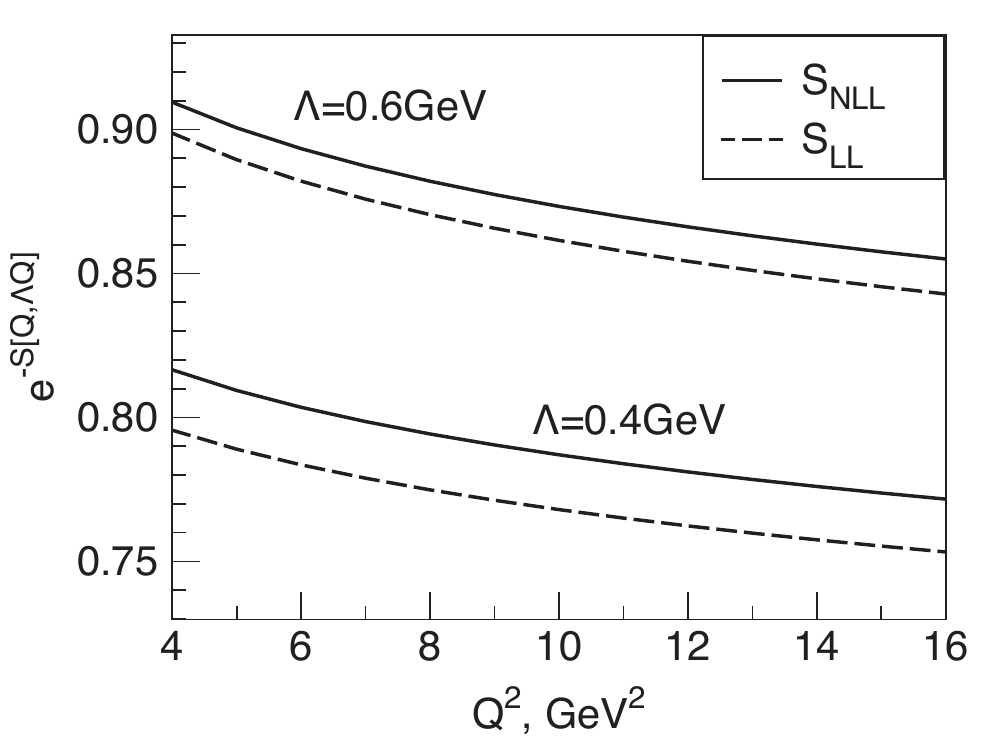}
\end{center}
\caption{Effect on the FFs from the leading logarithmic summation (LL) for different values of
$\Lambda$. }%
\label{sudakovs}%
\end{figure}

First, it is interesting to investigate how strong suppression is obtained from
the resummed Sudakov logarithms. In Fig.\ref{sudakovs} 
we demonstrate the results for the leading Sudakov logarithm factor
$e^{-S(Q,\mu_{h},\mu_{I})}$ taking $\mu_{h}=Q$ and $\mu_{I}=\sqrt{\Lambda Q}$.
We use two different values for the soft scale $\Lambda=\{0.4,0.6\}$GeV and consider
$Q^{2}=4-16~$GeV$^{2}$. For our numerical estimate, we used formula Eq.~(\ref{S}) with
the two-loop running coupling[ $N_{f}=4$ and $\alpha_{s}(2~$GeV$)=0.31$]. We
observe that the Sudakov factor provides a reduction of around $10\%-25\%$ depending on
the choice of $\Lambda$, and changes quite slowly in the given range of
$Q^{2}$. Therefore we can conclude that the soft spectator scattering contribution
may provide quite a substantial effect over an extended range of $Q^{2}$ if \ the
SCET$_{I}$ FFs $f_{1,2}$ are not too small.

However,  the full next-to-leading evolution includes also single logarithms
described by the kernels $U_{A,B}$. In the case of $F_{2}$, the evolution effect
from $U_{B}\left[  \tau~;\mu_{h},\mu_{I}\right]  $ in Eq.~(\ref{F2LL}) is given by
the approximate solution of Eq.~(\ref{Uapp}) and does not depend on the gluon momentum fraction
$\tau$. Using Eq.~(\ref{Uapp}) we can write
%new in v9
\begin{eqnarray}
F_{2}^{(s)}(Q)&=&\frac{m^{2}_{N}}{Q^{2}} e^{-S(Q;\mu_{h},\mu_{I})}\int_{0}^{1}d\tau   
U_{B} \left[\tau~;Q,\mu_{I}\right]  \ f_{2}(\tau;Q_{I},\mu_{I}) 
\nonumber \\
& \approx& 
\frac{m^{2}_{N}}{Q^{2}}
e^{-S(Q;\mu_{h},\mu_{I})}U_{B}^{\text{app}}\left[  Q,\mu_{I}\right]
\int_{0}^{1}d\tau~f_{2}(\tau;Q_{I},\mu_{I}),
\end{eqnarray}
Therefore in the ratio $F_{2}/F_{1}$ the leading and next-to-leading Sudakov
logarithms cancel and we obtain that this quantity depends only on the 
SCET$_{I}$ FFs:
%new in v9
\begin{equation}
\frac{Q^{2}F_{2}^{(s)}}{F_{1}^{(s)}}\simeq\frac{U_{B}^{\text{app}}\left[  Q,\mu
_{I}\right]  }{U_{A}\left[  Q,\mu_{I}\right]  }\frac{m^{2}_{N}\int_{0}^{1}d\tau
~ f_{2}(\tau;Q_{I},\mu_{I})}{~f_{1}(Q_{I},\mu_{I})}.
\label{F2toF1}
\end{equation}
The ratio of the kernels $U_{A,B}$ in Eq.~(\ref{F2toF1}) changes slowly, for instance,
\begin{equation}
0.93\leq\frac{U_{B}^{\text{app}}\left[  Q,\mu_{I}\right]  }{U_{A}\left[
Q,\mu_{I}\right]  }\leq0.95\text{ \ \ \ for }4~\text{GeV}^{2}\leq Q^{2}%
\leq16~\text{GeV}^{2}\text{ and }\Lambda=400~\text{MeV.}%
\end{equation}
%new in v9
For large $Q$ values, when $|Q_{I}|\sim\sqrt{\Lambda Q}\rightarrow\infty$ we
expect the asymptotic $Q^{2}F_{2}/F_{1}\rightarrow \text{const}$ as
it follows from SCET counting rules. It is clear that
such asymptotic could be reached only at very large values of $Q^{2}$.
Therefore it is not surprising that the ratio, measured recently up to 
$Q^{2}\lesssim8.5~$GeV$^{2}$ \cite{Puckett:2010ac}, shows a behavior which drops 
less fast in $Q^2$, when compared with the expected power $Q^{-2}$. 
For such values of $Q^{2}$ the
ratio Eq.~(\ref{F2toF1}) is defined practically only by the ratio of the SCET form
factors $f_{1,2}$ depending on $Q_{I}$. But the hard-collinear scale in this region
is approximately $Q_{I}\sim\sqrt{\Lambda Q}\simeq0.9-1.3$ GeV, i.e. quite
small in order to expect the asymptotic behavior.

We obtained that the effect from the Sudakov suppression in the region of moderate
space-like $Q^{2}$ can be estimated as $\sim10\%-25\%$. However the situation
can be different for timelike momenta $q^{2}>0$. In this case, the Sudakov factor
after analytical continuation from spacelike to timelike region may produce
a substantial enhancement. Properties of the timelike processes have been studied in
many publications  (see, for instance, \cite{Parisi79, Magnea90, Radyushkin00}).
It is well known that analytic continuation of the Sudakov FF to the
timelike region produces enhanced $\pi^{2}$ terms. Such corrections were
resummed for different processes \cite{Parisi79, Magnea90, Ahrens08}. In order
to perform such resummation it was suggested to perform the matching at a
timelike renormalization point $-\mu_{h}^{2}$ \cite{Ahrens08}. Then the
timelike Sudakov factor $e^{-S_{\text{TL}}}$ accumulates the large $\pi^{2}$
contributions together with the Sudakov logarithms. Using this recipe we must
compute $S_{\text{TL}}\equiv S[-q^{2}-i\varepsilon,-\mu_{h}-i\varepsilon,\mu]$
in the timelike region. This can be done with the help of analytical continuation
of the running coupling which to our accuracy reads
\cite{Ahrens08,Radyushkin99}:%
\begin{equation}
\frac{\alpha_{s}(\mu^{2})}{\alpha_{s}(-\mu^{2})}=1-ia(\mu^{2})+\frac{\beta
_{1}}{\beta_{0}}\frac{\alpha_{s}(\mu^{2})}{4\pi}\ln\left[  1-ia(\mu
^{2})\right]  +\mathcal{O}(\alpha_{s}^{2}),\label{alfaS}%
\end{equation}
where $a(\mu^{2})=\beta_{0}\alpha_{s}(\mu_{h}^{2})/4$ $\sim\mathcal{O}(1)$ for
moderate values of $Q^{2}$.

Existing experimental data for the ratio $R_{M}=|G_{M}(q^{2})|/G_{M}(Q^{2})$ show
a considerable enhancement of timelike FFs over their spacelike
counterparts: $R_{M}$ $\simeq1.5-$ $2$ over the range $Q^{2}\sim10~$GeV$^{2}$.
Despite that the extraction of the absolute value of the time like FF
$|G_{M}(q^{2})|$ involves considerable assumptions about the behavior of the
timelike electric FF $G_{E}$ and probably includes large systematic
errors, the timelike enhancement is considered as a well established fact. In
Ref.~\cite{Radyushkin00} it was suggested that "soft terms" accompanied by the
Sudakov double logarithms could play an important role in a so-called,
$K$-factor type enhancement to hadronic FFs in the timelike region.
Using the results of Eqs.~(\ref{F1LL}, \ref{F2LL}) with resummed Sudakov logarithms we
can easily estimate such an effect in our approach.

\begin{figure}[h]
\begin{center}
\includegraphics[
height=6cm
]{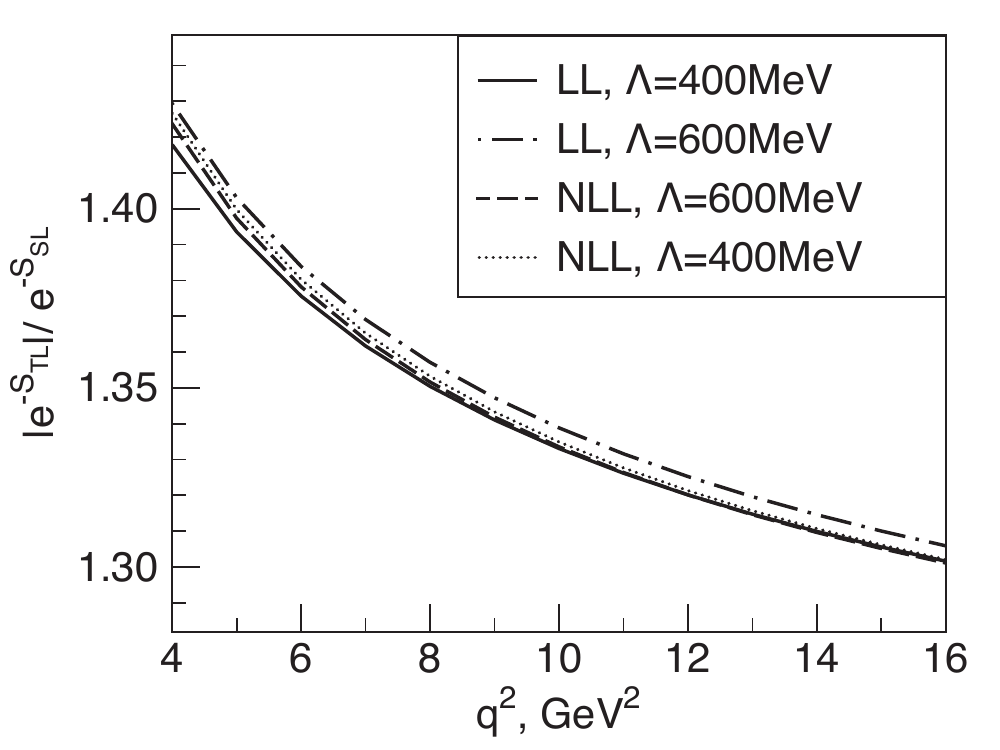}\includegraphics[
height=6cm
]{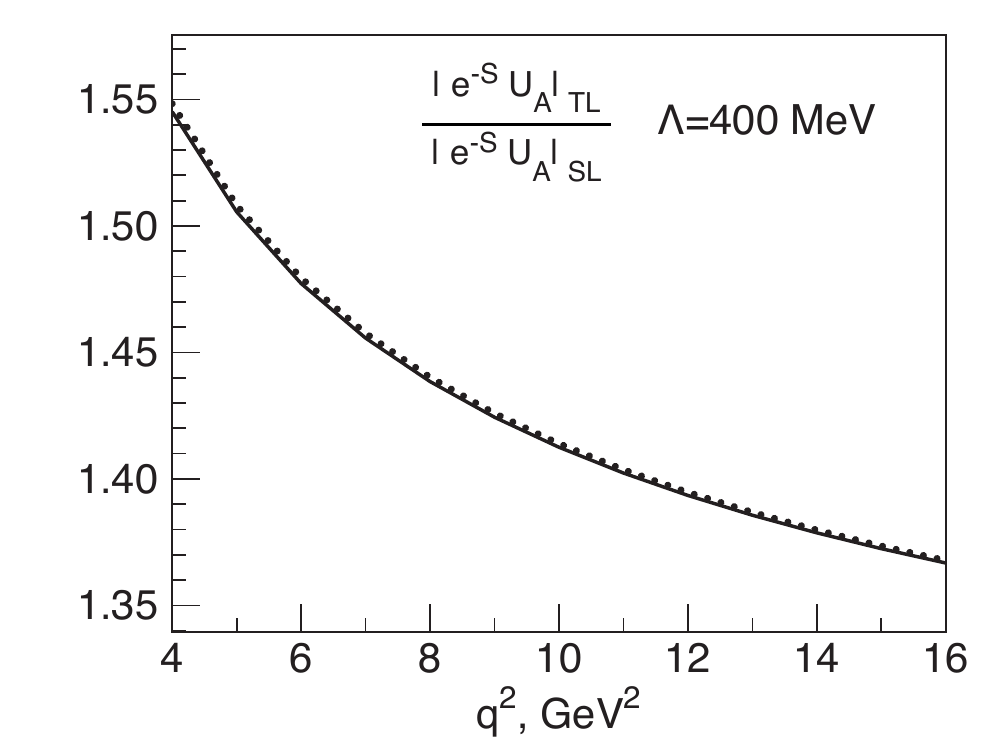}
\end{center}
\caption{Left panel: Contribution of the timelike (TL) to spacelike (SL) ratio of 
Sudakov logarithms $|e^{-S_{\text{TL}}}|/e^{-S_{\text{SL}}}$ as function of $q^{2}$ 
for different choices of $\Lambda$. Solid and dash-dotted curves correspond to 
LL approximation with $\Lambda=400$ and $600$~MeV respectively; 
dashed and dotted curves describe
next-to-leading approximation. Right panel: complete NLL evolution (including
kernel $U_{A}$) with $\Lambda=400$MeV. }
\label{rm}
\end{figure}

In order to study the qualitative effect of the SCET$_{I}$ evolution we consider
the ratio of the Dirac timelike (TL) and spacelike (SL) FFs. Let us
introduce:
\begin{equation}
R_{1}=\frac{|F_{1}(q^{2})|}{F_{1}(Q^{2})}\simeq\frac{|e^{-S_{\text{TL}}}%
|}{e^{-S_{SL}}}\frac{|U_{A}^{\text{TL}}|}{U_{A}}\frac{~\left\vert f_{1}(q_{I}%
,\mu_{I})\right\vert }{~f_{1}(Q_{I},\mu_{I})~}.\label{R12}%
\end{equation}
where we used $q_{I}$ in order to specify timelike momentum transfer
$|q_{I}|\sim\sqrt{\Lambda q}$ in the numerator. We assume that the soft spectator
scattering mechanism dominates also in timelike kinematics and appropriate
quantities are related by analytical continuation. At present we do not know
the SCET FFs $f_{1,2}$. However, we can study the pQCD evolution effect
\ from the resummed logarithms. In Fig.\ref{rm} we demonstrate the timelike (TL) to spacelike (SL) 
ratio $|e^{-S_{\text{TL}}}|/e^{-S_{\text{SL}}}$ which represents the Sudakov logarithms in the
ratio $R_{1}$. 
One can see that the obtained ratio $|e^{-S_{\text{TL}}}|/e^{-S_{\text{SL}}}$ very
weakly depends on the choice of $\Lambda$ and provides an almost $30\%-40\%$ 
enhancement effect of the timelike FFs relative to their spacelike counterparts. 
When we combine the Sudakov evolution with the $U_{A}-$kernel of Eq.~(\ref{UA}) we obtain
the results shown in Fig.\ref{rm} (right panel). We see that single logarithms
increase the ratio by $5\%-8\%$. But this effect is only a small fraction of the
full evolution, i.e. non-Sudakov logarithms cannot provide substantial
enhancement. Therefore we can conclude that the soft spectator scattering
mechanism plays an important role in the discussion of the timelike FF's.
Sudakov logarithms appearing in this case provide an important enhancement in the
region of moderate values of timelike momentum transfers $q^{2}$.  
This enhancement is in qualitative agreement with
the extracted absolute value $|G_{M}|$. Moreover, taking account of  the simple
relation of the pQCD evolution in the spacelike and timelike regions, we can assume
that the enhancement in the timelike region suggests an additional, indirect
confirmation of the dominance of the soft spectator scattering mechanism in
the spacelike region. This might be true if the SCET FFs $f_{1,2}$
are not modified very strongly after analytical continuation from space like
to timelike regions.

\section{Conclusions}
\label{sec6}

In the present work, we studied the soft rescattering contribution to the nucleon Dirac and Pauli FFs. 
This work is motivated by phenomenological studies of nucleon FFs suggesting that in the $Q^2$ range 5 - 10~GeV$^2$, the nucleon FFs are not yet dominated by a hard scattering mechanism involving three active quarks, interacting via hard two-gluon exchange.
In the soft rescattering picture studied here, as first suggested by Feynman, the highly virtual photon interacts with one active quark, whereas the other spectator quarks remain soft. 
Such a picture is characterized by two large scales~: 
the hard scale $Q^2$, representing the virtuality of the hard photon probe, and the hard-collinear 
scale $\Lambda Q$ (with $\Lambda$ a soft scale of order~$\sim 0.5 $~GeV), corresponding to the virtuality of the active, so-called hard-collinear quark.

By way of example, we started our investigation by calculating within perturbation theory the soft rescattering contributions to the nucleon FFs.  Within such perturbative calculation, 
the three collinear quarks in the initial nucleon wave function are connected to the active hard-collinear quark and two remaining soft quarks through (hard-collinear) two-gluon exchange. Analogously, the hard collinear quark after the interaction with the hard photon also scatters with the remaining two soft quarks through 
(hard-collinear) two-gluon exchange. 
For the Dirac FF $F_1$, this analysis overlaps with previous work in the literature, whereas  for the Pauli FF $F_2$ it has been performed for the first time here. 
We have demonstrated that such a specific two-loop contribution to the nucleon FFs gives the same  scaling behavior as the hard region, involving hard two-gluon exchange, i.e., 
$F_1 \sim 1/Q^4$, and $F_2 \sim 1/Q^6$. Furthermore, the perturbative calculation 
suggests a factorization formula for the FFs in terms of nucleon distribution amplitudes, describing how the collinear quarks make up the initial and final nucleon, a hard scattering process on the active quark, and a soft correlation function describing the propagation of the remaining two soft spectator quarks. 

The specific perturbative calculation demonstrates that a description of the soft rescattering mechanism could be carried out in general in two steps. First, one integrates over hard fluctuations (of order $Q^2$), leaving only hard-collinear virtualities (of order $\Lambda Q$) and soft virtualities 
(of order $\Lambda ^2$).  For large enough scale $Q$, such that $\Lambda Q \gg \Lambda^2$, 
one can then further use perturbation theory and also factorize hard-collinear fluctuations leaving at the end only collinear and soft modes, describing the soft QCD dynamics. 

The possibility of such a two-step factorization, with the aim of developing a systematic approach of the soft contribution in the case of nucleon form factors, was addressed for the first time in this paper. A similar approach has also been considered recently for inclusive cross sections 
in \cite{Chay:2010hq}. 
The first step corresponds with the matching of full QCD onto the soft collinear effective field theory at a factorization scale $\mu^2_I = Q^{2}$, and denoted by SCET$_I$. Technically, we have demonstrated this step by calculating the leading-order hard coefficient functions in front of the operators constructed from SCET$_I$ fields, corresponding with the Dirac and Pauli FF structures. These leading-order hard coefficient functions involve the emission of 
hard-collinear transverse gluons, comoving with the active quark. We subsequently resummed the 
large logarithms of order $\sim \ln Q^2 / \mu_I^2$, 
which appear when evolving the SCET$_I$ operators from the 
hard scale $Q^2$ down to the scale $\mu_I^2\sim Q\Lambda$. 
Both for the leading e.m. current operator structure, corresponding with the Dirac FF $F_1$, 
and the subleading operator structure, corresponding with the Pauli FF $F_2$, we solved the renormalization group equations for the 
corresponding coefficient functions, and obtained the NLL solution. 
This provides a practical check that to NLL accuracy the first-step factorization (so-called SCET$_I$ factorization) for both $F_1$ and $F_2$ indeed holds. 

We next discussed the further matching of the SCET$_I$ theory to the effective theory involving only collinear and soft particles (so-called SCET$_{II}$), defined at a factorization scale 
$\mu^2_{II} = Q\Lambda$. As a first step to arrive at such a full factorization 
formula for the soft rescattering contribution, we analyzed in this work 
the leading terms in the effective theory.  
The factorization formula involves two so-called jet functions, 
describing the amplitude for the transition of 
three collinear quarks into a hard-collinear (active) quark and two soft quarks; a soft correlation function describing 
the soft rescattering of the two soft  spectator
quarks in the background soft-gluon fields emitted by the hard-collinear (active) quark; and the two nucleon distribution amplitudes, 
describing how the three initial and final collinear quarks make up the nucleons. 
The jet functions  can be computed performing the matching from SCET$_{I}$ operators onto SCET$_{II}$  at the factorization scale $\mu_{II}^2=Q\Lambda$. 
Also here large logarithms $\sim \ln \Lambda Q / \mu_{II}^2$ arise, when we evolve the factorization scale $\mu_{II}^2$ down to value of order $\Lambda^{2}$.  They can be resummed again using RG equations. We leave this consideration  to a future work.  

For the Pauli FF $F_2$ we also discussed that an analysis is more involved as there may be a double counting between the hard and soft rescattering mechanisms. Furthermore the matching from SCET$_I$ onto SCET$_{II}$ does not yield a well defined expression for the Pauli FF, due to end-point singularities, which calls for a more refined treatment for $F_2$ in a future publication. 

The SCET$_I$ factorization formulas allowed us already to discuss some phenomenological consequences in this work. For the soft rescattering contribution to the $Q^{2} F_2 / F_1$ ratio, we found that the ratio of the next-to-leading order evolution kernels changes only by a few percent in the range $Q^2 \simeq 4 - 16$~GeV$^2$, and is mainly dominated by SCET$_I$ FFs defined at a corresponding scale 
$Q^2_I \sim \Lambda Q \simeq 0.8 - 1.6$~GeV$^2$. Such scale is quite small to expect the asymptotic constant behavior. 
The experimental data for the $Q^{2}F_2/F_1$ ratio in this $Q^2$ range indeed show rising behavior, in agreement with the above analysis. 
A second phenomenological consequence of our framework was discussed for the ratio of the spacelike to timelike FF $F_1$. We showed that the resummed Sudakov logarithms provide a 30\% - 40 \% enhancement to this ratio in the range of momentum transfers $q^2$ around 10 GeV$^2$. This enhancement is in qualitative agreement with the empirical extracted ratio for the absolute value of the dominant FF $G_M$ in the timelike as compared to the spacelike region.
A more detailed phenomenological analysis requires us to parametrize the SCET$_I$ FFs, 
which is equivalent to using the SCET$_{II}$ factorization formula, and express them in terms of DAs, jet functions, and a two-quark soft correlation function, as outlined in this work. 
Such an analysis remains a challenge for a future work.

%%%PRDv2
\section*{ Appendix A.  Leading order coefficient functions}

Here we discuss in detail   calculation of leading-order hard coefficient functions. 
First,  Eq. (\ref{Jem:match}) can be rewritten in compact form in momentum
space,%
\begin{equation}
\bar{q}(0)\gamma^{\mu}q(0)=~\text{tr}\left[  C^{\mu}(Q,\mu)~O^{q}(0)\right]
+\int_{0}^{1}d\tau~\text{tr}\left[  ~C_{\bar{n}}^{\mu}(\tau,Q,\mu)~O_{\bar{n}%
}^{q}(\tau)+C_{n}^{\mu}(\tau,Q,\mu)~O_{n}^{q}(\tau)~\right]
+~...\label{Jem:mom}%
\end{equation}
where we used translation invariance and defined the momentum space
coefficients as%
\begin{equation}
C^{\mu}(Q,\mu)~=\int d\hat{s}_{1}d\hat{s}_{2}~\tilde{C}^{\mu}(\hat{s}_{1}%
,\hat{s}_{2})~e^{i(P^{\prime}\cdot\bar{n})s_{2}-i(P\cdot n)s_{1}},
\end{equation}%
\begin{equation}
C_{\bar{n}}^{\mu}(\tau,Q,\mu)~=\int d\hat{s}_{1}d\hat{s}_{2}d\hat{s}%
_{3}~e^{-i(P\cdot n)s_{1}}~e^{i(P^{\prime}\cdot\bar{n})~\tau s_{3}%
+i(P^{\prime}\cdot\bar{n})~\bar{\tau}s_{2}}~\tilde{C}_{\bar{n}}^{\mu}(\hat
{s}_{1},\hat{s}_{2},\hat{s}_{3}),
\end{equation}%
\begin{equation}
C_{n}^{\mu}(\tau,Q,\mu)=\int d\hat{s}_{1}d\hat{s}_{2}d\hat{s}_{3}%
~e^{i(P^{\prime}\cdot n)s_{2}}~e^{-i(P\cdot n)~\tau s_{3}-i(P\cdot
n)~\bar{\tau}s_{1}}~\tilde{C}_{n}^{\mu}(\hat{s}_{1},\hat{s}_{2},\hat{s}_{3}),~
\end{equation}
Here $P$ and $P^{\prime}$ denote the total hard-collinear momentum of the external
state for each jet, and $(P\cdot n)=(P^{\prime}\cdot\bar{n})=Q$ is the large
component of each momentum. The variable $\tau$ is the fraction of large
momentum component $(P^{\prime}\cdot\bar{n})$ [$(P\cdot n)$] carried by
the hard-collinear gluon $\mathcal{A}_{\bot}^{\prime}$ ($\mathcal{A}_{\bot}$),
$\bar{\tau}\equiv1-\tau$. The objects $O_{\bar{n}}^{q}(\tau)$ and $O_{n}%
^{q}(\tau)$ denote the Fourier transformed SCET operators:%
\begin{equation}
O_{\bar{n}}^{q}(\tau)=\int\frac{d\hat{s}_{3}}{2\pi}~e^{-is_{3}(P^{\prime}%
\cdot\bar{n})\tau}~\tilde{O}_{\bar{n}}^{q}(0,0,s_{3}),~\ \ \ \ O_{n}^{q}%
(\tau)=\int\frac{d\hat{s}_{3}}{2\pi}~e^{is_{3}(P\cdot n)\tau}~\tilde{O}%
_{n}^{q}(0,0,s_{3}).
\end{equation}

The tree-level coefficient functions in momentum space \ can be obtained from an
analysis of the matrix elements in QCD and SCET. \ In order to compute
$C^{\mu}(Q,\mu)$ defined in Eq.~(\ref{Jem:mom}) consider the matrix element of the e.m.
current between collinear quark states:%
\begin{equation}
\left\langle p^{\prime}\left\vert \bar{q}(0)\gamma^{\mu}q(0)\right\vert
p\right\rangle =\left\langle p^{\prime}\left\vert \text{tr}\left[  C^{\mu
}(Q,\mu)~O^{q}(0)\right]  \right\vert p\right\rangle .~\label{C:calc}%
\end{equation}
The subleading term in Eq.~(\ref{Jem:mom}) does not contribute in this case. Then
for the matrix element at LO we obtain:%
\begin{equation}
\text{{\it lhs} of Eq.~(\ref{C:calc}):~\ }\left\langle p^{\prime}\left\vert \bar
{q}(0)\gamma^{\mu}q(0)\right\vert p\right\rangle _{\text{LO}}=\bar{u}^{\prime
}\gamma^{\mu}u=\bar{\xi}^{\prime}\gamma_{\bot}^{\mu}\xi+\mathcal{O}%
(1/Q),\label{CARHS}%
\end{equation}%
\begin{equation}
\text{{\it rhs} of Eq.~(\ref{C:calc}):}~~\left\langle p^{\prime}\left\vert \text{tr}\left[
C^{\mu}(Q,\mu)~O^{q}(0)\right]  \right\vert p\right\rangle =\text{tr}\left[
C^{\mu}(Q,\mu)~\bar{\xi}^{\prime} \otimes \xi\right]  ,\label{CALHS}%
\end{equation}
where $\xi^{\prime},\xi$ without subscript $hc$ denote large components of
Dirac spinors (\ref{xi-out}) and (\ref{xi-in}). Comparison (\ref{CARHS}) and
(\ref{CALHS}) yields%
\begin{equation}
C^{\mu}(Q,\mu)=\gamma_{\bot}^{\mu}~\delta_{F^{\prime}F}+\mathcal{O}(\alpha
_{s}).
\end{equation}
where $F^{\prime},F$ describe quark color indices.

In order to compute the subleading coefficient functions one has to consider the
matrix element with the quark-gluon external state. We consider an outgoing gluon
with hard-collinear momentum $q^{\prime}$ collinear to $p^{\prime}$ and for
simplicity we neglect the transverse components of the outgoing momenta. Then
we can compute the leading-order contribution to $C_{\bar{n}}^{\mu}(\tau,Q,\mu)$.

\begin{figure}[th]
\begin{center}
\includegraphics[
natheight=0.919300in,
natwidth=2.637700in,
height=0.9193in,
width=2.6377in
]{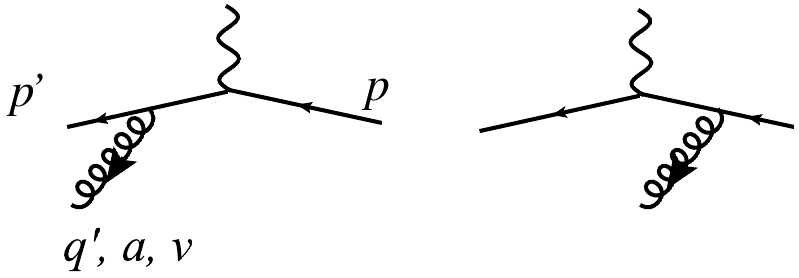}
\end{center}
\caption{QCD tree diagrams required for matching onto the subleading 3-particle
operator}%
\label{qcd-matching-to-3particle-diagram}%
\end{figure}

We start by considering the QCD calculation. The corresponding diagrams are shown in
Fig.\ref{qcd-matching-to-3particle-diagram}. 
For the first graph we have
\begin{eqnarray}
D_{1}=(ig)~\bar{u}(p^{\prime}) \setbox0=\hbox{${A} $} \dimen0=\wd0
\setbox1=\hbox{/} \dimen1=\wd1 \ifdim\dimen0>\dimen1 \rlap{\hbox to
\dimen0{\hfil/\hfil}} {A} \else \rlap{\hbox to \dimen1{\hfil${A} $\hfil}} /
\fi ^{\prime}\frac{i(\hat p^{\prime}+\hat q^{\prime})}{(p^{\prime}+q^{\prime
})^{2}}\gamma^{\mu}u(p) &\simeq& ~(ig) \bar{\xi}^{\prime} \setbox0=\hbox{${A} $}
\dimen0=\wd0 \setbox1 = \hbox{/} \dimen1=\wd1 \ifdim\dimen0>\dimen1
\rlap{\hbox to \dimen0{\hfil/\hfil}} {A} \else \rlap{\hbox to
\dimen1{\hfil${A} $\hfil}} / \fi ^{\prime} \left[  \frac{ \setbox0=\hbox{$n$}
\dimen0=\wd0 \setbox1=\hbox{/} \dimen1=\wd1 \ifdim\dimen0>\dimen1
\rlap{\hbox to \dimen0{\hfil/\hfil}} n \else \rlap{\hbox to
\dimen1{\hfil$n$\hfil}} / \fi }{2}\frac{i}{(p^{\prime}+q^{\prime})\cdot
n}+\frac{ \setbox0=\hbox{$\bar n$} \dimen0=\wd0 \setbox1=\hbox{/} \dimen1=\wd1
\ifdim\dimen0>\dimen1 \rlap{\hbox to \dimen0{\hfil/\hfil}} \bar n
\else \rlap{\hbox to \dimen1{\hfil$\bar n$\hfil}} / \fi }{2} \frac
{i}{(p^{\prime}+q^{\prime})\cdot\bar{n}}\right]  \gamma^{\mu}\xi\nonumber\\
&=&~(ig)\ \bar{\xi}^{\prime} \left[  \setbox0=\hbox{$ {A} $} \dimen0=\wd0
\setbox1=\hbox{/} \dimen1=\wd1 \ifdim\dimen0>\dimen1 \rlap{\hbox to
\dimen0{\hfil/\hfil}} {A} \else \rlap{\hbox to \dimen1{\hfil$ {A} $\hfil}} /
\fi _{\bot}^{\prime} \frac{i\bar{n}^{\mu}}{(p^{\prime}+q^{\prime})\cdot\bar
{n}}\right]  \xi.
\end{eqnarray}
For clarity, we write $A_{\mu}$ for the external gluon line with momentum
$q^{\prime}$ instead of polarization $\varepsilon^{\ast}(q^{\prime})$. The
second diagram:%
\begin{eqnarray}
D_{2}=(ig)\bar{u}(p^{\prime})\gamma^{\mu}\frac{i(\hat p-\hat q^{\prime}%
)}{(p-q^{\prime})^{2}} \setbox0=\hbox{$ {A} $} \dimen0=\wd0 \setbox1=\hbox{/}
\dimen1=\wd1 \ifdim\dimen0>\dimen1 \rlap{\hbox to \dimen0{\hfil/\hfil}} {A}
\else \rlap{\hbox to \dimen1{\hfil$ {A} $\hfil}} / \fi ^{\prime}%
u(p)
&\simeq& (ig)~\bar{\xi}^{\prime}\gamma^{\mu}\left[  \frac{
\setbox0=\hbox{$\bar n$} \dimen0=\wd0 \setbox1=\hbox{/} \dimen1=\wd1
\ifdim\dimen0>\dimen1 \rlap{\hbox to \dimen0{\hfil/\hfil}} \bar n
\else \rlap{\hbox to
\dimen1{\hfil$\bar n$\hfil}} / \fi }{2} \frac{i}{-(q^{\prime}\cdot\bar{n}%
)}+\frac{ \setbox0=\hbox{$n$} \dimen0=\wd0 \setbox1=\hbox{/} \dimen1=\wd1
\ifdim\dimen0>\dimen1 \rlap{\hbox to \dimen0{\hfil/\hfil}} n
\else \rlap{\hbox to \dimen1{\hfil$n$\hfil}} / \fi }{2}\frac{i}{(p\cdot
n)}\right]  ~ \setbox0=\hbox{$ {A} $} \dimen0=\wd0 \setbox1=\hbox{/}
\dimen1=\wd1 \ifdim\dimen0>\dimen1 \rlap{\hbox to \dimen0{\hfil/\hfil}} {A}
\else \rlap{\hbox to \dimen1{\hfil$ {A} $\hfil}} / \fi ^{\prime}\xi
\nonumber \\
&=&~\bar{\xi
}^{\prime}\left[  ~(ig)~ \setbox0=\hbox{${A}$} \dimen0=\wd0 \setbox1=\hbox{/}
\dimen1=\wd1 \ifdim\dimen0>\dimen1 \rlap{\hbox to \dimen0{\hfil/\hfil}} {A}
\else \rlap{\hbox to \dimen1{\hfil${A}$\hfil}} / \fi _{\bot}^{\prime}%
\frac{in^{\mu}}{(p\cdot n)}\right]  \xi.
\end{eqnarray}
Therefore the sum reads%
\begin{equation}
D_{1}+D_{2}=\bar{\xi}^{\prime}\left[  ~(-g)~ \setbox0=\hbox{${A}$}
\dimen0=\wd0 \setbox1=\hbox{/} \dimen1=\wd1 \ifdim\dimen0>\dimen1
\rlap{\hbox to
\dimen0{\hfil/\hfil}} {A} \else \rlap{\hbox to \dimen1{\hfil${A}$\hfil}} /
\fi _{\bot}^{\prime}~\left\{  \frac{n^{\mu}}{(p\cdot n)}+\frac{\bar{n}^{\mu}%
}{(p^{\prime}+q^{\prime})\cdot\bar{n}}\right\}  \right]  \xi, \label{D12}
\end{equation}
which involves the transverse gluon field and the longitudinal
projection of the e.m. current, as required. It is easy to see that
the obtained kinematical structure is e.m. gauge invariant. This term must be
compared with the SCET matrix element,
\begin{equation}
~\left\langle p^{\prime},q^{\prime}\left\vert \int_{0}^{1}d\tau~\text{tr}%
\left[  ~C_{\bar{n}}^{\mu}(\tau,Q,\mu)~O_{\bar{n}}^{q}(\tau)\right]
\right\vert p\right\rangle =\int_{0}^{1}d\tau~~\text{tr}~[C_{\bar{n}}^{\mu
}(\tau,Q,\mu)~\left\langle p^{\prime},q^{\prime}\left\vert ~O_{\bar{n}}%
^{q}(\tau)\right\vert p\right\rangle ],\label{me1}%
\end{equation}
with%
\begin{equation}
\left\langle p^{\prime},q^{\prime}\left\vert ~O_{\bar{n}}^{q}(\tau)\right\vert
p\right\rangle \simeq\bar{\xi}^{\prime} \setbox0=\hbox{${A}$} \dimen0=\wd0
\setbox1=\hbox{/} \dimen1=\wd1 \ifdim\dimen0>\dimen1 \rlap{\hbox to
\dimen0{\hfil/\hfil}} {A} \else \rlap{\hbox to \dimen1{\hfil${A}$\hfil}} /
\fi _{\bot}^{\prime}\xi~~\delta\left(  \frac{\left(  q^{\prime}\cdot\bar
{n}\right)  }{(p^{\prime}+q^{\prime})\cdot\bar{n}}-~\tau\right)  .
\end{equation}
Substituting this into Eq.~(\ref{me1}) and comparing this with the QCD result
of Eq.~(\ref{D12}) we obtain%
\begin{equation}
C_{\bar{n}}^{\mu}(\tau,Q,\mu)=-\frac{1}{Q}\left(  n^{\mu}+\bar{n}^{\mu
}\right)  ~\hat{1}~\delta_{F^{\prime}F},
\end{equation}
where the symbol $\hat{1}$ denotes the unity operator in Dirac space. Notice that the obtained
coefficient function does not depend on the momentum fraction $\tau$ at the LO
level. A calculation of the second term with $C_{n}^{\mu}$ can be done in 
an analogous way. The result can also be obtained without 
explicit calculations by invoking time reversal invariance which demands the result to 
be symmetric under $n\leftrightarrow\bar{n}$.  

%%%%

\section*{ Appendix B. Correspondence between QCD and SCET calculations}

\begin{figure}[th]
\begin{center}
\includegraphics[
height=0.9556in,
width=1.1026in
]{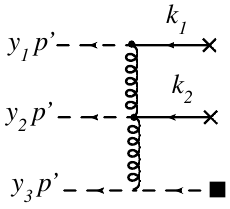}
\end{center}
\par
\caption{ One of the diagrams describing subprocess
(\ref{hs-to-3col}) in SCET$_{I}$. The soft fields are considered as external
fields,  and outgoing quarks are collinear. }%
\label{half-diagram}
\end{figure}

In order to illustrate the correspondence of SCET with QCD we here perform the
calculation of the jet functions discussed in Sec.~\ref{sec2}. We start from the
calculation of $\mathbf{J}^{\prime}$ defined in (\ref{Jdef}). In order to have
a direct correspondence with the expression (\ref{Jalf})%
\[
J_{(\alpha)}^{\prime}\left[  y_{i},\omega_{i}\right]  =\frac{1}{Q^{3}}\frac
{1}{y_{1}\bar{y}_{3}^{2}}\frac{1}{(\omega_{1}+\omega_{2})^{2}
(-\omega_{1})  }\left[  ~\bar{\xi}_{3}^{\prime}\right]  _{\alpha_{3}%
}~\left[  ~\bar{\xi}_{2}^{\prime}\gamma_{\bot}^{i}\right]  _{\alpha_{2}%
}\left[  \bar{\xi}_{1}^{\prime}\gamma_{\bot}^{i}\right]  _{\alpha_{1}},
\]
we consider the appropriate subprocess%
\begin{equation}
{\xi}_{hc}(p^{\prime}-k_{1}-k_{2})+q_{s}(k_{1})q_{s}%
(k_{2})\overset{\text{SCET}_{I}}{\rightarrow}\xi_{c}(y_{1}p^{\prime})~\xi
_{c}(y_{2}p^{\prime})~\xi_{c}(y_{3}p^{\prime}).\label{hs-to-3col}%
\end{equation}
described by the matrix element
\begin{equation}
J_{(\alpha)}^{\prime}~q_{\alpha_{1}}(k_{1})q_{\alpha_{2}}(k_{2})=\left\langle
y_{1}p^{\prime},y_{2}p^{\prime},y_{3}p^{\prime}\right\vert T\left(  \bar{\xi
}_{hc}^{\prime}W^{\prime}(0),\mathcal{L}_{\xi q}^{(1)},\mathcal{L}_{\xi
q}^{(1)},\mathcal{L}_{\xi}^{(0)}\right)  \left\vert 0\right\rangle .
\end{equation}
The soft quark fields $q_{s}$ are considered as external. In order to
reproduce the expression in Eq.~(\ref{Jalf}) we need the diagram shown in
Fig.\ref{half-diagram}. 

From the Lagrangians $\mathcal{L}_{\xi}^{(0)}$ and $\mathcal{L}_{q\xi}^{(1)}$ one
can easily define the Feynman rules. They were already presented in the
literature (see, e.g., Refs.~\cite{Bauer2000, Pirjol2002, Beneke:2005gs}). For the
convenience of the reader we reproduce the relevant vertices here. Taking account 
of only the required leading-order terms one obtains%
\begin{equation}%
\begin{tabular}
[c]{l}%
{\includegraphics[
natheight=0.583800in,
natwidth=0.947800in,
height=0.5838in,
width=0.9478in
]{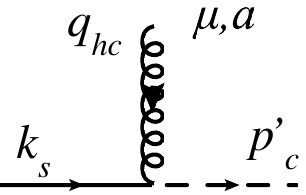}}%
\end{tabular}
\ \simeq ig\left[  T^{a}\right]  _{AB}\left[  \gamma_{\bot}^{\mu}\right]
_{\alpha\beta}~,~\
\begin{tabular}
[c]{l}%
{\includegraphics[
natheight=0.499000in,
natwidth=0.837100in,
height=0.499in,
width=0.8371in
]{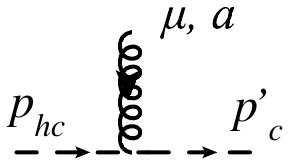}}%
\end{tabular}
\ \simeq ig\left[  T^{a}\right]  _{AB}\frac{ \setbox0=\hbox{$\bar n$}
\dimen0=\wd0 \setbox1=\hbox{/} \dimen1=\wd1 \ifdim\dimen0>\dimen1
\rlap{\hbox to
\dimen0{\hfil/\hfil}} \bar n \else \rlap{\hbox to \dimen1{\hfil$\bar
n$\hfil}} / \fi }{2}n^{\mu}.
\end{equation}%
\begin{equation}%
\begin{tabular}
[c]{l}%
{\includegraphics[
natheight=0.667600in,
natwidth=1.017900in,
height=0.6676in,
width=1.0179in
]{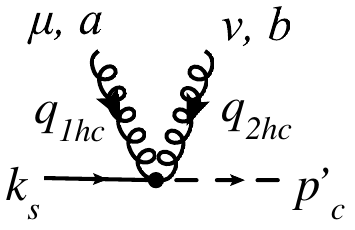}}%
\end{tabular}
\ \simeq-ig^{2}\left(  \left[  T^{a}T^{b}\right]  _{AB}\left[  \gamma_{\bot
}^{\mu}\right]  _{\alpha\beta}\frac{\bar{n}^{\nu}}{(q_{2}\cdot\bar{n}%
)}+\left[  T^{a}T^{b}\right]  _{AB}\left[  \gamma_{\bot}^{\nu}\right]
_{\alpha\beta}\frac{\bar{n}^{\mu}}{(q_{1}\cdot\bar{n})}\right)  ~.
\end{equation}
Assuming the same choice of momenta as in Fig.~\ref{d-quark-2-loop}, we obtain
the following analytical expression%
\begin{eqnarray}
J_{(\alpha)}^{\prime}q_{\alpha_{1}}(k_{1})q_{\alpha_{2}}(k_{2})  & \sim & 
\bar{\xi}_{1}^{\prime} \gamma_{\bot}^{\nu}T^{b}q(k_{1}) \bar{\xi}_{2}^{\prime
}\left\{  -\frac{\bar{n}^{\mu}\gamma_{\bot}^{\nu}}{\left(  p_{4g}\cdot\bar
{n}\right)  }T^{b}T^{a}\right\}  q(k_{2}) \bar{\xi}_{3}^{\prime}\frac{
\setbox0=\hbox{$\bar n$} \dimen0=\wd0 \setbox1=\hbox{/} \dimen1=\wd1
\ifdim\dimen0>\dimen1 \rlap{\hbox to \dimen0{\hfil/\hfil}} \bar n
\else \rlap{\hbox to \dimen1{\hfil$\bar n$\hfil}} / \fi \setbox0=\hbox{$n$}
\dimen0=\wd0 \setbox1=\hbox{/} \dimen1=\wd1 \ifdim\dimen0>\dimen1
\rlap{\hbox to \dimen0{\hfil/\hfil}} n \else \rlap{\hbox to
\dimen1{\hfil$n$\hfil}} / \fi }{4}\left\{  T^{a}n^{\mu}\right\}  \frac
{1}{(p_{3g}\cdot n)}\frac{1}{p_{3g}^{2}}\frac{1}{p_{4g}^{2}} \nonumber \\
&  \sim & \frac{1}{(k_{1}^{+}+k_{2}^{+})^{2}}~\frac{1}{\left(  -~k_{1}%
^{+}\right)  }~\frac{1}{Q^{3}}\frac{1}{y_{1}\bar{y}_{3}^{2}}~\bar{\xi}%
_{1}^{\prime}\gamma_{\bot}^{i}q(k_{1})~\bar{\xi}_{2}^{\prime}\gamma_{\bot}%
^{i}q(k_{2})~[\bar{\xi}_{3}^{\prime}]_{\alpha_{3}} \nonumber \\
&  = & \int d\omega_{1,2}~J_{(\alpha)}^{\prime}\left[  y_{i},\omega_{i}\right]
~~q_{\alpha_{1}}(k_{1})~q_{\alpha_{2}}(k_{2})~\delta(\omega_{1}-k_{1}%
^{+})\delta(\omega_{2}-k_{2}^{+}).
\end{eqnarray}
with the same $J_{(\alpha)}^{\prime}$ as in Eq.~(\ref{Jalf}). Consider the
color structure which we ignored in the calculation in Sec.~\ref{sec2}. Projecting
the color indices of the outgoing collinear quarks onto the colorless nucleon, 
we obtain~:
\begin{equation}
\frac{\varepsilon^{i^{\prime}j^{\prime}k^{\prime}}}{3!}\left[  T^{b}\right]
_{i^{\prime}i}\otimes\left[  T^{b}T^{a}\right]  _{j^{\prime}j}\otimes\left[
T^{a}\right]  _{k^{\prime}k}=\frac{2}{27}\varepsilon^{ijk}.
\end{equation}
The resulting antisymmetrical tensor $\varepsilon^{ijk}$ is then contracted with the
color indices of the soft fields yielding the soft operator in Eq.~(\ref{Snbold}). 

Consider now the helicity flip FF $F_{2}$. Again, from the QCD calculation we obtained
the result of Eq.~(\ref{F2:J'andJet}):
\begin{equation}
J_{(\alpha)}^{\prime}\left[  y_{i},\omega_{i}\right]  =\frac{1}{Q^{2}}%
\frac{\left[  \bar{\xi}_{1}^{\prime}\gamma_{\bot}^{\alpha}\right]
_{\alpha_{1}}\left[  \bar{\xi}_{2}^{\prime}\gamma_{\bot}^{\alpha}\right]
_{\alpha_{2}}~\left[  \bar{\eta}_{3}^{\prime}~\hat{n}\right]  _{\alpha_{3}}%
}{~\bar{y}_{1}y_{3}(\omega_{2}+\omega_{3})\omega_{3}^{2}},\label{J'F2}%
\end{equation}
In this case, we have to consider the presence of the small component $\bar{\eta}%
_{3}^{\prime}$. In SCET this field is eliminated by the equation of motion, 
yielding%
\begin{equation}
\bar{\eta}_{c}^{\prime}\simeq-~\bar{\xi}^{\prime}i\hat{D}_{\bot c}\frac{
\setbox0=\hbox{$\bar n$} \dimen0=\wd0 \setbox1=\hbox{/} \dimen1=\wd1
\ifdim\dimen0>\dimen1 \rlap{\hbox to \dimen0{\hfil/\hfil}} \bar n
\else \rlap{\hbox to \dimen1{\hfil$\bar n$\hfil}} / \fi }{2}(i\bar{n}%
\cdot\overleftarrow{D}_{c})^{-1}.
\end{equation}
In order to have a connection with the QCD~result of Eq.~(\ref{F2:J'andJet}), we must
substitute
\begin{equation}
\bar{\eta}_{c}^{\prime}=\bar{\xi}_{c}^{\prime}i\hat{D}_{\bot c}(i\bar
{n}\overleftarrow{D}_{c})^{-1}\frac{ \setbox0=\hbox{$\bar n$} \dimen0=\wd0
\setbox1=\hbox{/} \dimen1=\wd1 \ifdim\dimen0>\dimen1 \rlap{\hbox to
\dimen0{\hfil/\hfil}} \bar n \else \rlap{\hbox to \dimen1{\hfil$\bar
n$\hfil}} / \fi }{2}\simeq\bar{\xi}_{c}^{\prime}\hat{A}_{c\bot}(i\bar{n}%
\cdot\overleftarrow{\partial})^{-1}\frac{ \setbox0=\hbox{$\bar n$}
\dimen0=\wd0 \setbox1=\hbox{/} \dimen1=\wd1 \ifdim\dimen0>\dimen1
\rlap{\hbox to
\dimen0{\hfil/\hfil}} \bar n \else \rlap{\hbox to \dimen1{\hfil$\bar
n$\hfil}} / \fi }{2}.
\end{equation}
From this expression one can see that such a state consists of a collinear quark
and a transverse gluon, as is shown in Fig.~\ref{scet-pauliff-jpr-diagram}.
The corresponding vertex is generated by the Lagrangian $\mathcal{L}_{\xi q}^{(1)}$,
includes 2 gluons, and can be associated with the following combination:%
\begin{eqnarray}
\bar{\xi}_{hc}~i\hat{D}_{\bot}Wq  &  \simeq & \bar{\xi}_{c}~g \setbox0=\hbox{$A$}
\dimen0=\wd0 \setbox1=\hbox{/} \dimen1=\wd1 \ifdim\dimen0>\dimen1
\rlap{\hbox to \dimen0{\hfil/\hfil}} A \else \rlap{\hbox to
\dimen1{\hfil$A$\hfil}} / \fi _{\bot c}^{\prime}~\left[  \frac{i}{\left(
\bar{n}\partial\right)  }~ig\bar{n}\cdot A_{hc}\right]  q=~\bar{\xi}%
_{c}^{\prime}~g \setbox0=\hbox{$A$} \dimen0=\wd0 \setbox1=\hbox{/}
\dimen1=\wd1 \ifdim\dimen0>\dimen1 \rlap{\hbox to \dimen0{\hfil/\hfil}} A
\else \rlap{\hbox to \dimen1{\hfil$A$\hfil}} / \fi _{\bot c}^{\prime}(\bar
{n}\cdot\overleftarrow{\partial})^{-1}\frac{ \setbox0=\hbox{$\bar n$}
\dimen0=\wd0 \setbox1=\hbox{/} \dimen1=\wd1 \ifdim\dimen0>\dimen1
\rlap{\hbox to
\dimen0{\hfil/\hfil}} \bar n \else \rlap{\hbox to \dimen1{\hfil$\bar
n$\hfil}} / \fi }{2}~g\hat{A}_{hc}\frac{ \setbox0=\hbox{$\bar n$} \dimen0=\wd0
\setbox1=\hbox{/} \dimen1=\wd1 \ifdim\dimen0>\dimen1 \rlap{\hbox to
\dimen0{\hfil/\hfil}} \bar n \else \rlap{\hbox to \dimen1{\hfil$\bar
n$\hfil}} / \fi \setbox0=\hbox{$n$} \dimen0=\wd0 \setbox1=\hbox{/}
\dimen1=\wd1 \ifdim\dimen0>\dimen1 \rlap{\hbox to \dimen0{\hfil/\hfil}} n
\else \rlap{\hbox to \dimen1{\hfil$n$\hfil}} / \fi }{4}q(k_{3}) \nonumber \\
&  = & \left[  \bar{\eta}_{c}^{\prime} 
\Dslash{A}_{\bot c}^{\prime}(\bar{n}\cdot\overleftarrow{\partial})^{-1}\frac{
\setbox0=\hbox{$\bar n$} \dimen0=\wd0 \setbox1=\hbox{/} \dimen1=\wd1
\ifdim\dimen0>\dimen1 \rlap{\hbox to \dimen0{\hfil/\hfil}} 
\bar n \else \rlap{\hbox to
\dimen1{\hfil$\bar n$\hfil}} / \fi }{2}\right]  ~\hat{A}_{hc}\frac{
\setbox0=\hbox{$\bar n$} \dimen0=\wd0 \setbox1=\hbox{/} \dimen1=\wd1
\ifdim\dimen0>\dimen1 \rlap{\hbox to \dimen0{\hfil/\hfil}} \bar n
\else \rlap{\hbox to
\dimen1{\hfil$\bar n$\hfil}} / \fi \setbox0=\hbox{$n$} \dimen0=\wd0
\setbox1=\hbox{/} \dimen1=\wd1 \ifdim\dimen0>\dimen1 \rlap{\hbox to
\dimen0{\hfil/\hfil}} n \else \rlap{\hbox to \dimen1{\hfil$n$\hfil}} /
\fi }{4}~q(k_{3}).
\end{eqnarray}
Therefore, we obtain for the diagram in Fig.\ref{half-diagram}
(again ignoring color structures)~:%
\begin{eqnarray}
J_{(\alpha)}^{\prime}~q_{\alpha_{2}}(k_{2})q_{\alpha_{3}}(k_{3})  &
\sim & \left[  \bar{\xi}_{1}^{\prime}\gamma_{\bot}^{\mu}\right]  _{\alpha_{1}
}\bar{\xi}_{2}^{\prime}\left\{  n^{\nu}\right\}  \frac{
\setbox0=\hbox{$\bar n$} \dimen0=\wd0 \setbox1=\hbox{/} \dimen1=\wd1
\ifdim\dimen0>\dimen1 \rlap{\hbox to
\dimen0{\hfil/\hfil}} \bar n \else \rlap{\hbox to \dimen1{\hfil$\bar
n$\hfil}} / \fi \setbox0=\hbox{$n$} \dimen0=\wd0 \setbox1=\hbox{/}
\dimen1=\wd1 \ifdim\dimen0>\dimen1 \rlap{\hbox to \dimen0{\hfil/\hfil}} n
\else \rlap{\hbox to \dimen1{\hfil$n$\hfil}} / \fi }{4}\frac{1}{k_{3}^{+}%
}\gamma_{\bot}^{\mu}q(k_{2})~\bar{\eta}_{3}^{\prime}\gamma^{\nu}\frac{
\setbox0=\hbox{$\bar n$} \dimen0=\wd0 \setbox1=\hbox{/} \dimen1=\wd1
\ifdim\dimen0>\dimen1 \rlap{\hbox to \dimen0{\hfil/\hfil}} \bar n
\else \rlap{\hbox to
\dimen1{\hfil$\bar n$\hfil}} / \fi \setbox0=\hbox{$n$} \dimen0=\wd0
\setbox1=\hbox{/} \dimen1=\wd1 \ifdim\dimen0>\dimen1 \rlap{\hbox to
\dimen0{\hfil/\hfil}} n \else \rlap{\hbox to \dimen1{\hfil$n$\hfil}} /
\fi }{4}q(k_{3})~\frac{1}{y_{3}Qk_{3}^{+}}\frac{1}{\bar{y}_{1}Q(k_{2}%
^{+}+k_{3}^{+})} \nonumber \\
& \sim & \frac{1}{Q^{2}}\frac{\left[  \bar{\xi}_{1}^{\prime}\gamma_{\bot}%
^{\alpha}\right]  _{\alpha_{1}}\left[  \bar{\xi}_{2}^{\prime}\gamma_{\bot
}^{\alpha}\right]  _{\alpha_{2}}~\left[  \bar{\eta}_{3}^{\prime}~\hat
{n}\right]  _{\alpha_{3}}}{\bar{y}_{1}y_{3}~(k_{2}^{+}+k_{3}^{+})\left(
k_{3}^{+}\right)  ^{2}} q_{\alpha_{2}}(k_{2})~q_{\alpha_{3}}(k_{3}) \nonumber \\
&  = & \int d\omega_{1,2}~J_{\mathbf{\alpha}}^{\prime}\left[  y_{i},\omega
_{i}\right]  ~~q_{\alpha_{1}}(k_{1})~q_{\alpha_{2}}(k_{2})~\delta(\omega
_{1}-k_{1}^{+})\delta(\omega_{2}-k_{2}^{+}),
\end{eqnarray}
with the same $J_{(\alpha)}^{\prime}$ \ as in Eq.~(\ref{J'F2}).

Using these two examples, we demonstrated that the SCET correctly reproduces the 
tree-level hard-collinear subprocesses computed before in QCD.

\end{document}